\documentclass[fleqn,usenatbib]{mnras}

\usepackage{newtxtext,newtxmath}

\usepackage[T1]{fontenc}

\DeclareRobustCommand{\VAN}[3]{#2}
\let\VANthebibliography\thebibliography
\def\thebibliography{\DeclareRobustCommand{\VAN}[3]{##3}\VANthebibliography}

\usepackage{enumitem}
\usepackage{graphicx}	
\usepackage{amsmath}	
\usepackage{amssymb}	

\title[AGN light curves cadence selection]{On possible proxies of AGN light curves cadence selection in future time domain surveys}

\author[A. B. Kova{\v c}evi{\'c} et al.]{
Andjelka B. Kova{\v c}evi{\'c},$^{1,2}$\thanks{E-mail: andjelka@matf.bg.ac.rs(ABK)}
Dragana Ili{\'c},$^{1,3}$
Luka {\v C}. Popovi{\'c},$^{1,2,4}$
Viktor Radovi{\'c},$^{1}$
Isidora Jankov,$^{1}$
Ilsang Yoon, $^{5}$
\newauthor{
Neven Caplar,$^{6}$
 Iva {\v C}vorovi{\'c}-Hajdinjak,$^{1}$
 Sa{\v s}a Simi{\'c}$^{7}$}
\\
$^{1}$Department of astronomy, Faculty of mathematics, University of Belgrade, Studentski trg 16, 11000 Belgrade, Serbia\\
$^{2}$Fellow of Chinese Academy of Sciences President's International Fellowship Initiative (PIFI) for visiting scientist\\
$^{3}$ Humboldt Research Fellow, Hamburger Sternwarte, Universit{\"a}t
Hamburg, Gojenbergsweg 112, 21029 Hamburg, Germany\\
$^{4}$Astronomical Observatory, Volgina 7, 11000 Belgrade, Serbia   \\
$^{5}$The National Radio Astronomy Observatory, 520 Edgemont Road Charlottesville, VA 2290, USA\\
$^{6}$Astrophysical Department, Princeton University,
4 Ivy Lane, Princeton, NJ, USA\\
$^{7}$Faculty of Science, University of Kragujevac, Radoja Domanovi\'ca 12,
34000 Kragujevac, Serbia 
}
\date{Accepted XXX. Received YYY; in original form ZZZ}

\pubyear{2021}

\begin{document}
\label{firstpage}
\pagerange{\pageref{firstpage}--\pageref{lastpage}}
\maketitle

\begin{abstract}
Motivated by upcoming photometric and spectroscopic surveys (Vera C. Rubin Observatory Legacy Survey of Space and Time (LSST),  Manuakea Spectroscopic Explorer), we design the statistical proxies to measure the cadence effects on active galactic nuclei (AGN) variability-observables (time-lags, periodicity, and structure-function (SF)).
 We constructed a multiple-regression model 
to statistically identify the cadence-formal error pattern knowing AGN time-lags and periodicity from different surveys. We defined the simple metric for the SF's properties, accounting for the 'observed' SF's deviation relative to those obtained from the homogenously-sampled light curves.
We tested the regression models on different observing strategies: the optical dataset of long light-curves of eight AGN with  peculiarities and the artificial datasets based on several idealized and LSST-like cadences. The  SFs metric is assessed on synthetic datasets.  
 The regression models (for both data types) predict similar cadences for time-lags and oscillation detection, whereas for light curves with low variability ($\sim 10\%$), cadences for oscillation detection differ. For  higher variability ($\sim20\%$), predicted cadences are larger than  for  $F_{var}\sim 10\%$. The predicted cadences  are decreasing with redshift.
  SFs with dense and homogenous cadences are more likely to behave similarly.   SFs with oscillatory signals are sensitive to the cadences, possibly impacting  LSST-like operation strategy.
  The proposed proxies can help to select spectroscopic and photometric-surveys cadence strategies, and they will be tested further in larger samples of objects.

\end{abstract}

\begin{keywords}
 galaxies: active--quasars: supermassive black holes-- methods:numerical-- surveys--telescopes
\end{keywords}



\section{Introduction}
The  extreme limits of  physical conditions in the universe are set by an event horizon of  a supermassive black hole (SMBH) in the centers of an active galactic nuclei (AGNs). 
Particularly, information {of} the {   size and structure of} broad line region (BLR) surrounding SMBH comes mainly from the optical reverberation mapping (RM) campaigns
 \citep{1998ApJ...501...82P,10.1086/423269,  2001A&A...376..775S,2004A&A...422..925S,
	2008A&A...486...99S,2010A&A...509A.106S,
	2010A&A...517A..42S,2012ApJS..202...10S,
	2013A&A...559A..10S,2016ApJS..222...25S,
	2017MNRAS.466.4759S,2019MNRAS.485.4790S,2008ApJ...689L..21B,2009ApJ...705..199B,
	2009ApJ...704L..80D,2011ApJ...743L...4B,2013ApJ...769..128B,2015ApJS..217...26B,
	2011A&A...528A.130P,2014A&A...572A..66P,
	2012ApJ...755...60G,2014ApJ...793..108W,2014ApJ...782...45D,10.1088/0004-637X/806/1/22,2016ApJ...825..126D,2018ApJ...856....6D,2016ApJ...818...30S,2017ApJ...851...21G, 2019ApJ...870..123E,10.1038/s41550-019-0979-5}. 
Optical RM has been used for  the mass determination of  SMBH in broad emission-line AGNs,   {so-called type 1 AGNs}. 
  {In contrast to studies that consider the nearby quiescent galaxies, for most AGNs spatially resolved observations are not possible. However, RM allows a time-domain investigation of the SMBH influence on its surroundings through the spectroscopic monitoring of the variability of continuum and the lagged response -reverberation - of broad emission lines.}
Most of the data suffer from 'Static Illusion', because what we can observe may change on timescales larger than centuries  \citep[$10^{4}-10^{7}$yr,][]{10.1007/978-94-010-0320-9}.

The problem is more apparent in the periodicity detection in the light curves of AGNs.
{Most} AGN light curves are consistent with random (aperiodic, stochastic, noise) variations  called 'red noise' caused by accretion processes which can mimic periodic behavior \citep[see e.g.][and reference therein]{10.1093/mnras/stw1412} {    which makes periodic oscillations more difficult to detect with classical statistical methods} \citep{2010MNRAS.402..307V}. 
{Most AGNs exhibit up to 0.2 mag rms stochastic variability in the optical band, with a long tail of extreme variability objects (>0.5 mag rms), so-called changing-look AGNs}
\citep[CLAGN,see e.g.][]{1984PAZh...10..803L, 1985A&A...146L..11K, 10.1088/0004-637X/796/2/134,10.3847/1538-4357/aab88b, 2019MNRAS.485.4790S, Ilic20}.
The RM campaigns have shown that {   for a higher accuracy of time-lag measurement, an optimal combination of the following parameters are needed: i) higher temporal resolution (i.e. cadences), ii) longer duration of light curve, iii) higher signal-to-noise (i.e. better flux measurement accuracy), and iv) higher level of variability \citep{10.3847/1538-4357/ab40cd}.}
Precise specification of required sampling rate depends on physical time scales of the source (e.g. minimum and maximum timescales of interest, such as the BLR light-crossing time) and the character of variations \citep{10.1086/420755}, which in most cases resemble red noise. 
{High accuracy of flux measurement allows us to better examine the level of variability. We expect an improved accuracy of  time-lag measurement  for objects with larger variability.}
Finally, the number of data points for a given time lag is important, as denser light curve sampling improves {   the prominence of light curve features such as `peaks' and `valleys', and may help} understand any possible  trend in the light curve. Thus cadence can be {   probed} by {   a statistical metric} which will correlate with measured time-lag errors \citep{10.3847/1538-4357/ab40cd}.
{Similarly, the detection} of possible periodic {   oscillations} in the RM light curves 
{   would be} improved with {denser} sampling rates, {higher} flux measurement accuracy, larger variability \citep{2012ApJ...759..118B, 2015Natur.518...74G, 2015Natur.525..351D, 2015ApJ...814L..12J, li16,   bon16, 2017MNRAS.470.1198D, 2018MNRAS.476.4617C, 2017Ap&SS.362...31K, 2018MNRAS.475.2051K, 10.3847/1538-4357/aaf731,10.1093/mnras/staa737}.   As in case of time-lags, a metric based on the {measured} light-curve properties could be used to estimate cadences necessary for the detection of periodic oscillations.

{Motivated by upcoming photometric surveys over broad range of wavelengths which will {periodically} observe a {large} fraction of the sky with good photometric accuracy (e.g., the Vera C. Rubin Observatory Legacy Survey of Space and Time, LSST), \citet{10.1088/0004-637X/747/1/62} show that  photometric reverberation mapping of the BLR is feasible. They prove that  the different variability properties of continuum and line processes can be  separated  at the light curve level. {Hence, in this work,} our empirical analysis of different LSST {cadences} builds on this photometric reverberation mapping proof-of-concept. {However,
{the statistical analysis of the photometric RM data} is {not limited to the LSST survey} and  can be adopted easily by other {surveys for example, the forthcoming Manuakea Spectroscopic Explore (MSE).}}  } 
{{The MSE} will expand the observed space in both debth, time and wavelength \citep{2019arXiv190404907T}. Apart from having a 11.25m telescope aperture, another special feature of the design is the state-of-art multi-object-spectrograph that will cover also the infrared spectral band, and thus be able to provides simultaneous coverage of C IV and H$\beta$ for quasars up to $\sim 2.5$. This would be particularly interesting for the reverberation mapping of quasar, since the MSE will significantly improve the depth and redshift range of H$\beta$ line coverage \citep{2019BAAS...51c.274S}. It would be important to {assess} the cadence strategies for the MSE reverberation campaign of about $\sim$5000 quasars, which will be spanned over a period of several years. {Therefore, to prepare the operation and maximize the science output, for the future large scale time domain surveys, we need to develop} universal statistical proxies for evaluation of different cadence strategies.}

In order to  contribute towards quantification of  the products  observing strategies in the era of large synoptic time domain surveys such as LSST, we examine the relationship between  the AGN variability related  observables (time lag, periodic oscillations and structure function) and {the characteristics of the 'LSST-like' observations and survey cadences. }
 {The selection of cadences in different {observing} strategies of the LSST-like surveys will affect the accuracy of time-lag estimates from AGN light curves, as well as the detection of periodic oscillations. The {latter} defines our ability to detect candidates of Close Binary Super Massive Black Holes (CB-SMBH) which will be the targets of the next generation of nano-Hertz gravitational observatories \citep{10.1007/s00159-019-0115-7}. With the anticipated number of millions of AGNs that will be observed and newly discovered by LSST, it is not viable to spectroscopically follow up of all of them. Thus, being able to photometrically detect possible periodic signals will allow us to harness the power of the datasets LSST will provide and further contribute to the multimessenger astronomy. 
}
{One of the commonly used approaches for AGN variability study is} structure function (SF) {analysis} \citep[see e.g.,][and references therein]{10.1088/0004-637X/721/2/1014,10.3847/1538-4357/834/2/111}, firstly introduced
by \cite {10.1086/163418}.  
{Therefore, in this work we aim to show impacts  and associated relation of different { {observing cadence}, both from the previous realistic observations and the} proposed observing {cadence} from the { LSST Operations Simulator (OpSim) outputs} \citep{2020DPS....5211002J}, on  detection of AGN variability realted observables. }
 For this analysis, we employ two types of data: {a compilation of the data} from 2-3 decades long AGN monitoring campaigns; and {a suit of artificial light curves}. The uniform set of the  light curves of type 1 AGN was collected during  very long RM  campaigns (up to 3 decades) presented in  \citet{2001A&A...376..775S,2004A&A...422..925S,
2008A&A...486...99S,2010A&A...509A.106S,
2010A&A...517A..42S,2012ApJS..202...10S,
2013A&A...559A..10S,2016ApJS..222...25S, 2017MNRAS.466.4759S,2019MNRAS.485.4790S}.
{{The artificial light curve datasets comprises two subsets: simulations with} a) {idealized 1 day uniform observing} {cadence and} b) LSST OpSim outputs.}
{The idealized artificial data sets} are generated based on Damped Random Walk \citep[DRW,][]{10.1088/0004-637X/698/1/895}. {For testing  oscillation detection the  certain periods are introduced {in light curve simulations.}}. Their cadences  correspond to several idealized observing strategies. {Also, in similar manner we constructed light curves for}  different runs from the LSST OpSim outputs \citep{2020DPS....5211002J}.
By constructing proxy variables and then applying them to these two sets of light curves, one obtains statistical features that allow us to predict the suitability of different cadences for time-lag estimations, for very delicate detection of oscillation in light curves, as well as for reproducing the SF properties.
{   {The main goal} of this study is to put constraints on the cadence of the LSST and similar surveys required to achieve a requisite level of time lag uncertainty and periodicity detection, and reconstruction of SF properties}.

To find these constrains, we first develop a suite of {statistics} (i.e. metrics) to quantify {an efficacy of the LSST observation with} different cadences (or sampling rates) for estimation of time-lags, detection of CB-SMBH candidates, and SF properties. These are all explained in detail in Section 2, in which we also describe the used data, i.e. the compilation of observed very long AGN light curves, and the generation of artificial set of 10yr-long light curves. 
Then, in Section 3 we calculate the statistics and characteristics of  detectable time-lags and oscillatory signals, {   and give estimates for their error predictions usable {for future time domain surveys}.}. {In Section 4, we test} the proposed metric for {AGN light curve SF} on the set of artificial {light curves} with various observing {cadences}.
We compare different observing {cadences} in the frame of oscillatory detection and discuss the impact of observing cadence {on the variability detection}. We summarize the results and give our conclusion in Section 5.

 \section{Method and data}
 Large time-domain surveys aim to carefully design their observing strategies in order to meet the requirements of most science cases. As an example, the specific observing strategy
   that LSST will follow is not completely decided and might not be sufficient to fully probe all AGN variability of interest, e.g. high magnification events \citep{10.1093/mnras/staa1208}.  These potential strategies are \citep{1708.04058,brandt18, 2020DPS....5211002J}: 
   \begin{itemize}
    \item  uniform in both cadence and filters within the high level constraints. This would emphasize detection of longer time scale brightness fluctuation events that can be followed in all LSST bands.
    \item  “rolling” cadences {are} intended to follow up shorter time scale events.
    {It is planned that $ 90\%$ of the observing time will be applied on the  18 000 $\mathrm{deg}^{2}$ wide-fast-deep survey (WFD).
The idea of rolling cadence is to split the WFD and
focus on distinct sky segments in different years (a specific area is to be observed with an
increased cadence while other areas are to be observed with a decreased cadence), and then “roll” those areas around the sky. At the end of the survey, the sky is observed
reasonably uniformly \citep{1708.04058}. 
For example, a  rolling cadence can  divide WFD in half so that the northern part gets more observational visits in odd years ($2k+1, k=0,1,2,3,4$) and the southern region in even years ($2k,k=1,2,3,5$) or vice versa.  Full range  variants of LSST rolling cadences where  WFD region is divided in 2, 3, and 6
declination bands  are  given in \citet[][see their Fig. 23]{pstn-051.lsst.io/PSTN-051.pdf}. Also these rolling cadence variants scale the rolling weight to be 80, 90, and
99$\%$; a larger weight results in more visits in the emphasized declination band and fewer outside the band. An example of   rolling cadence is given in Fig. \ref{fig:rollbase}. We can see the comparison between the rolling cadence (left) and the
baseline cadence (right). The shown rolling cadence (FBS 1.6 realisation,
rolling\_fpo\_2nslice1.0) splits the WFD ($-62^{\circ}$
< $\delta$ <  $2^{\circ}$) into two distinct declination bands and alternate
between them in different years during the survey (top and middle figure).
The rolling weight for this cadence is 99\%.  At the end of the
10 years operations (bottom figure), the total number of visits  is very similar to the
baseline survey strategy.
 The LSST rolling cadences cluster observations  according to complex criteria rather than to simply choose the preferential number of observations and seasons in years. In what follows, we will refer to the latter as 'variable cadence'.}
   
\end{itemize}
Therefore, it is important to define {metrics}, which can be used to characterize the performance of different observing {strategies} and {assess a} quality of {the detection} of oscillatory signal, as described.
 
In order to compare different observing strategies applicable to LSST, we: (i) construct an ensemble of {metrics}, (ii) compile three decades long light curves from real monitoring campaigns, and iii) generate synthetic data points assuming both hypothetical observing strategy, as well as {the ones that} are observed by LSST. 
In this section we present the details of the method used for  calculating {the} cadence metrics, and the data used to obtain these  empirical metrics.

\subsection{Metrics based on statistical proxies of time lag and periodicities uncertainties}\label{lagmetr}
{In what follows, we will assume that  the  time sampling (or cadence) of the light curve is $\Delta t^{c}$, the total time span of the light curve is $\Delta T$, the time lag is $\tau$, and an underlying periodicity is $P$}.
{There are two important {aspects in the AGN light curve} that affect the error of the {rest-frame} time lag ($\tau$)  and period of underlying {oscillation (P)} in the light curve: 
\begin{enumerate}
    \item  the  {fractional variability} (or amplitude of flux variation) $F_\mathrm{var}$ relative to the measured flux (or photometric) error $\sigma$: 
    $\frac{F_\mathrm{var}}{\sigma}$, { where}
$$ F_{\rm var}= [\sqrt{\sigma(F)^2 -e^2}]/F_{\rm mean} $$
{ and $e^2$ being the mean square value of the individual measurement
uncertainty $e(i)$ for N observations, i.e. $e^2=\frac{1}{N}\sum_i^N
e(i)^2$ \citep{10.1086/169036, 10.1086/312996} }. 
    \item the ratio of observed time scale $\mathcal{T}_\mathrm{obs}\in(\tau, P)$  and light curve sampling time $\Delta t^{c}$:  $\frac{\mathcal{T}_\mathrm{obs}}{\Delta t^{c}}$. 
\end{enumerate}
{In general, one can expect that the error of the time lag or oscillation decreases with increasing $\frac{F_{var}}{\sigma}$ and $\frac{\mathcal{T}_\mathrm{obs}}{\Delta t^{c}}$}.

Then, by assuming that there is no correlation 
between the amplitude of flux variation and the measured time scale $\mathcal{T}_\mathrm{obs}$,  we propose a proxy, $\phi_{\mathcal{T}}$ for the error of measured quantity, $\mathcal{T}_\mathrm{obs}$ as follows}:
\begin{equation}
\log \phi_{\mathcal{T}}=\log\frac{\sigma_{\mathcal{T}_\mathrm{obs}}}{\mathcal{T}_\mathrm{obs}}\propto A+C_{1}\frac{F_\mathrm{var}}{\sigma}+C_{2}\frac{ \mathcal{T}_\mathrm{obs}}{(1+z)\Delta t^{c}},
\label{eq:89}
\end{equation}
\noindent {where we assume that the error ${\sigma_{\mathcal{T}_\mathrm{obs}}}$ of {the time} lag or {periodicity inferred from the observed light curve} will be {increasing} with increasing redshift of the object {for the same observed time scale and sampling time}.


The model will provide the coefficient $A, C_{1}, C_{2}$ of the error proxy. If  errors of time lag or periodicity obtained from RM campaigns correlate with the proxy  $\phi_{\mathcal{T}}$, as defined in Eq. \ref{eq:89} then that relationship can be used to predict the minimum temporal sampling (cadence) required to recover the measured quantity  within specified accuracy.}

\subsection{Structure function  metric}

We adopt the first-order SF method \citep[see discussion in][and references therein]{10.3847/0004-637x/826/2/118}, defined as
\begin{equation}
SF(\Delta t)=\sqrt{\frac{1}{N_{\Delta t \mathrm{pairs}}}\sum_{1}^{N_{\Delta t}\mathrm{pairs}}(y(t)-y(t+\Delta t))^{2}},
\label{eq:33}
\end{equation}
\noindent {where  collection of measured data $y=\left\{y_i\right\}, i=1,n$ (e.g., magnitudes) at times $t=\left\{t_i\right\}, i=1,n$ with $\Delta t=|t_{i+1}-t_{i}|$
and $N_{\Delta t\mathrm{pairs}}$ is a  number of data pairs with time {separation} $\Delta t$.}
The idea of a potential metric is to estimate deviations between the SFs {for densely and uniformly} sampled light curve ($SF_{\rm conti}$) and {   those with gaps or variable cadence, termed "gappy" light curve} ($SF_{\rm gappy}$).

Assuming two SF curves,  both of which {constructed with the same bins of time lag}, we can {define the deviation of SF as follows,} 
\begin{equation}
M=SF_{\rm conti} -SF_{\rm gappy}.
\label{eq:44}
\end{equation}

Thus M will be the curve representing the deviations of the $SF_{\rm gappy}$ (based on the gaped light curve) from the $SF_{\rm conti}$.
We can also define the metric for {an ensemble of $k$ simulated}  light curves {providing} $SF^{i}_{\rm conti} $and $SF^{i}_{\rm gappy}, i=1, k$ where k is the number of simulated light curves. 
If we calculate these deviations curves for different redshift bins we can  average the deviations curves for each redshift bin $z$.
Then the {metric} given by Eq.\ref{eq:44} becomes:
 
 \begin{equation}
 M^{z}=\frac{1}{N_z}\sum SF^{i}_{\rm conti} -SF^{i}_{\rm gappy}=\frac{1}{N_z}\sum^{N_{z}}_{i=1} M^{i},
\label{eq:66}
\end{equation}
\noindent where {$M^{z}$ is an averaged deviation curve for redshift bin $z$},  $N_z$ is the number of deviations curves $M^{i}$ within the redshift bin z.
We can plot this {averaged SF deviations} curves {on the redshift and characteristics time scale domain}.
We note that {one can use other} metrics based on machine learning methods  to measure similarity between two curves as  e.g., Machalonobis,  Minkowski, cross correlation, etc.

\begin{table*}[ht!]
	\caption{Summary of characteristics of the objects. The columns are: (1) object name, (2) AGN taxonomy, (3) redshift, (4) monitoring time-baseline, (5) mean { relative} error in flux measurements (\%), (6) {variability parameter (\%)}, (7) {rest-frame time lag {in light-days (ld)}} (corrected for time dilation), (8) {mean sampling} time, (9) error of {rest-frame time lag}, (10) {rest-frame oscillation} period, (11) error of {the rest-frame oscillation} period. As reference, {H$\beta$ line was used for spectral time lag measurement}.}             
	\label{tab:data}      
	\centering          
	\begin{tabular}{c c c c c c c c c c c}     
		\hline
		\hline       

object    & type    & $z$    & Base[yr]    & $\sigma[\%]$    & $F_\mathrm{var}[\%]$    & $\tau_\mathrm{rest}[\mathrm{ld}]$    & $\Delta t^{c}[\mathrm{days}]$    & $\mathrm{err} \tau[\%]$     &  $P_\mathrm{rest}[\mathrm{yr}]$    & $\mathrm{err} P_\mathrm{rest}[\%]$\\
(1)&(2)&(3)&(4)&(5)&(6)&(7)&(8)&(9)&(10)&(11)\\
\hline          
NGC 3516    & CLAGN    & 0.0088    & 22    & 4    & 15.8    & 9.6    & 69.8    & 204.35    & -    & -\\
NGC 7469    & Sy1.0    & 0.0163    & 19    & 5    & 23    & 20.7    & 32.2    & 32.8    & 7.006    & 41.46\\
E1821+643    & quasar    & 0.297    & 24    & 5    & 7    & 90.98    & 64.3    & 0.065    & 9.84    & 33.84\\
Arp 102B    &  DPL    & 0.0242    & 26    & 5   & 21    & 14.65    & 96.9    & 130.2    & -    & -\\
Ark 564    & NLSy1    & 0.0247    & 11    & 5    & 7    & 3.9    & 40.5    & 658.7    & -    & -\\
3C 390.3    & DPL   & 0.0561    & 12    & 5   & 38    & 90.9    & 128.6    & 27.6    & 9.56    & 0.94\\
NGC 4151    & Sy1.5-1.8    & 0.0033    & 10    & 5    & 42    & 4.98    & 25.3    & 558.2    & 13.71    & 27.02\\
NGC 5548    & Sy1.0-1.8   & 0.0172    & 6    & 5    & 33    & 48.2    & 28.35    & 38.1    & 13.075    & 16.71\\

\hline                  
\end{tabular}
\end{table*}

\subsection{Data}

We used two type of data sets to {test our} metrics: compiled {observations of} three decade long light curves and artificially generated {light curves}.
{ Initially, the artificial light curves
are generated to be five times longer than the operation period of ten years \citep{10.1088/0004-637X/771/1/9}, which also satisfies the condition that they are
  $\sim 10 $ or more times longer than their characteristic timescale \citep{10.1051/0004-6361/201629890}.
Then the fake curves have been cut to a ten-year baseline.}

\subsubsection{Compiled data set of observed light curves}
We select a sample of objects to be consistent in the sense of instruments used, length of monitoring campaigns, calibration and flux measurements, as well as statistical tools used {to infer} time lag and periodicity.
{We} use the optical RM data from very long monitoring campaigns of \citet{2001A&A...376..775S,2004A&A...422..925S,
	2008A&A...486...99S,2010A&A...509A.106S,
	2010A&A...517A..42S,2012ApJS..202...10S,
	2013A&A...559A..10S,2016ApJS..222...25S, 2017MNRAS.466.4759S,2019MNRAS.485.4790S}, {   for which the baseline length is {comparable to that of} the upcoming large sky surveys, such as LSST} (see Table \ref{tab:data}). 
It is important to point out that these  data have specific  characteristics which can be very difficult to simulate. For example, NGC 5548 shows the time-lag variability, Arp 102B and 3C 390.3  are classified as double peaked line (DPL) emitters but their oscillatory characteristics are quite different  \citep{2018MNRAS.475.2051K}, E1821+643 shows extremely low variability as a binary SMBH  candidate \citep{2016ApJS..222...25S}, Ark 564 is as a narrow-line Sy 1 (NLSy1) object with very low-variability.  
In Shapovalova et al. series of papers \citep[see e.g.,][]{2016ApJS..222...25S}, we introduced the time lag determination based on Gaussian process light curve modeling as a novel tool.
 Moreover, the periodicity detection for several of these objects {   was made possible} by our new tool - 2DHybrid method \citep{2018MNRAS.475.2051K}. Since {commonly used} periodicity detection methods are not designed for red noise light-curves we applied our 2DHybrid method, which results are given in Table  \ref{tab:data} \citep[see also][]{2018MNRAS.475.2051K}. {Some notable features of our 2D hybrid method are:  enhancement of apparent  resolution by spreading peaks over the second dimension, and establishment of direction of changes in signal through correlation coefficients  \citep{10.1515/astro-2020-0007}. }
Even though some data was added from other RM campaigns, the Shapovalova et al. dataset served as the backbone for periodicity detection  \citep{2018MNRAS.475.2051K}. { Table 1. lists the object basic information and spectral characteristics calculated for the H$\beta$ line. 
 
Two parameters, the level of variability and relative photometric error, used for the calculations of the statistical proxy given in Eq.(1) are taken from the Shapovalova et al. campaign. }

\subsubsection{Artificial set of light curves with ideal and LSST OpSim cadences}
\cite{10.1088/0004-637X/698/1/895} found that the optical variability could be represented by a stochastic model based on Damped random walk (DRW) {process}. The model {incorporates a} characteristic amplitude $\tilde{\sigma}$, which affects exponentially-decaying variability with time scale $\tau$ around the mean magnitude $m_{0}$. The model specifications $\tilde{\sigma}$ and $\tau$ are related to the SMBH mass $M_{\rm{BH}}$ and/or luminosity $L$ of the AGN \citep{10.1088/0004-637X/698/1/895,10.1088/0004-637X/779/2/187}.
{ The following summarizes approximation of AGN properties:
taking into account that the currently limited monochromatic luminosity range  $10^{42}\leq L_{5100}\leq10^{46} \mathrm{ergs\, s}^{-1}$ is somewhat uncertain \citep{10.1086/509650},  the monochromatic luminosities at 5100 \AA }  are chosen randomly from the range 
 $\log L\in[42.2,47] $
{ to allow for  more luminous objects}. The SMBH mass $M_{\rm{BH}}$  is determined {by} L,  Eddington luminosity 
 $$L_\mathrm{Edd}=1.25 \cdot 10^{38} \frac{M_{\rm{BH}}}{M_{\odot}[\rm{erg\, s^{-1}}]} $$
 \citep{10.1086/342878} and an  Eddington ratio \citep{10.1088/0004-637X/690/1/20}. Then, the characteristic radius of the BLR is approximated {by} the empirical radius-luminosity relationship \citep{10.1088/0004-637X/767/2/149}.

  For the simulations, DRW scales  ($\tilde{\sigma}$ and $\tau$) are drawn based on luminosity from the distributions given in Equations 22 and 25 by \cite{10.1088/0004-637X/698/1/895}.

{The sequence of  AGN light curve points $p_i$ comes from the DRW model which is recursive in the  flux dimension and iterative in the time dimension \citep{10.1088/0004-637X/698/1/895}:
\begin{align}
\begin{split}
p_{i+1}=\mathcal{G}\Big(p_{i}e^{\frac{-\Delta t}{\tau}} +m_{0}(1-e^{\frac{-\Delta t}{\tau}}),
\tilde{\sigma} \sqrt{\frac{\tau (1-e^{\frac{-2\Delta t}{\tau}})}{2}}\Big) \\
\end{split}
\label{eq:11}
\end{align}
\noindent where $\mathcal{G}$ is the Gaussian distribution, $\Delta t=t_{i+1}-t_{i}$ is the time interval, $m_{0}=23$ mag, $p_{0}=\mathcal{G}(m_{0}, \tilde{\sigma}\sqrt{\frac{\tau}{2}})$  at $t_0$.}

{For photometric uncertainty we adopt the photometric error model of LSST \citep{10.3847/1538-4357/ab042c}:
\begin{equation}
\sigma^{2}_{LSST}=\sigma^{2}_{sys}+\sigma^{2}_{rand}
\end{equation}
\noindent where $\sigma_{sys}=0.005$ is  the systematic error due to imperfect modeling of point source, 
$$\sigma_{rand}=(\frac{1}{25}-\gamma)X+\gamma X^{2}$$ 
is the random photometric error where 
$X=10^{0.4(m-m_{5})}$, $m_{5}=24.7$ and $\gamma=0.039$ for {$r$-band} \citep[see][and their Table 2]{10.3847/1538-4357/ab042c}.
Finally, the observed light curve is obtained from 
\begin{equation}
y_{i} = p_{i} + \mathcal{G}(0,\sigma_{LSST}(p_{i})).
\label{eq:period}
\end{equation}}
{
The parameters $\tau, \tilde{\sigma}$ and period of  the simulated light curves are corrected for the $(1+z)$ where z is the redshift.
 The simulation of the  flux in some emission line $l$, which is emitted by the BLR, is based on the  linear approximation}:
 
 \begin{equation}
 f^{l}(t)=(f^{c}*\xi)(t)=\int^{\infty}_{-\infty} f^{c}(\tau)\xi(t-\tau)d\tau
 \label{eq:transfer}
 \end{equation}
\noindent {where the transfer function $\xi$ defines, essentially,  the
geometry of the BLR region as seen by the observer and $f^{c}$ is a flux originating closer to the SMBH and driving $f^{l}$. For simplicity, we consider
that $\xi\propto \mathcal{G}(R_{\rm BLR}, 0.25R_{\rm BLR})$ \citep{10.1088/0004-637X/747/1/62}, where $R_{\rm BLR}$ is already derived from the mass of SMBH, and the input continuum flux $f^{c}$ is the realization of Eq. \ref{eq:11}. Unlike the spectroscopic approach, this method {does not} allow the emission line and continuum light curves to be extracted, since the data consists of their combined signal. In reality under certain condition,  separation of these processes is possible so the lag can be measured  \citep{10.1088/0004-637X/747/1/62}. }
{Particularly, we assume that  continuum and emission line through given filter  have been determined in advance either using  \citet{10.1088/0004-637X/747/1/62} method or known somehow else.
Thus, we simulated disentangled  continuum and emission line using   Eq. \ref{eq:period} and   Eq. \ref{eq:transfer}, respectively, while taking into account dimension of BLR inferred at the beginning of our procedure.}

{For simulating underlying periodic signal, we assumed that the inferred {SMBH mass} is the total mass of the hypothetical binary system at mutual distance of $\sim 10$ ld and that the amplitude is about $14\%$ resembling the case of PG1302-102
\citep{2015Natur.525..351D}.
Also the modulation of the signal can be approximated to the first order by amplitude of
$v \cos \psi \sin i/c$ where $\psi$ is the orbital phase, $v$ is the velocity of the secondary component and $i$ is the {inclination angle} \citep{2015Natur.525..351D}.  For near-equal mass binaries, some studies show that the mass accretion rates fluctuate periodically, but they resemble a series of sharp bursts, differing from  sinusoid-like shape \citep{2015Natur.525..351D}.}

{   These continuous artificial light curves were  sampled according to several observing strategies:
\begin{enumerate}[labelwidth=*]
    \item {Idealized  observing strategies were constructed to assist in the interpretation of the  results related to SF which are obtained from LSST-cadences, serving as approximations ranging from reasonable to excellent observing strategies}:
    \begin{enumerate}[label={\arabic*},labelwidth=*]
    \item ideal light curves cadence: {uniform} (1-day cadence) during 10 yrs; 
    \item "gappy" light curves cadence: series of 3 months /6 months/9 months of {uniform} (1-day cadence) observations per year during 10 yrs campaign. There are no observations in  gaps; 
    \item variable-cadence light curve: in the first year only three months are observed with 1-day cadence, and in the next years 3 months with 1-day cadence are observed, followed by 6 months of 30-day cadence, and a gap of 3 months. 
    {The preferential clustering of observations is based on the criterion of a number of observations and alternating seasons (years), which is more straightforward than the LSST rolling cadence motif.}
    \end{enumerate}
    \item OpSim runs with three different observing strategies:
     \begin{enumerate}[label=\Roman*,labelwidth=*]
      \item  observing some selected LSST fields by taking around 90  epochs during 10 yrs, 
      \item observing  fields with  about 1500 epochs during 10 yrs,  
      \item observing fields with  200 epochs over 10 yrs of survey.
      \end{enumerate}
\end{enumerate}
}

\begin{figure*}
	\centering
	\includegraphics[clip,trim=20 30 55 60,width=0.33\textwidth]{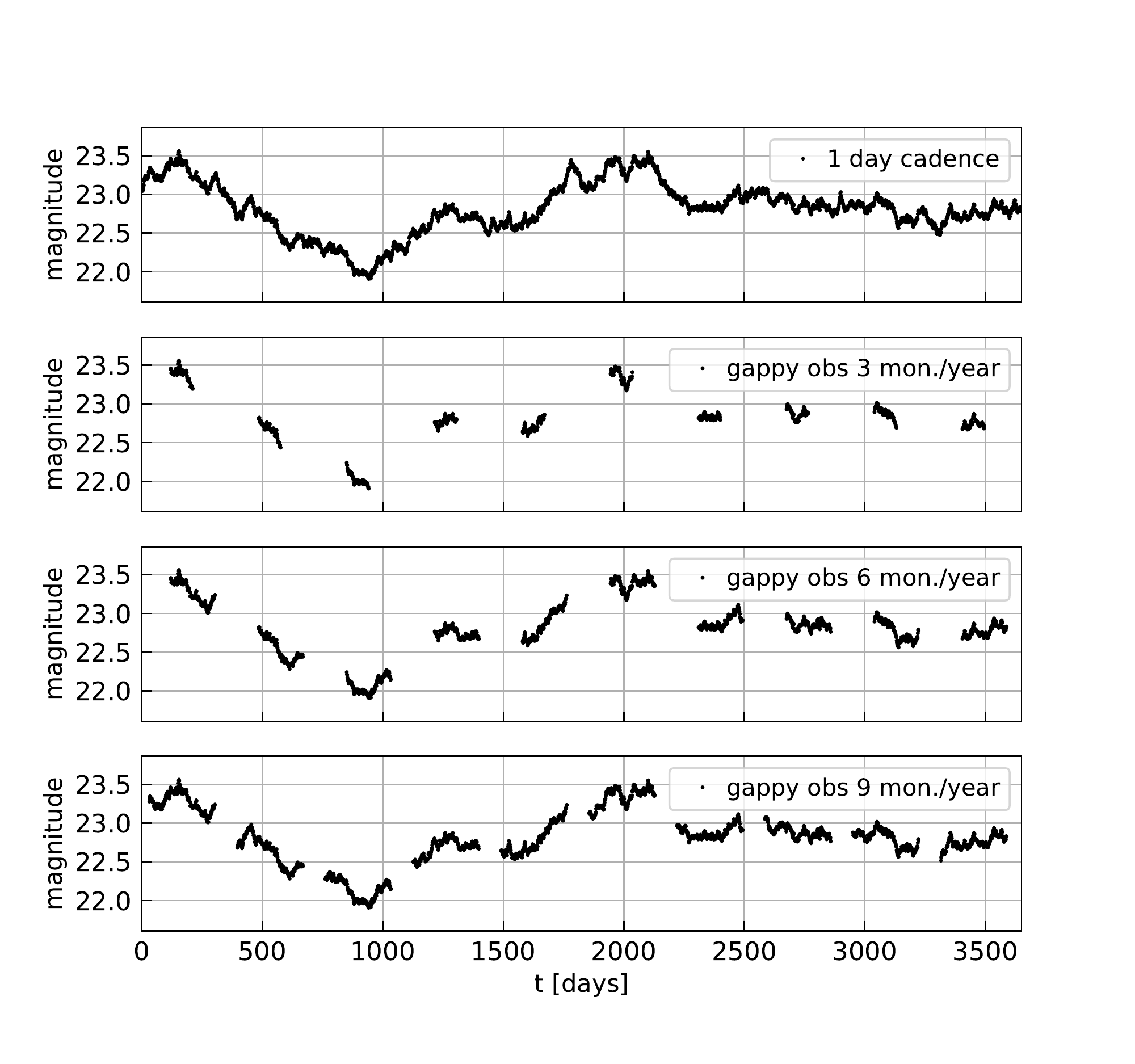}
	\includegraphics[clip,trim=20 30 55 60,width=0.33\textwidth]{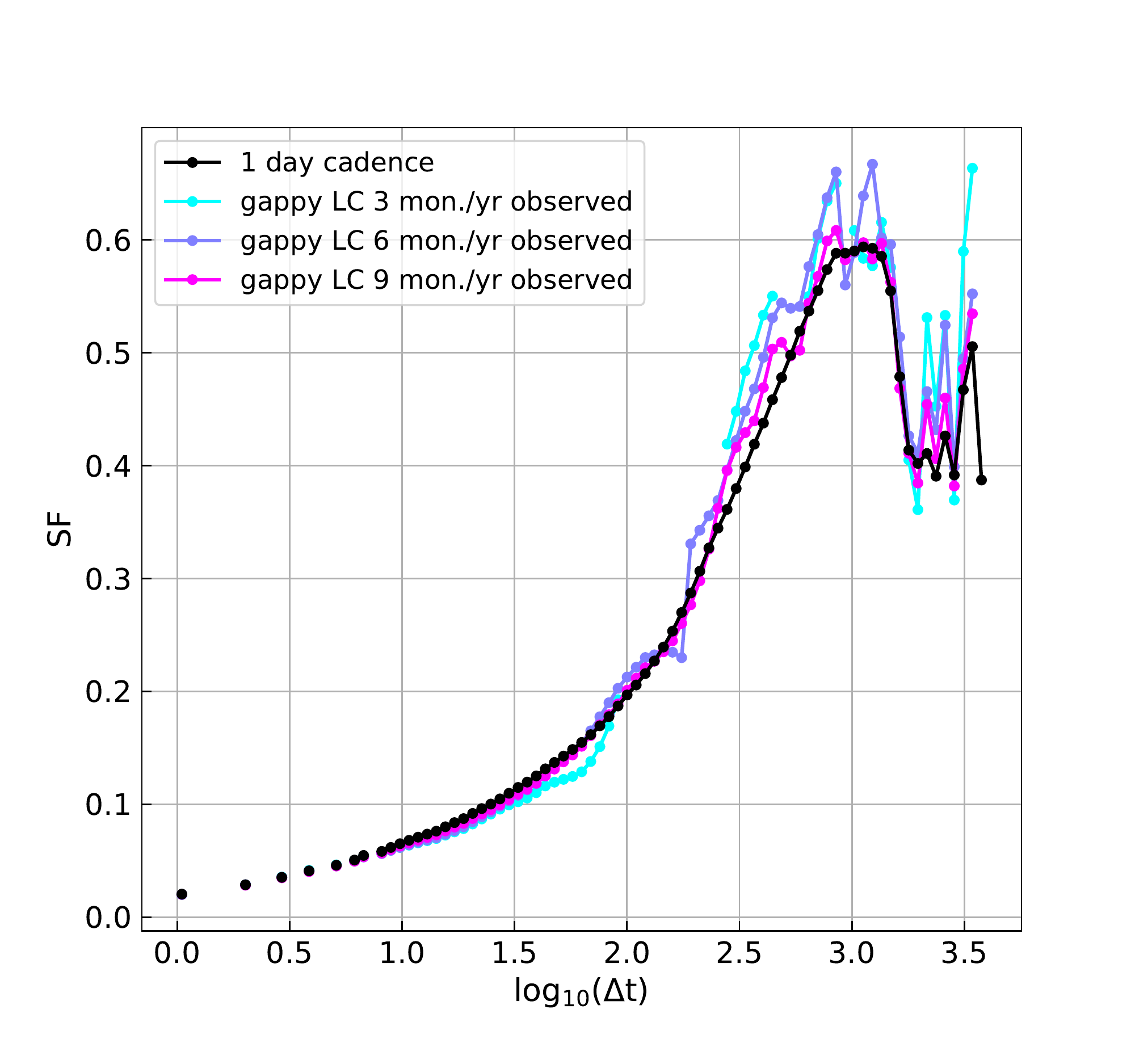}
	\includegraphics[clip,trim=10 30 55 60,width=0.33\textwidth]{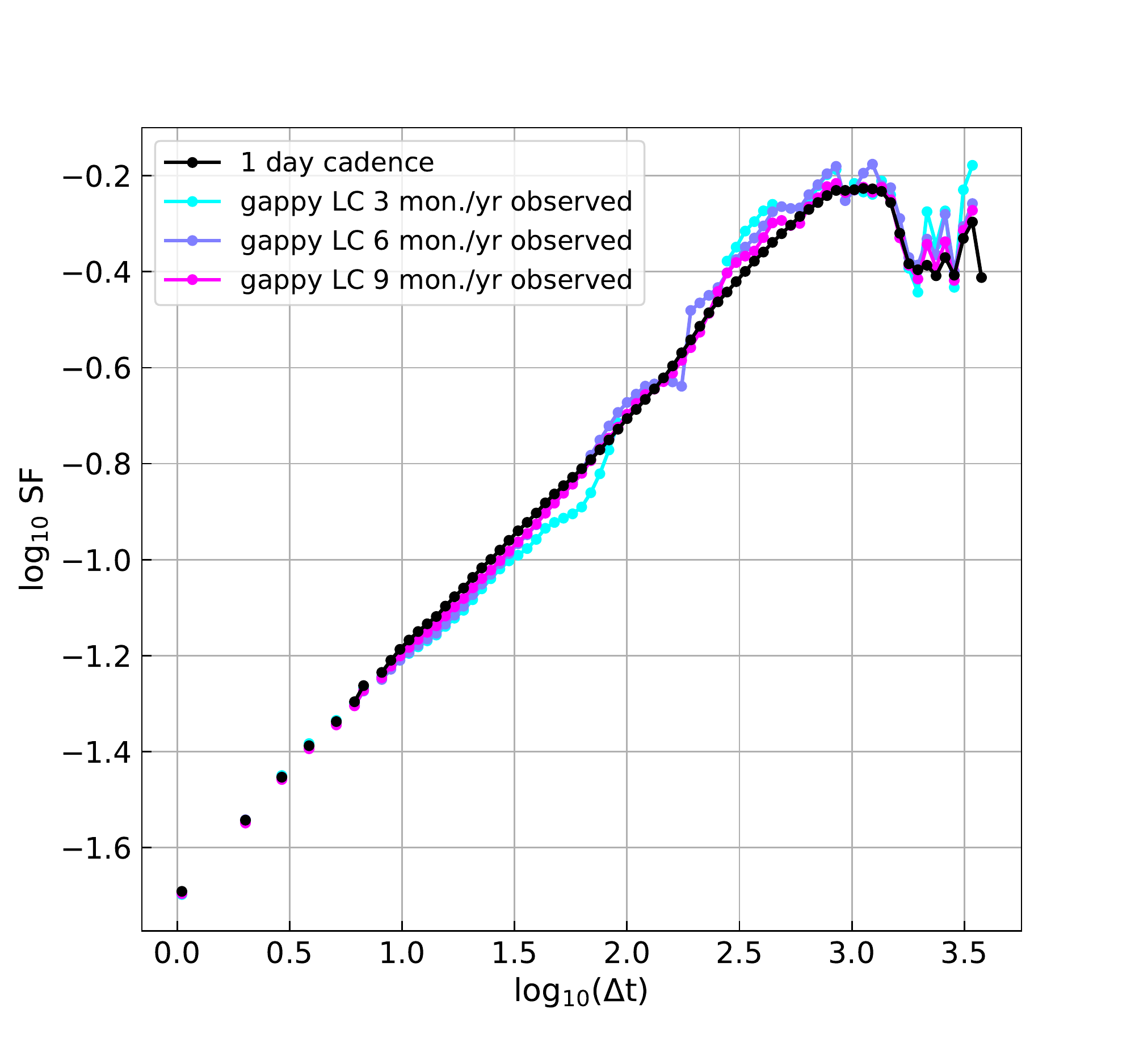}
	\caption{\textit{Left}: {   Artificial} AGN light curve {   based on DRW model with added oscillatory signal}. {{Generated light curve corresponds to AGN with black hole mass of $2.8 \times 10^{8} \mathrm{M_{\odot}}$ and bolometric luminosity of $8.5 \times 10^{45}\ \mathrm{erg \ s^{-1}}$. The period is 4.3 yr.}} {   Upper panel gives the ideal light curve with homogeneous 1-day cadence, the bottom three panels give "gappy" light curve  assuming respectively that 3, 6 or 9 months per year are observed.} For all idealized rolling cadences during observed periods the sampling rate is 1 day. \textit{Middle}: SFs calculated for the light curves given in left panels. Black curve is the SF calculated for the {   ideal} light curve with homogeneous 1-day cadence. Blue, violet, and pink stand for SFs calculated for gappy cadences of 3 months/yr, 6 months/yr and 9 months/yr respectively. \textit{Right}: The same plot as in the middle panel but in logarithmic scale.}
	\label{fig:fig10}%
\end{figure*}

\begin{figure*}
	\centering
	\includegraphics[trim=20 85 40 90,width=0.33\textwidth]{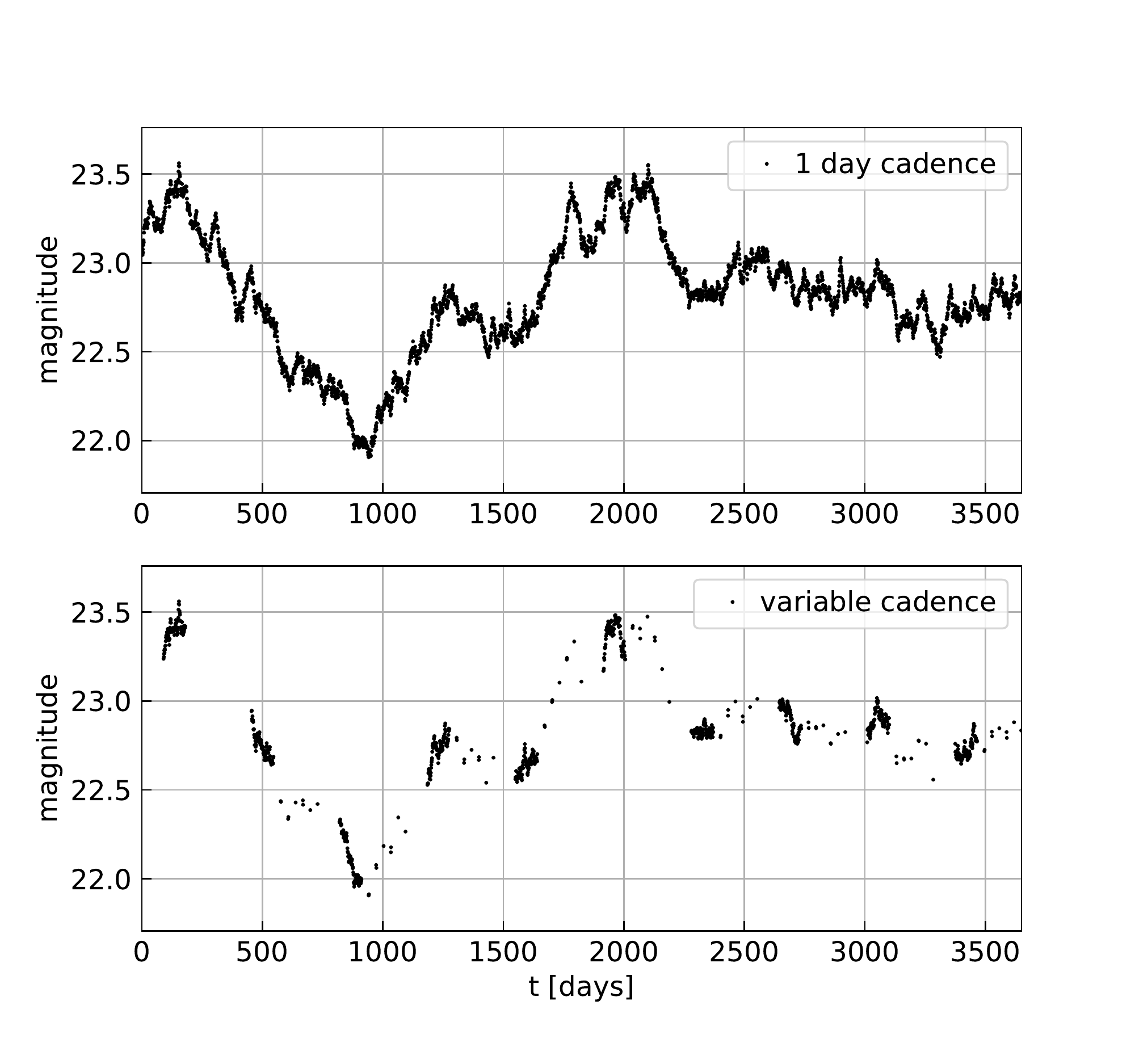}
	\includegraphics[trim=20 85 40 90,width=0.33\textwidth]{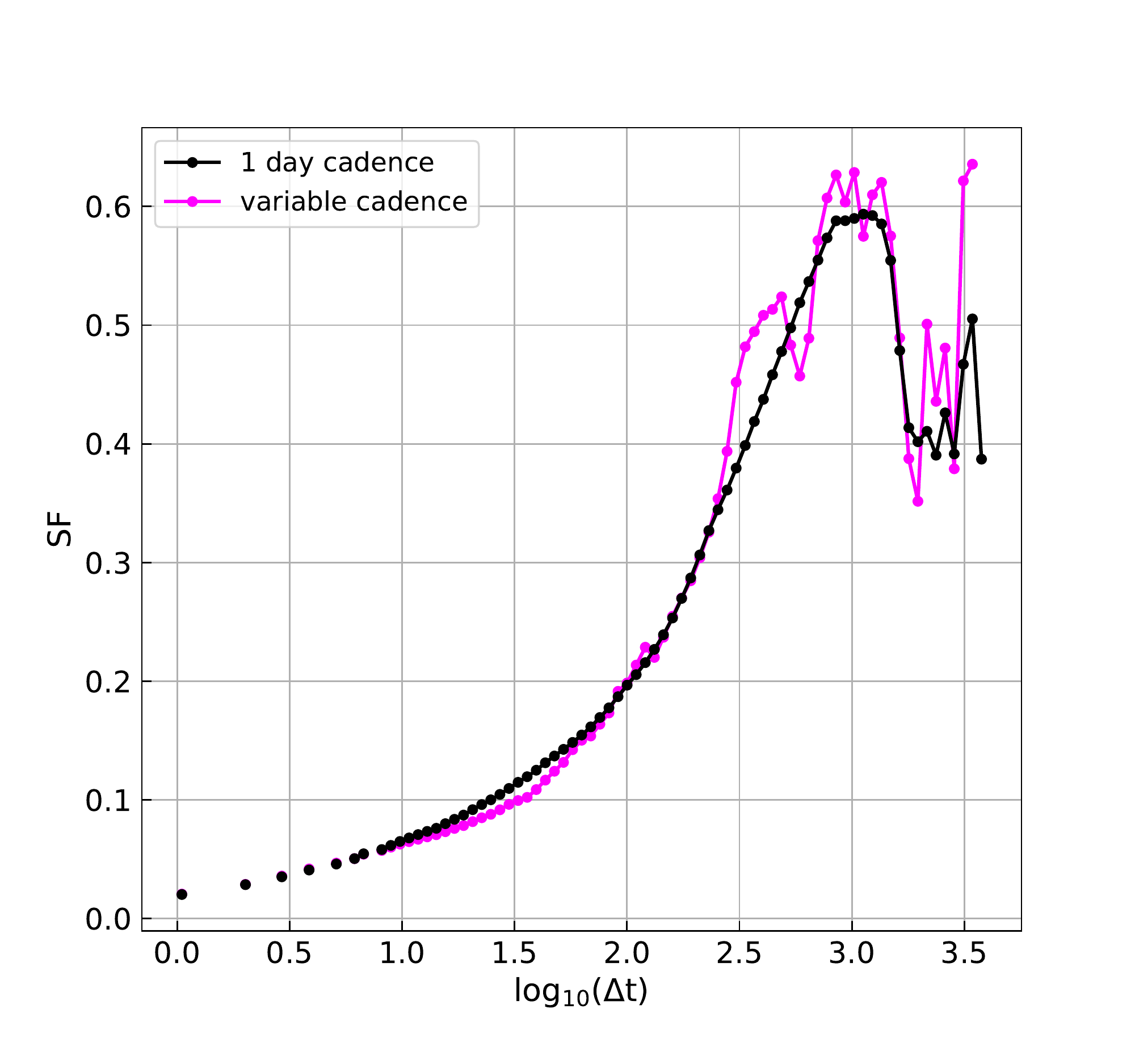}
	\includegraphics[trim=20 85 40 90,width=0.33\textwidth]{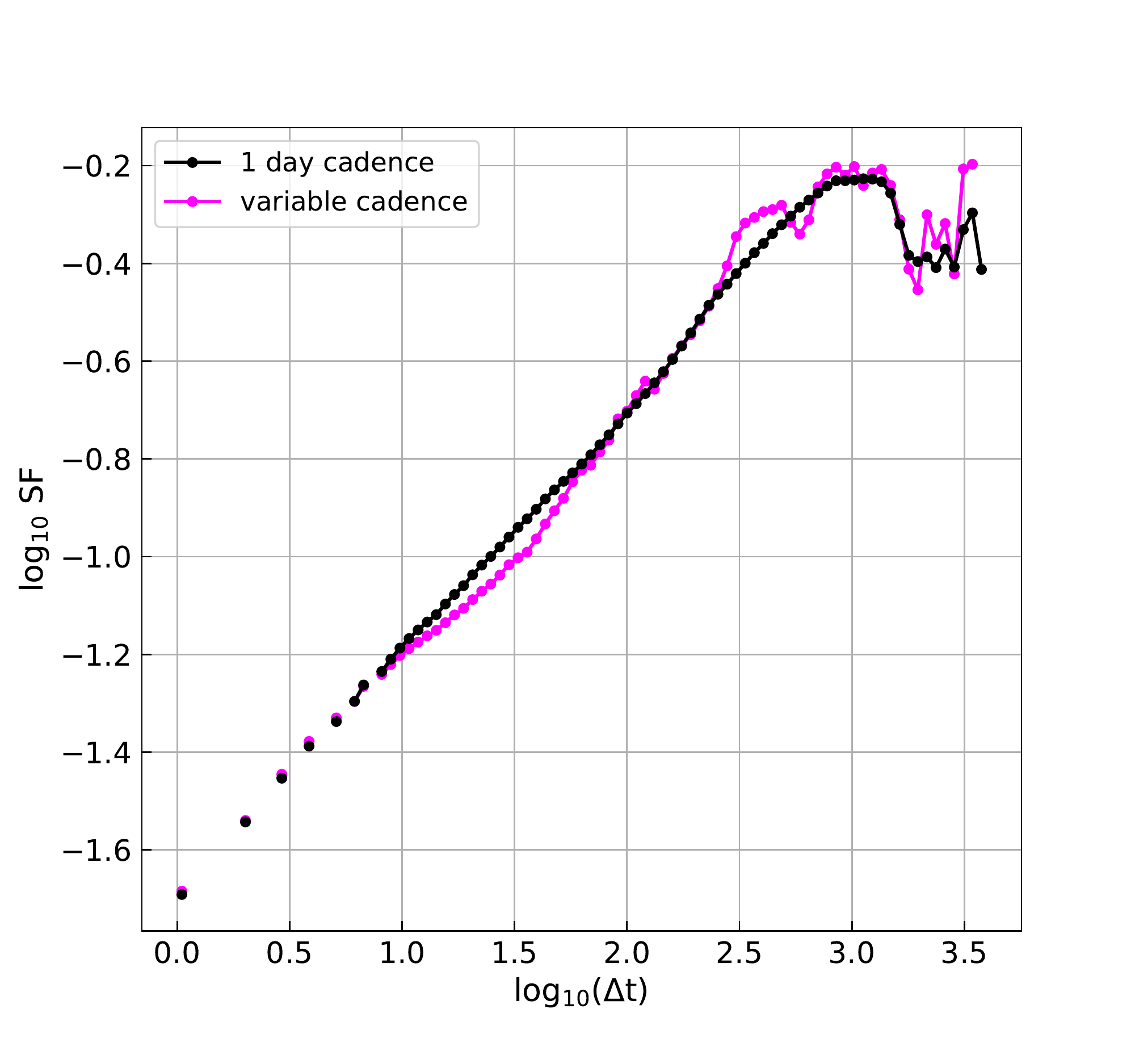}
	\caption{ \textit{Left}: {   Artificial} AGN light curve {   based on DRW model with added oscillatory signal}, but now with variable cadence. Prior to light curve generation itself, we used the same seed in the pseudorandom number generator to obtain the exact same light curve as the one in Fig. \ref{fig:fig10}. Upper panel gives the ideal light curve with {uniform} 1-day cadence, and the bottom panel gives the variable-cadence light curve assuming that 3 months per year are observed with 1 day cadence and 6 months per year with 30-day cadence. \textit{Middle}: SFs calculated for the light curves given in the left panel. Pink curve is the SF of variable-cadence light curve,  and black stand for the SF calculated for ideal light curve with {uniform} 1-day cadence. \textit{Right}: The same plot as in the middle panel, but in logarithmic scale.}
	\label{fig:fig11}%
\end{figure*}

\noindent Fig. \ref{fig:fig10} (left panel) shows a realization of the artificial light curve with a periodic signal of 4.3 yr (see Eq. \ref{eq:period}) and different cadences. The upper panel shows a light curve with {uniform} 1-day cadence {case i-(1)}, {   and the bottom three panels give light curves with variable} cadence  case i-(2). 
The generated artificial light curves with variable cadence (denser and sparser, as described above) are shown in the left panel in Fig. \ref{fig:fig11}.

{   For further testing of the proposed metrics on the LSST observing strategies,} we created artificial set of objects based on OpSim realizations (their designations listed in the last column of Table \ref{tab:Opsimgen}).  {For these artificial objects, the redshift is kept as low as $z=0.05$, since H$\alpha$ and H$\beta$  lines for photometric RM at low redshifts ($z < 0.02-0.05$) are prominent and easy to monitor; but for objects at higher redshift where Balmer lines continually shift to longer wavelengths, broad-bands filters (g,r) often observe  the continuum  contaminated with broad emission lines \citep{10.1088/0004-637X/747/1/62, 10.1088/2041-8205/750/2/L43, 10.1088/0004-637X/756/1/73, 10.3847/1538-4357/ab40cd}} 

\begin{figure}
	\centering
	\includegraphics[clip,trim=35 8 10 8,width=0.43\textwidth]{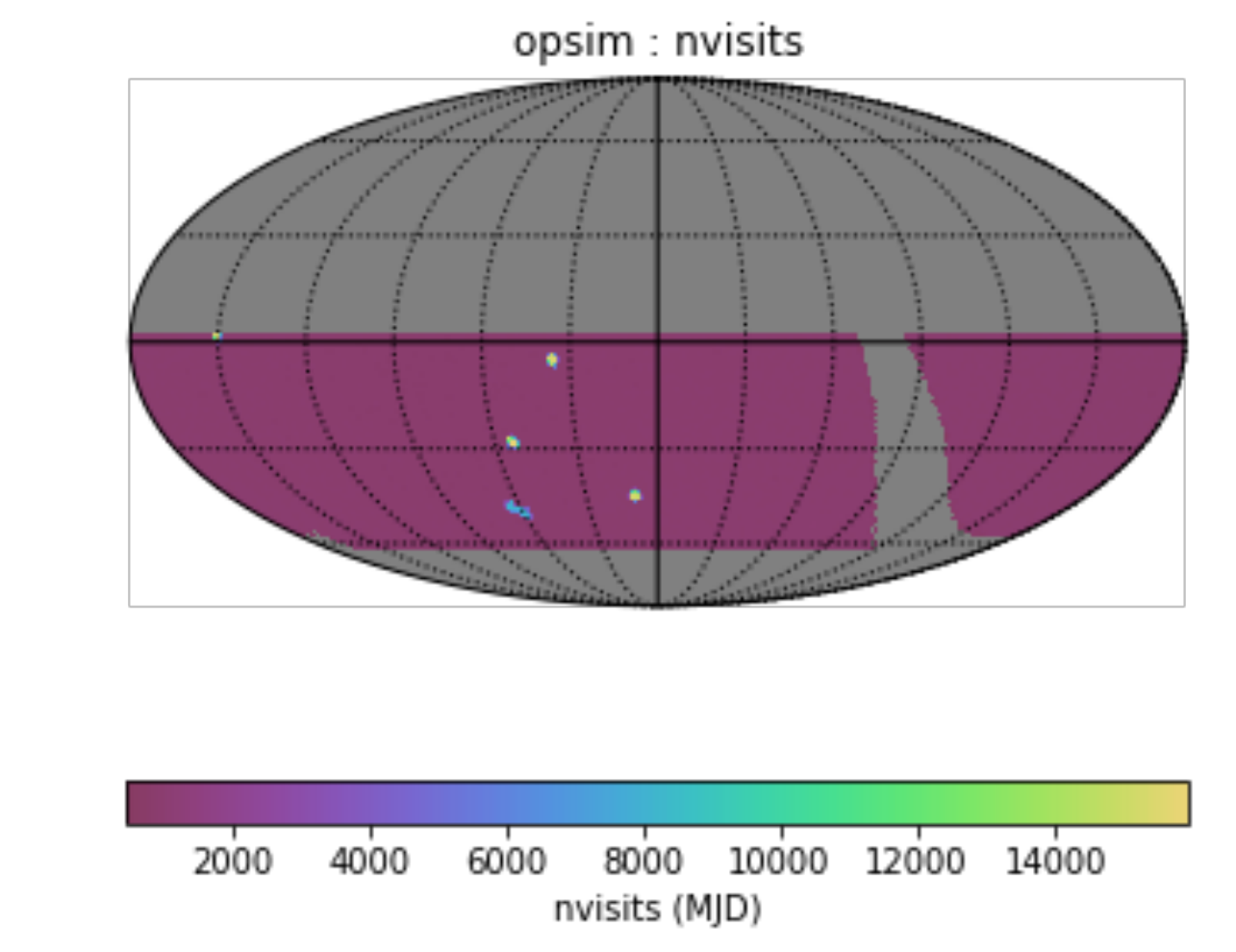}
	\caption{Number of visits for each equatorial coordinate pair during the whole scope of the LSST mission (10-year period) for the $r$ band. Color bar represents number of visits. DDF fields are easily distinguished with largest number of visits (yellow and blue color). Plot is obtained using the Python-based LSST Simulation Framework. }
	\label{fig:fig0}%
\end{figure}

 We selected several OpSim {runs} which are {thought to be relevant for} AGN research, such as the AGN Deep Drilling Fields (DDF), but in general, this concept could be applied on any OpSim run, {therefore we have also used rolling cadences. Rolling cadences are non-uniform observing strategy where some region of the sky is emphasized in one year, and then minimized in the next. } 
 
The DDF simulations have different {observing cadence} for the five DDFs, while a standard baseline observing  strategy is applied for the rest of the sky. {The AGN DDF OpSim run} takes shorter DDF sequences more often (around $2.5$~\% of visits are spent on DDFs) \footnote{https://pstn-051.lsst.io/PSTN-051.pdf}.

We used FBS 1.5\footnote{http://astro-lsst-01.astro.washington.edu:8081} AGN DDF runs in $g$ and $r$ bands for testing the properties of structure function.  The $r$ band has a total number of 491676 visits during the 10-year period of LSST mission, while the $g$ has 223871 visits. 

In Fig. \ref{fig:fig0} the total number of visits per equatorial coordinate is shown for the $r$ band. The noticeable difference between number of the visits in DDFs and the rest of the sky is clearly visible. This allow us to choose one point \citep[for definition of point see, e.g. ][]{10.3847/1538-4365/ab72f2} from DDF and one point {outside} the DDF in order to probe the difference. 

{ We demonstrate our analysis using two different OpSim cadence releases, OpSim 1.5 and 1.6. From the OpSim 1.5 we used AGN DDF cadences and from the OpSim 1.6 we used rolling cadences. Results based on the newest cadence release, OpSim 1.7, can be found at our online supplementary channel} (\url{https://github.com/LSST-sersag/agn_cadences}). Also, our repository could be efficiently used for different OpSim realisations.  There are no discrepancies between results obtained from these two realisations.

Based on described procedure, we generated the set of artificial objects using the cadences from the OpSim.
Further, we applied our methods to determine the time lag and periodicity. Both, the model input and measured values are given in Table \ref{tab:Opsimgen}. Using these values we {performed multiple linear regression} as in the case of our true monitoring campaign.  

\begin{table*}
\small
	\caption{Summary of characteristics of the artificial objects  {(multicolumn Input values, columns 1-8) and recovered RM and periodicity values (multicolumn Recovered values, columns 9-12)}. The columns are: (1) object ID, (2) luminosity, (3) SMBH mass, (4) BLR dimension, (5) mean relative error in { magnitude measurements (in \%)}, (6) {variability parameter (in \%)},  (7) {imparted period of oscillations in the light curve given in rest frame assuming that the mass is a total mass of binary at mutual distance of  10 ld}, (8) cadence, (9) detected {rest-frame time lag}, (10) error of {rest-frame time lag}, (11) {rest-frame oscillation} period, (12) error {of the rest-frame oscillation} period, (13) designation of the Opsim realization. {All objects are at redshift 0.05}. }             
	\label{tab:Opsimgen}      
	\centering          
	\begin{tabular}{l ll ll l l c c l l  l l l}     
		\hline
		
\multicolumn{8}{c }{Input values} 
&                                            
\multicolumn{5}{c}{Recovered values} \\\hline   		
ID  &L & M$_{\rm{BH}}$ &$R_{BLR}$  & $\sigma$ &$F_\mathrm{var}$  &$P_\mathrm{rest}$ &$\Delta t^{c}$& &$ \tilde{\tau}$   &$\delta\tilde{\tau}$   & $\tilde{P}_\mathrm{rest}$ &$\delta\tilde{P}_\mathrm{rest}$&OpSim\_cad\\ 
&[$10^{44} \mathrm{erg\, s^{-1}}$] & [$10^{6} M_{\odot}$] &[ld]&$[\%]$ &$[\%]$  &[yr] &[days]& & [ld] &[ld] &  [yr] &[yr]&\\
(1)& (2) & (3) &(4)&(5) &(6) &(7)&(8)& & (9) &(10) &  (11) &(12)& (13)\\
\hline
I &85.5&282.7& 36.03 &0.16 &14.9 &4.28 &43& &39.05&5 & 3.93&1.4& roll\_cad\_0.8\_g\_ra\_0\_de$-$10\\
II &1156.7& 1605.4 &144.5&0.23&09.3 &1.8&1.4& &138.1 &4.8 &1.85&0.31&agn\_g\_ra\_9.0\_de$-$44 (DDF)\\
III &199 &496.5 &56.5 &0.19 &16.2 &3.23 &18 &&54.3 &5 &3.2&0.45&roll\_cad\_0.8\_ra\_0.0\_de
$-$30\\
IV &80.4 &271.3 &34.87 &0.14&20.4 &4.4 &17& &31.4 &4.7&4.4&0.5&roll\_cadence\_0.8\_ra\_0.0\_de$-$50\\
V &12.1 & 76.6&12.68 & 0.22&18.8&8.2&18&&16.1 &4.9&6.9&0.7&
roll\_cad\_0.8\_r\_ra\_0.0\_de$-$30\\
VI &391 &778.8 & 81.02&0.23&15.3 &2.58 &19 &&84.5 &4.8 &2.12&0.31& roll\_cad\_0.8\_z\_ra\_0.0\_de$-$10\\
VII &15 &88.5 &14.24&0.18&13.4 &7.66 &60& &16.2&4.8 &7.9&0.76&roll\_cad\_0.8\_u\_ra\_0.0\_de$-$30\\
\hline                  
\end{tabular}
\end{table*}

\section{Results}
{The results are presented in three main sections that seek to capture the different perspectives of the relation between cadence estimates and AGN variability-related observables. A more general issue of the number of binary AGN candidates detected by LSST like survey allows us to infer general constraints on cadences compatible with multiple regression predictions. The cadence estimates for time-lag and oscillation measurement use real -world and LSST-like objects samples.

{We were particularly interested in assessing the capabilities of the cadences to accommodate underlying oscillations not only  in the light curves, but also in the SFs.
 SF is the RMS scatter of magnitude (flux) differences
 calculated as a function of temporal separation $\Delta t$, behaving approximately as a power law with respect to time 
$SF \propto (\Delta t)^{\gamma}$ \citep{10.3847/1538-4357/abc698/pdf}.
Thus, for coarser cadences of the real-world sample with large time separation $\Delta t$ (compare columns (8) in Tables \ref{tab:data} and \ref{tab:Opsimgen}), as epochs
in the light curve cease to be correlated, calculated SF will stabilize to
a constant value-the asymptotic SF \citep{10.3847/1538-4357/abc698/pdf} and any underlying information is lost then. Contrary, finer LSST cadences (column (8) in Table \ref{tab:Opsimgen}) preserve underlying information on oscillation.
Thus, to understand the factors affecting cadence estimates for SF with underlying oscillation, we tested SF-metric  with idealized and LSST-like data set.} }

\subsection{Cadence estimates for time-lag measurement}

\begin{figure*}
	\centering
	\includegraphics[clip,trim=100 35 25 30,width=0.49\textwidth]{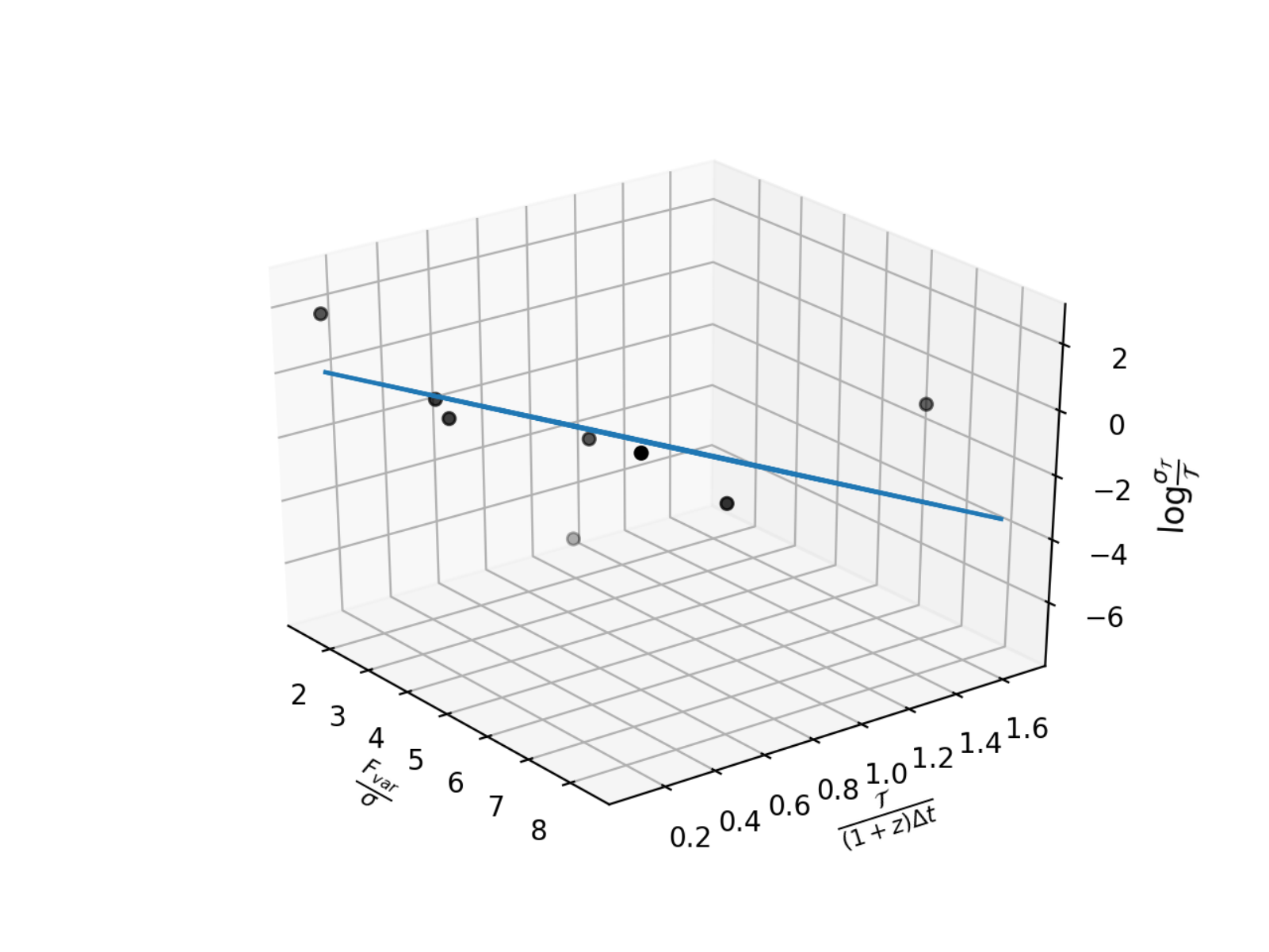}
	\includegraphics[clip,trim=100 30 25 30,width=0.49\textwidth]{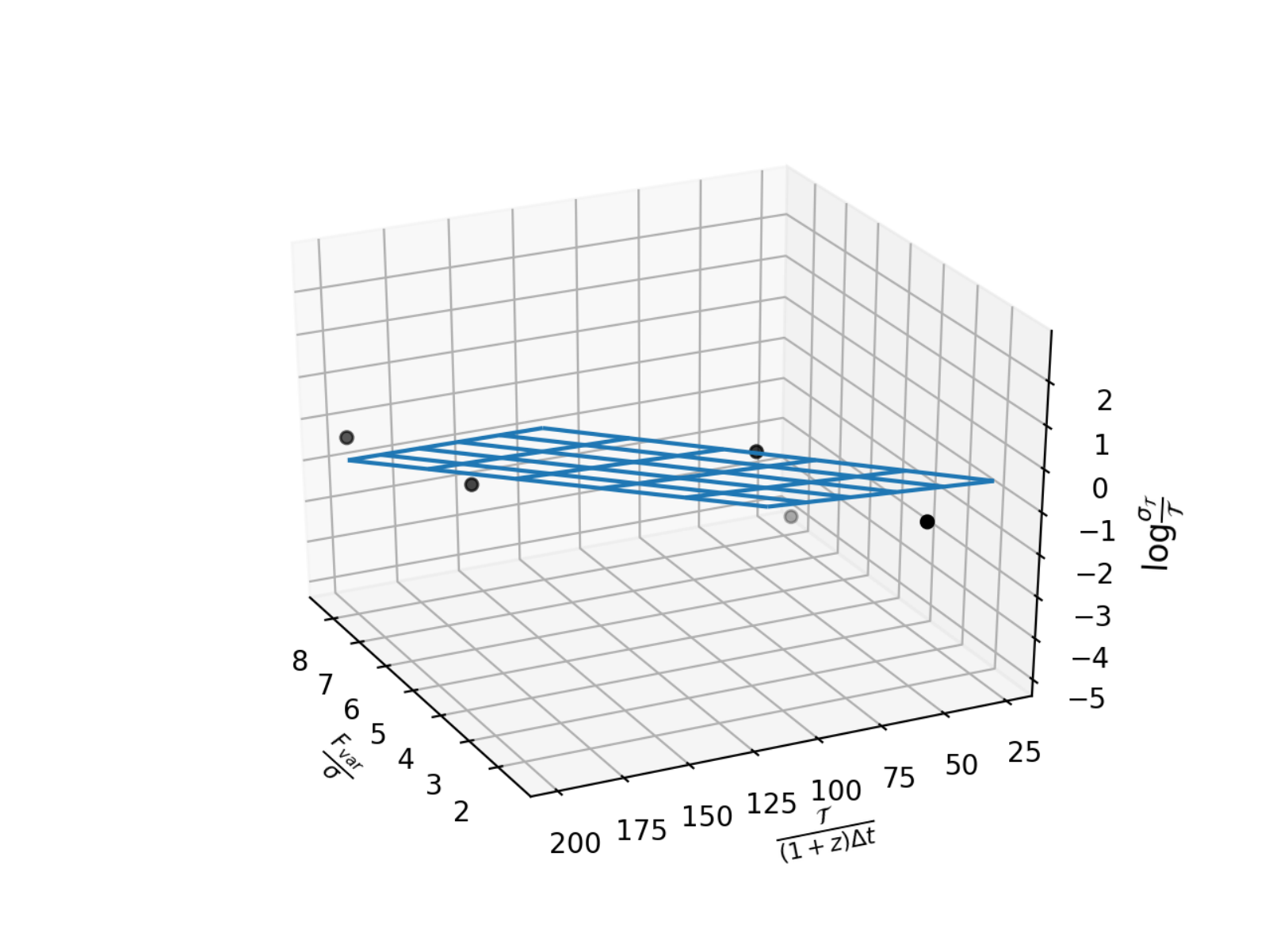}
	\caption{{3D plot of fitted multi-linear regression model (blue plane) for} empirical proxies for uncertainties  in time lags (left) and in periodicities (right). Filled circles denotes data given in Table \ref{tab:data}. {\textit{Left}  $C_{1}= 0.34 \pm 0.20, C_{2}= -3.75\pm 1.38$;  \textit{Right}  $C_{1}= -0.64 \pm 0.13, C_{2}=0.018\pm0.0068$.} }
	\label{fig:fig1}%
\end{figure*}

{{A success of the recovery of time lag and periodicity} depends on the {observing cadence} \citep{10.1086/420755} as well as the cadence of observations. }
{ We thus {learned} the multiple regression of the }proxy variable for the relative uncertainty in time lags given by Eq. \ref{eq:89} and plot its relationship with two independent variables (see Fig. \ref{fig:fig1} left plot) based on our data given in Table \ref{tab:data}. The multiple linear regression (blue plane) is tilted along both independent variables. {In other words, the proxy of uncertainty is decreasing along the first  and increasing along the second independent variable, corresponding to  the first and second term on the right hand side of Eq. \ref{eq:89}.}
For this analysis, the time lags have been restricted to the lags of H$\beta$ only.
Each object has the same weight in the regression but  we do not, in fact, expect the luminosity and lag to be the same in campaigns during different years as for NGC 5548. The left panel of Fig. \ref{fig:fig1} shows each individual data point for the relative uncertainty in time lag  from the monitoring campaigns included here.  {Blue plane shows the best multiple-linear regression} model to the relationship given by  Eq. \ref{eq:89}.

\begin{table}
	\caption{Comparison of the prediction of  cadences  from models derived from empirical   $\Delta t_{\rm E}$ and  artificial $\Delta t_{\rm A}$  data  sets,   assuming the flux errors of $5\%$, $0.01\%$, respectively; and time lag uncertainty  $\sim10\%$. Luminosities 
	$L[\mathrm{erg}\, \mathrm{s}^{-1}] $ are calculated from given fiducial time lags $\tau$ using  the R-L relation \citep{10.1088/0004-637X/767/2/149}.   }             
	\label{tab:cadlag}      
	\centering                          
	\begin{tabular}{c c c c cc}        
		\hline\hline                 
	$F_\mathrm{var}[\%]$& $\tau[\mathrm{ld}]$ & $L[\mathrm{erg}\, \mathrm{s}^{-1}] $& z&$\Delta t_{\rm E}[\rm days] $&$\Delta t_{\rm A}[\rm days] $ \\   
	(1)& (2) & (3)& (4)&(5)&(6) \\    %
		\hline                        
		& & & 0 & 63&83.9\\      
		 &100  &  $5.8\cdot 10^{46}$   & 1& 31.2&48.9\\
		  &  &     & 4& 12.6&20.1\\
		   10&  &     & 7& 7.9&12.4 \\
		\cline{2-6}
		   		 &  && 0 & 6.2&9.9\\      
		   &  &    & 1& 3.2 &4.83\\
		   & 10 &   $7.75\cdot 10^{44}$   & 4& 1.3& 1.89\\
		   &  &     & 7&0.8&0.9 \\
		   
		   \hline
			 &  & & 0 & 95&110\\         
		    &  100&  $5.8\cdot 10^{46}$   & 1& 47.7&95\\
		     &  &     & 4& 19.1&37.2 \\
		   20    &  &     & 7& 11.9& 23.5\\
		   \cline{2-6}
		       		   		 &  & & 0 & 10&12.9\\      
		       & 10 & $7.75\cdot 10^{44}$    & 1& 4.8 &9.64\\
		       &  &     & 4& 1.9&3.8 \\
		       &  &     & 7& 1.2& 2.2\\		       
		\hline                                   
	\end{tabular}
\end{table}

\begin{table}
	\caption{The same as Table  \ref{tab:cadlag}   but for the fiducial  luminosities  $L[\mathrm{ergs}\, \mathrm{s}^{-1}] $ at given  redshift  \citep{10.1086/505646/pdf} while  fiducial time lags $\tau$  are obtained using  the R-L relation \citep{10.1088/0004-637X/767/2/149}.  }             
	\label{tab:cadlag1}      
	\centering                          
	\begin{tabular}{c c c c cc}        
		\hline\hline                 
		$F_\mathrm{var}[\%]$& $\tau[\mathrm{ld}]$ & $\log L[\mathrm{ergs}\, \mathrm{s}^{-1}] $& z&$\Delta t_{\rm E} [\mathrm{days}]$&$\Delta t_{\rm A} [\mathrm{days}]$ \\    
			(1)& (2) & (3)& (4)&(5)&(6) \\    %
		\hline                        
		&11.5 & $45$ & 1 & 3.67&5.8\\      
		10	&39.2  &  $ 46$   & 2& 8.33&13.36 \\
		& 72.4 &  $ 46.5$  & 3& 11.45&18.29 \\
		\cline{2-6}
		&11.5 & $45$ & 1 & 5.6&10.9\\      
		20	&39.2 &  $ 46$   & 2& 12.7&24.4 \\
		& 72.4 &  $ 46.5 $  & 3& 17.5&34.2 \\
		\cline{2-6}
		
		\hline                                   
	\end{tabular}
\end{table}

\begin{table}
	\caption{The same as Table  \ref{tab:cadlag}   but   for the periodicity. }             
	\label{tab:perlag}      
	\centering                          
	\begin{tabular}{c c c cc }        
		\hline\hline                 
		$F_\mathrm{var}[\%]$& $P[\mathrm{yr}]$ & z&$\Delta t_{\rm E} [\mathrm{days}]$ &$\Delta t_{\rm A} [\mathrm{days}]$\\    
			(1)& (2) & (3)& (4)&(5) \\    %
		\hline                        
		&  &  0 & 29.2&80.3\\      
		& 5 &      1&14.6 &62.1 \\
		&  &      4& 7.3&47.5\\
		10&     &  7& 3.7 &32.8\\
		 \cline{2-4}
		&  &  0 & 18.3 &62.1\\      
		& 3 &      1& 11 &54.7\\
		&  &      4& 4 &36.5\\
		&  &      7& 2.2&25.6\\
		
		\hline
		&  &  0 & 91.3&92\\         
		& 5 &   1& 62.1&73\\
		&  &     4& 25.6 &29.2\\
		20    &  &    7& 14.6&18.25\\
		\cline{2-4}
		&  &  0 & 65.7&69.4\\      
		& 3 &      1& 36.5&40.2 \\
		&  &      4& 14.6&18.25\\
		&  &      7& 9.1 &11\\		       
		\hline                                   
	\end{tabular}
\end{table}


{One can now ask the question: What would be a suitable cadence for detecting certain AGN observable at the level of formal error of $10\%$ if setting the fiducial light curve variability and flux error.}
Based on the  multiple regression model best-fit of empirical data (E) and artificial data (A), and for the time lag error of $10\%$,   we can vary parameters in {both independent variables in order to get cadence estimates}.

The resulting prediction  (see Table \ref{tab:cadlag}) shows detailed information on each cadence which includes  assumed variability, redshifts, time lags and luminosities. A close inspection of  Table \ref{tab:cadlag} shows 
{   that required cadences are smaller for objects with  larger redshift, {and for smaller time lags}}
as expected. Also, two times larger object light curve variability ($F_\mathrm{var}\sim 20\%$) allows  larger cadences for two different fiducial time lags (luminosities) of potential targets. 

\begin{figure*}
	\centering
	\includegraphics[clip,trim=100 30 30 30,width=0.49\textwidth]{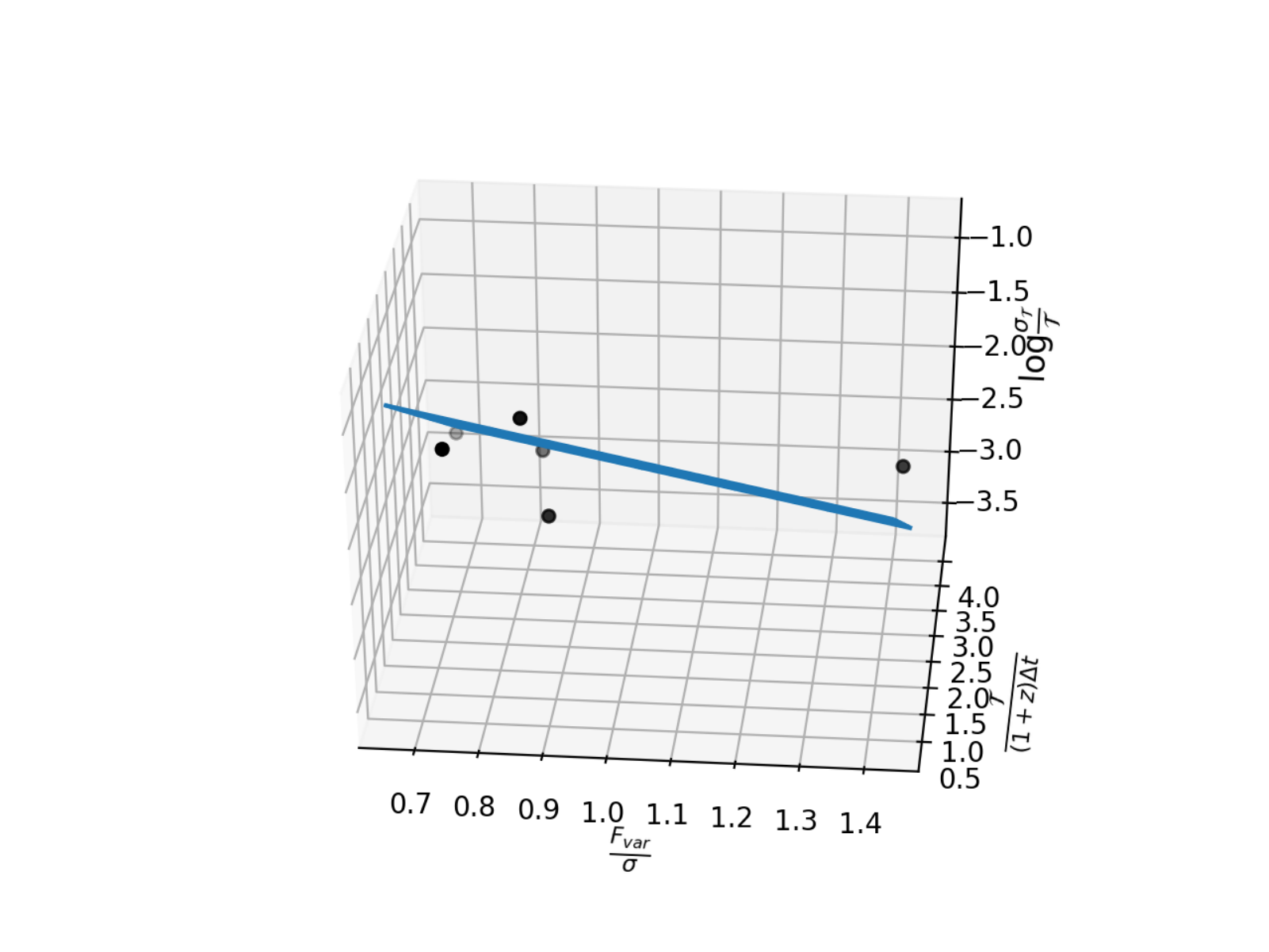}
	\includegraphics[clip,trim=110 35 35 70,width=0.49\textwidth]{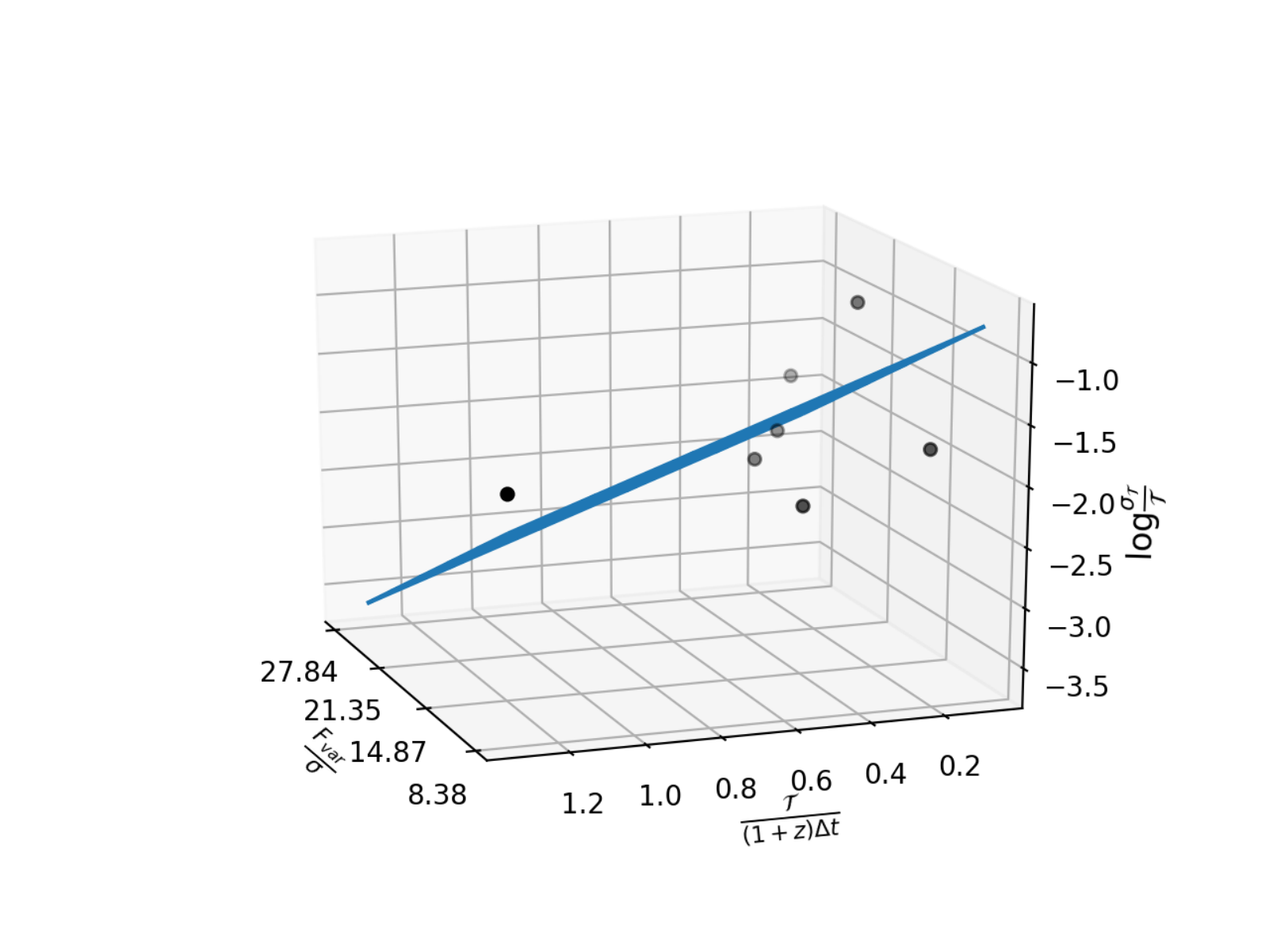}
	\caption{{Same as the comment for Fig. \ref{fig:fig1} but } for artificial set of {   light curves with OpSim cadences}. Filled circles denote data given in Table \ref{tab:Opsimgen}. {\textit{Left}: $C_{1}=-1.09\pm0.29, C_{2}=-0.49 \pm0.12$, $\sigma$ is of the order of 0.01; \textit{Right}: $C_{1}=-0.0008 \pm 0.0001, C_{2}= -1.089\pm 0.5$, $\sigma$ is of the order of 0.005.}}  
	\label{fig:fig1opsim}%
\end{figure*}

Cadence of {the order of tens of} days is qualitatively sufficient for time lag estimates of 100
days for light curves of smaller and larger variability. Also, ten times smaller time lags would require ten times
smaller cadence.
A relationship between AGN luminosity and redshift (L-z) can  also be combined with the R-L relation to give a rough estimate of the required cadence for different luminosities up to $z\sim 3$ (Table \ref{tab:cadlag1}).  For given redshift, the luminosity is taken from empirical  L-z relation reported in \citep{10.1086/505646/pdf}, while  R-L relation  \citep{10.1088/0004-637X/767/2/149} provides corresponding time lag.
It is expected that with larger lags {(i.e. luminosities)} cadences increase.
  Similarly, we estimated the  proxies (see the left panel in Fig. \ref{fig:fig1opsim}) for the set of data (A) based on OpSim cadences (see Table \ref{tab:Opsimgen}, last column). 
 
 {Cadence  predictions obtained from  the two model versions  are of the same of order so that  the two models  are in relative  good agreement (see Table \ref{tab:cadlag} and \ref{tab:cadlag1}, last columns).}

  {Even though  LSST OpSim cadences have gaps, we recovered lags which  are consistent within $3\sigma$ with the $R_{\rm BLR}$ of the input model. The success rates can be boosted by using   deep learning to deal with  gapped light curves \citep[][]{10.3847/2041-8213/ab3581}. }

\begin{figure*}
	\centering
	\includegraphics[width=0.49\textwidth]{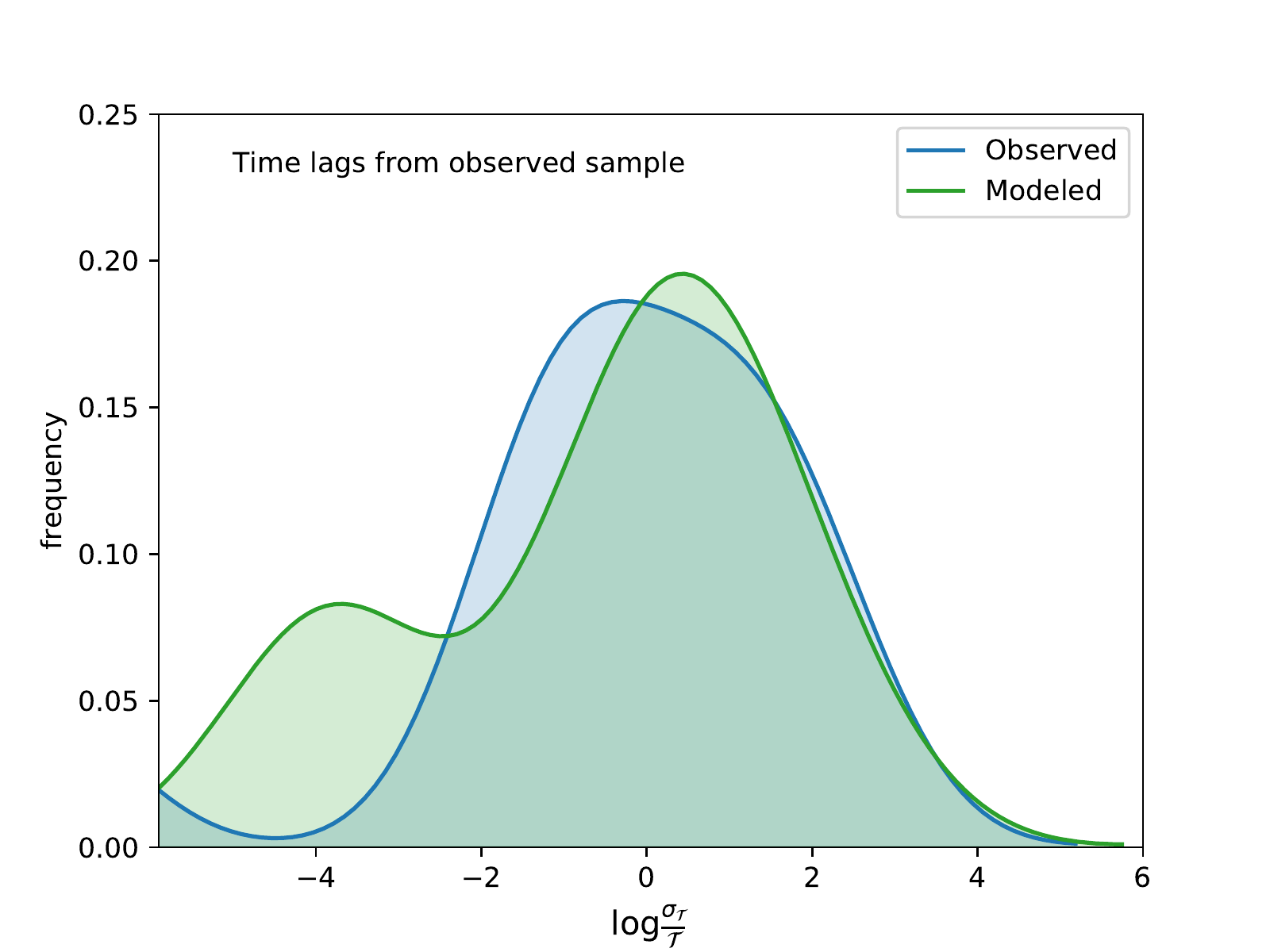}
	\includegraphics[width=0.49\textwidth]{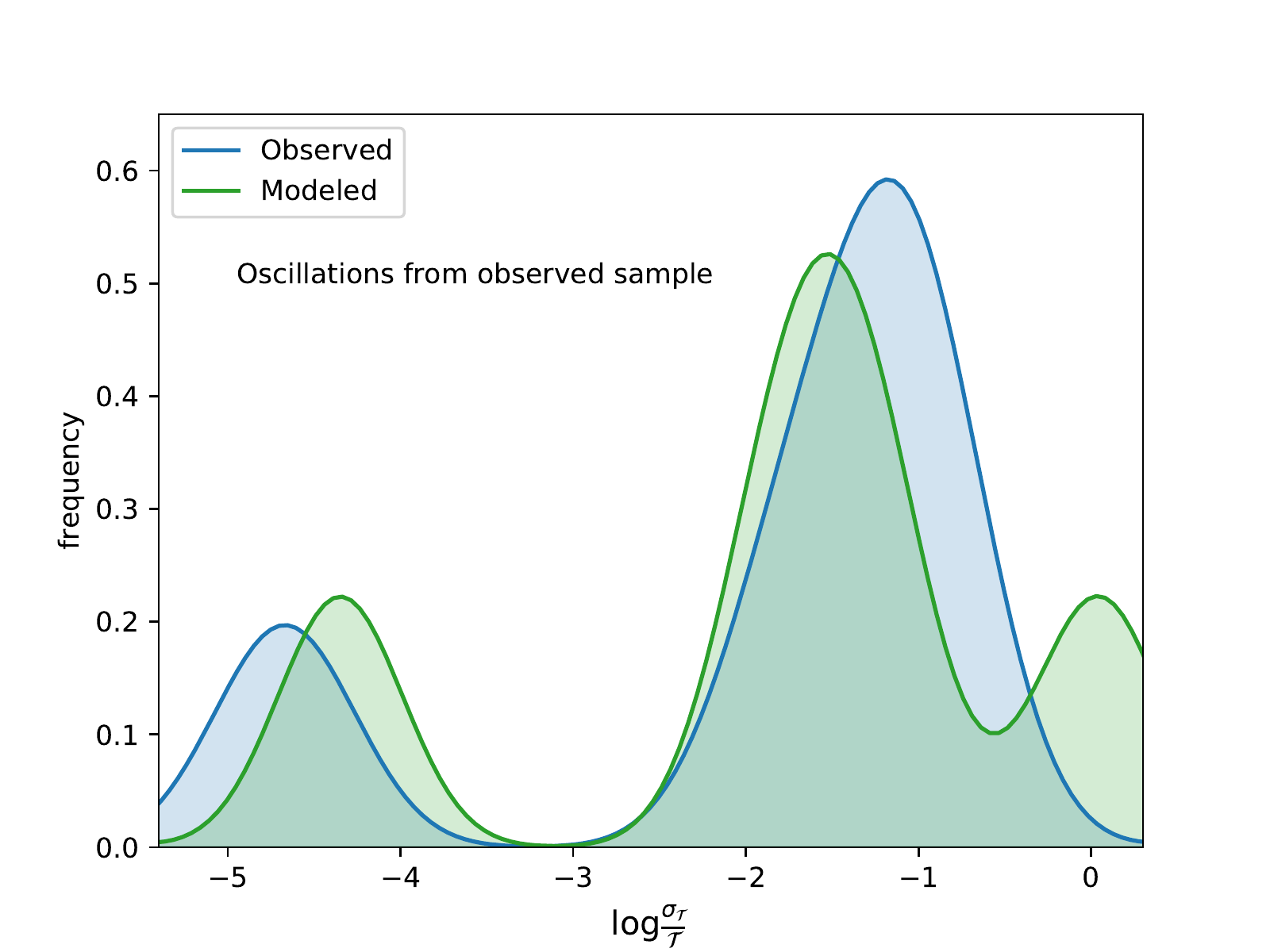}
	\caption{{Probability density functions of formal errors  from observed data and their predicted  values from mutliple regression for  time lags (left) and oscillations (right).} }
	\label{fig:fig1dist}%
\end{figure*}

\begin{figure*}
	\centering
	\includegraphics[width=0.49\textwidth]{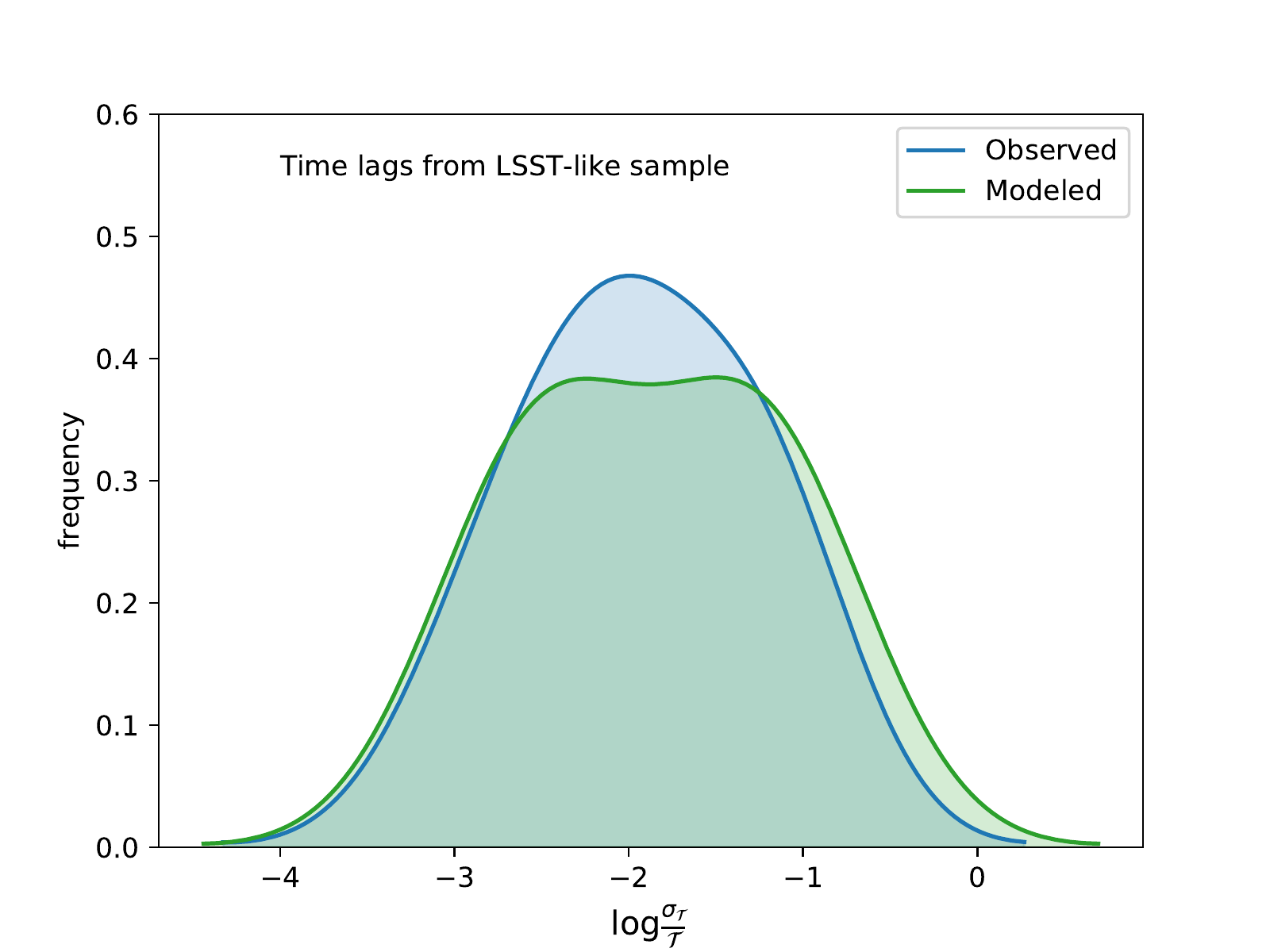}
	\includegraphics[width=0.49\textwidth]{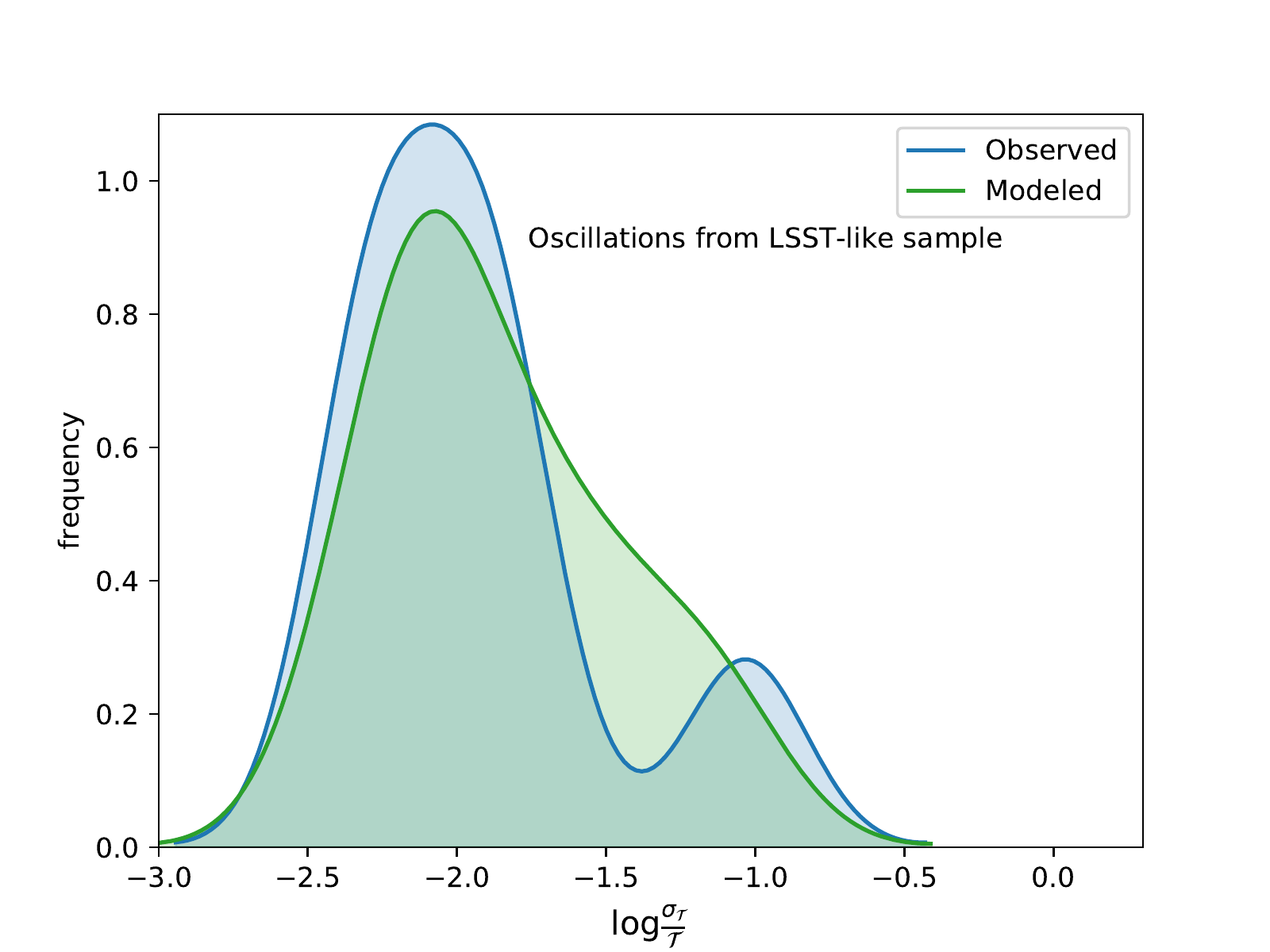}
	\caption{{The same as Fig. \ref{fig:fig1dist} but for artificial data which are also referred as 'observed'.  } }
	\label{fig:fig2dist}%
\end{figure*}

{
To compare multiple regression model prediction of $\log\frac{\sigma_{\mathcal{T}}}{\mathcal{T}}$ with those values obtained from  data sets we used density estimator. Such estimator is  an algorithm which takes a dataset and produces an estimate of probability distribution which that data is drawn from.
The inferred distributions from model results and data we will call descriptors and they are compared against each other.
Particularly, we implemented Kernel density estimation (KDE) which uses mixture consisting of one kernel component per point in the considered data set, resulting in an essentially non-parametric estimator of density \citep[see e.g.,][]{10.1080/24709360.2017.1396742}. There are several versions of kernel density estimation implemented in Python (notably in the SciPy and StatsModels packages). The density estimate at a point 
y  within a group of points $\left\{x_{i}\right\}, i=1,N$ is given by: 
\begin{equation}
f_{K}{y}=\sum^{N}_{i=1} K(y-x_{i};h)    
\end{equation}
\noindent where in our case 
$$ K(x;h)\propto e^{-\frac{x^{2}}{2h^{2}}}.$$}
The left plot in Fig. \ref{fig:fig1dist} displays the probability  density of  formal errors
of time lags inferred from the observed data and  from multiple regression predictions. Larger discrepancies can occur in the left tail of  observed formal errors. Similarly, the left plot in Fig.  \ref{fig:fig2dist} shows the same information but for artificial data set. The model much better perform on this data set. 

\subsection{Cadence estimates for oscillation detection}
Now we consider the cadence required for reliable detection of oscillation, i.e. periodicity in light curves. Here we will  repeat the procedure as for the time lag cadence. 
The relationship between the uncertainties for {detected} periods (see Table \ref{tab:data}) and  proxy variable (Eq.\ref{eq:89}) is shown on the right panel of Fig. \ref{fig:fig1}.
    
Full details of the predicted cadences for two hypothetical oscillations {(3 and 5 years)}  in decade long light curve are given in Table \ref{tab:perlag}.

For larger number of cycles of underlying signal, {the predicted} cadences are {somewhat smaller}. For example, the light curve with $F_\mathrm{var}\sim 20\%$ and underlying {   rest-frame periodicity} of 3 years at redshift $z\sim 4$ would require cadence of $\sim 14.6$ days. Such cadence would be sufficient for detection of a 5 year {   rest-frame oscillation} (at similar level of light curve variability $F_\mathrm{var}\sim 20\%$), for an object at $z=7$ ({see Table \ref{tab:perlag})}.

We {performed} multiple regression (see right panel in Fig. \ref{fig:fig1opsim}) {on} the set of artificial objects designed to have cadences from several OpSim runs (see Table \ref{tab:Opsimgen}).

While results of   models based on artificial and empirical data  are in good agreement for $F_{var}\sim 20\%$, there are considerable differences for $F_{var}\sim10\%$, (details in Discussion section).

{
The right plot in Figs. \ref{fig:fig1dist} shows the probability  density of  formal errors
of oscillations inferred from the observed data and  from multiple regressions. 
The distribution is multimodal, but model is relatively close to the measured formal errors. However, the distribution of artificial errors is bimodal (not multimodal as in the case of observed data). The distribution of predicted formal errors is broadened to capture both peaks of measured errors.
The right plot in Fig.  \ref{fig:fig2dist} shows the same  but for artificial data set.  }

\subsection{ Structure Function Results}
{SF, as a classical method,
has been  used to detect the periodicity and {timescale of AGN variability} in different {observing} bands \citep{10.1007/s10509-017-3079-y, 2019PASP..131f3001M}. It is believed that SF is suitable to handle unevenly sampled time {series data}. Thus, we also {analyze} {the influence} of different  cadences on  the first-order SF method applied on AGN light curves with underlying oscillations. The ideal and LSST-like cadences were used for SF construction. }

\subsubsection{Simulated cadences}

Firstly, we {investigate} individual SFs for the "gappy" artificial light curves (left panel, Fig. \ref{fig:fig10}) and variable cadence (left panel, Fig. \ref{fig:fig11}), in order to compare cadence effects.

Right panel in  Fig. \ref{fig:fig10} shows that the {SFs} of the "gappy" light curve with {   continuous observations} of {9 months (red), 6 months (purple), 3 months (cyan), with respect to the SF of the reference light curve with homogeneous 1-day cadence (black).} {The SF for 3 and 6 months} {of observations} {shows a large deviation} SF of the ideal light curve of homogeneous 1-day cadence. 
 
For variable-cadence light curves, we can expect larger deviations of SF than in previous case (see middle panel in Fig. \ref{fig:fig11}), but periodic signal is clearly visible  at the largest timescales of the SF in comparison to the ideal light curve with homogeneous 1-day cadence. The right panel shows the same SF as in the middle panel but on logarithmic scale. Deviations at smallest timescales appear, whereas oscillatory patterns are persistent, but less prominent.

\begin{figure*}
	\centering
	\includegraphics[clip,trim=35 11 48 47,width=0.33\textwidth]{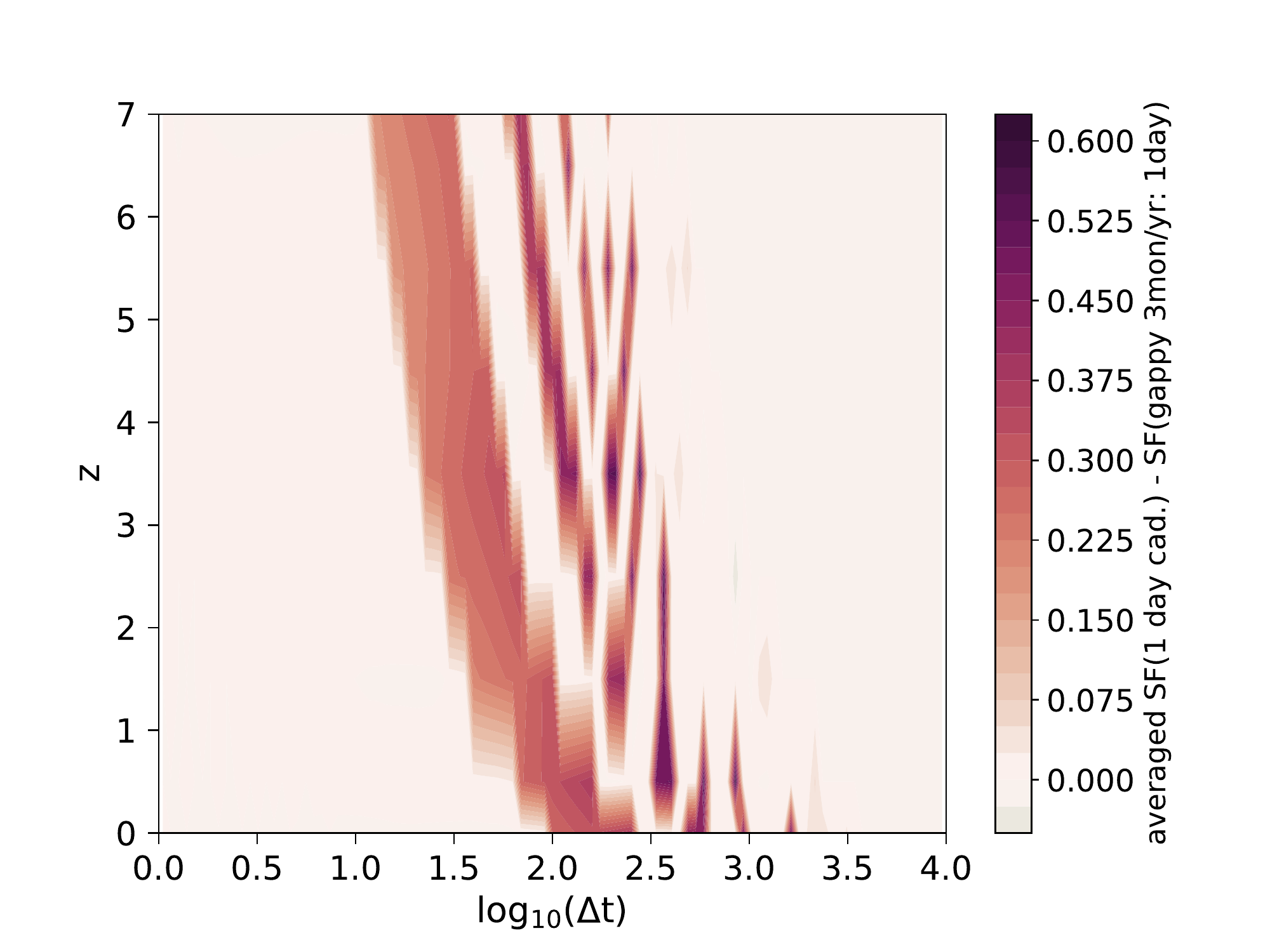}
	\includegraphics[clip,trim=35 11 48 47,width=0.33\textwidth]{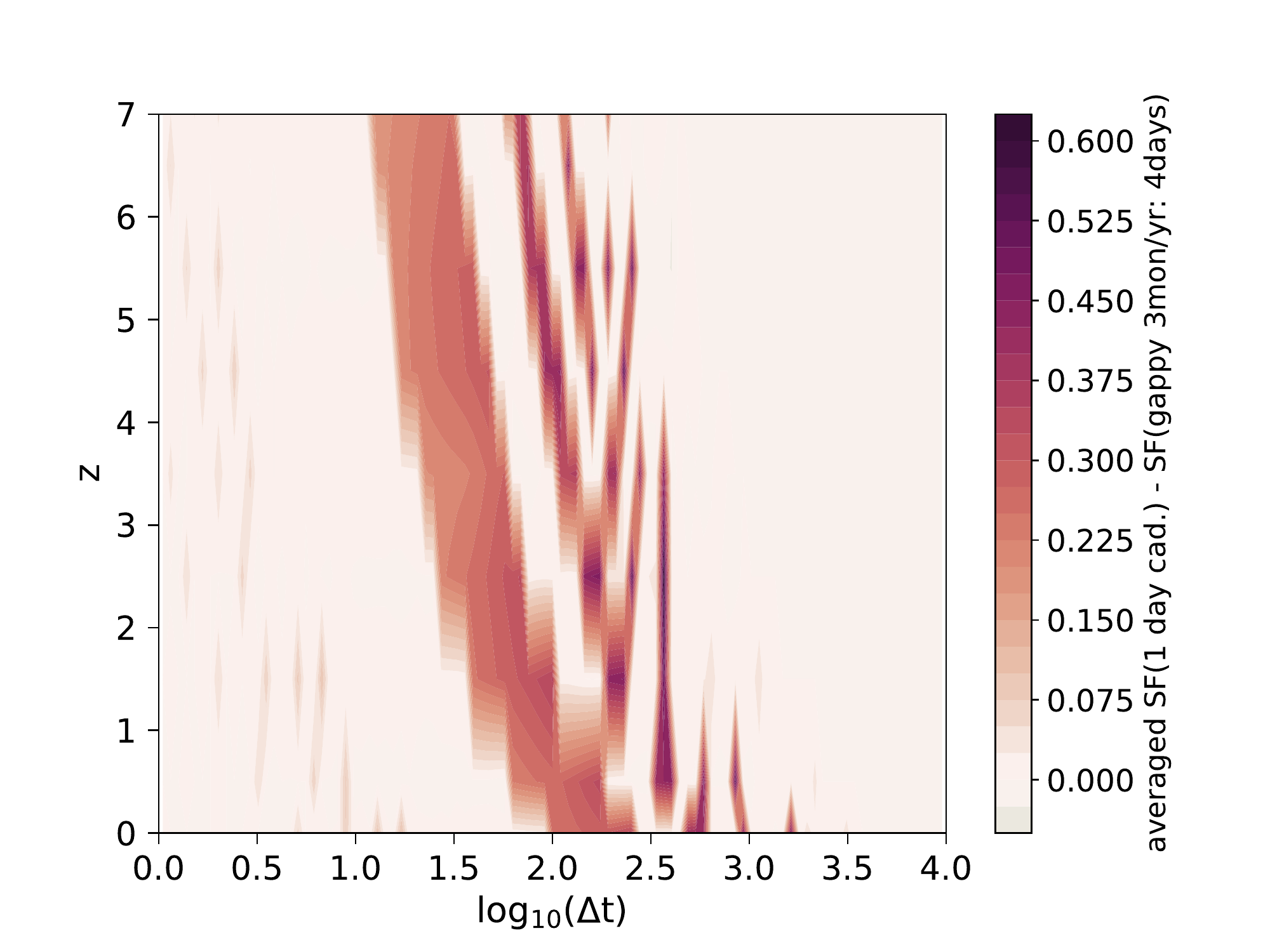}
		\includegraphics[clip,trim=35 11 48 47,width=0.33\textwidth]{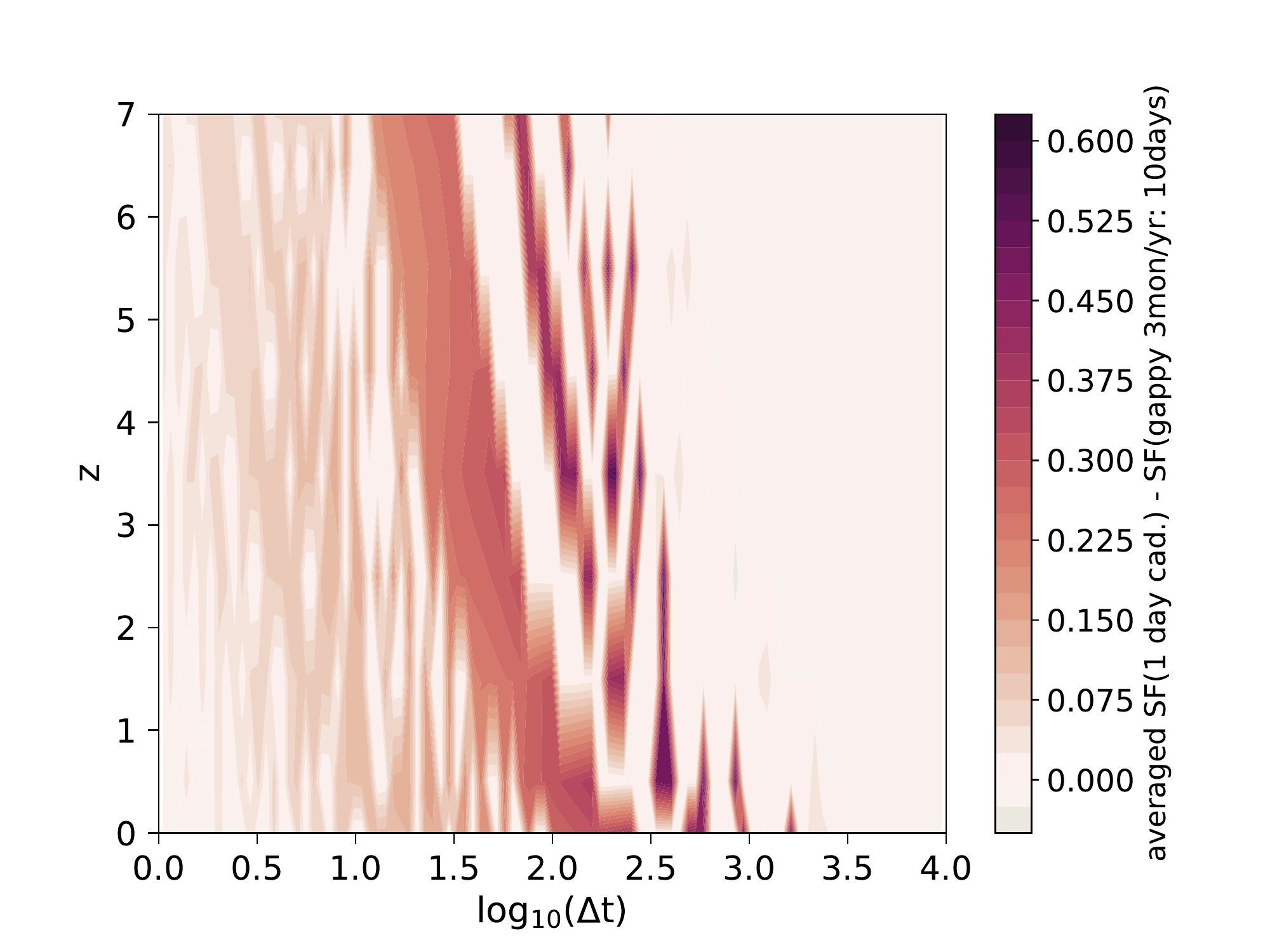}
	\caption{Heatmaps of deviation of SFs for  cadence 3 month/yr in the rest frame of quasar. From left to right: the sampling rate is 1 day in observed 3 months/yr;   4 days sampling rate in observed 3 months/yr; and 10 days sampling rate. Colorbar represents deviations. Positive deviations stand for SFs when values of homogeneous curve are larger than SF of gaped curve  in average per bin and vice versa. }
	\label{fig:fig2}%
\end{figure*}
%

\begin{figure*}
	\centering
	\includegraphics[clip,trim=35 11 27 47,width=0.33\textwidth]{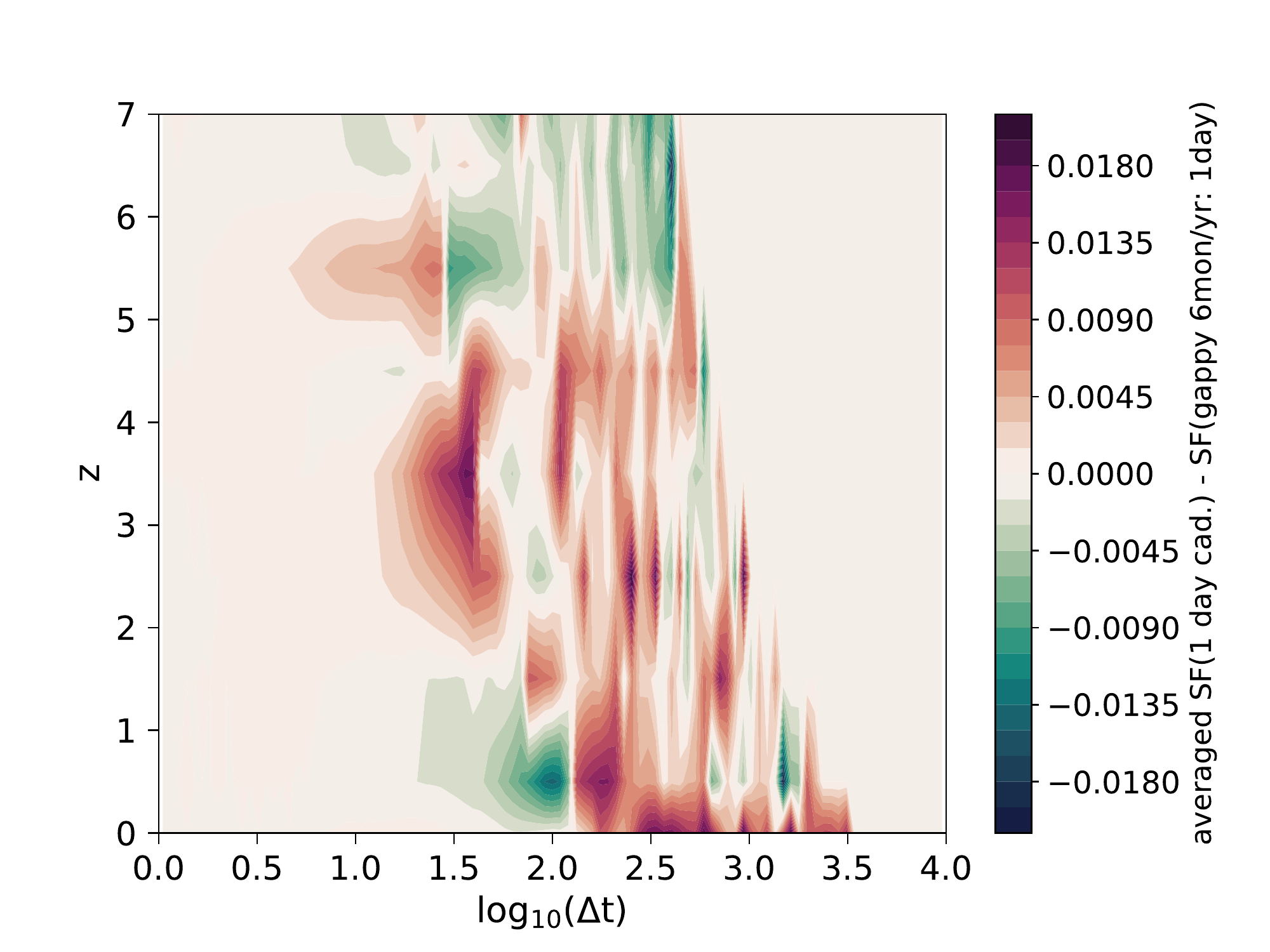}
	\includegraphics[clip,trim=35 11 27 47,width=0.33\textwidth]{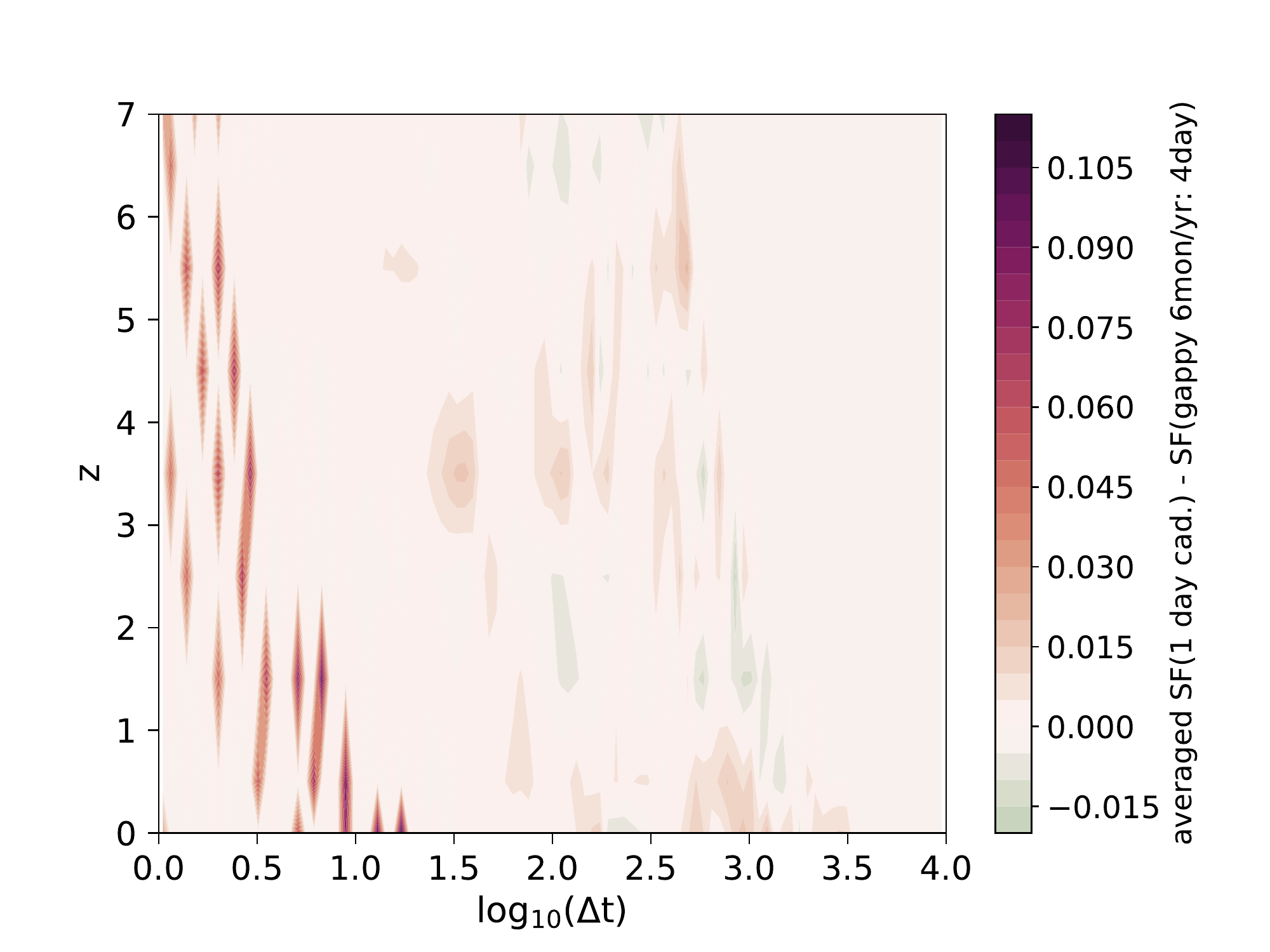}
	\includegraphics[clip,trim=35 11 27 47,width=0.33\textwidth]{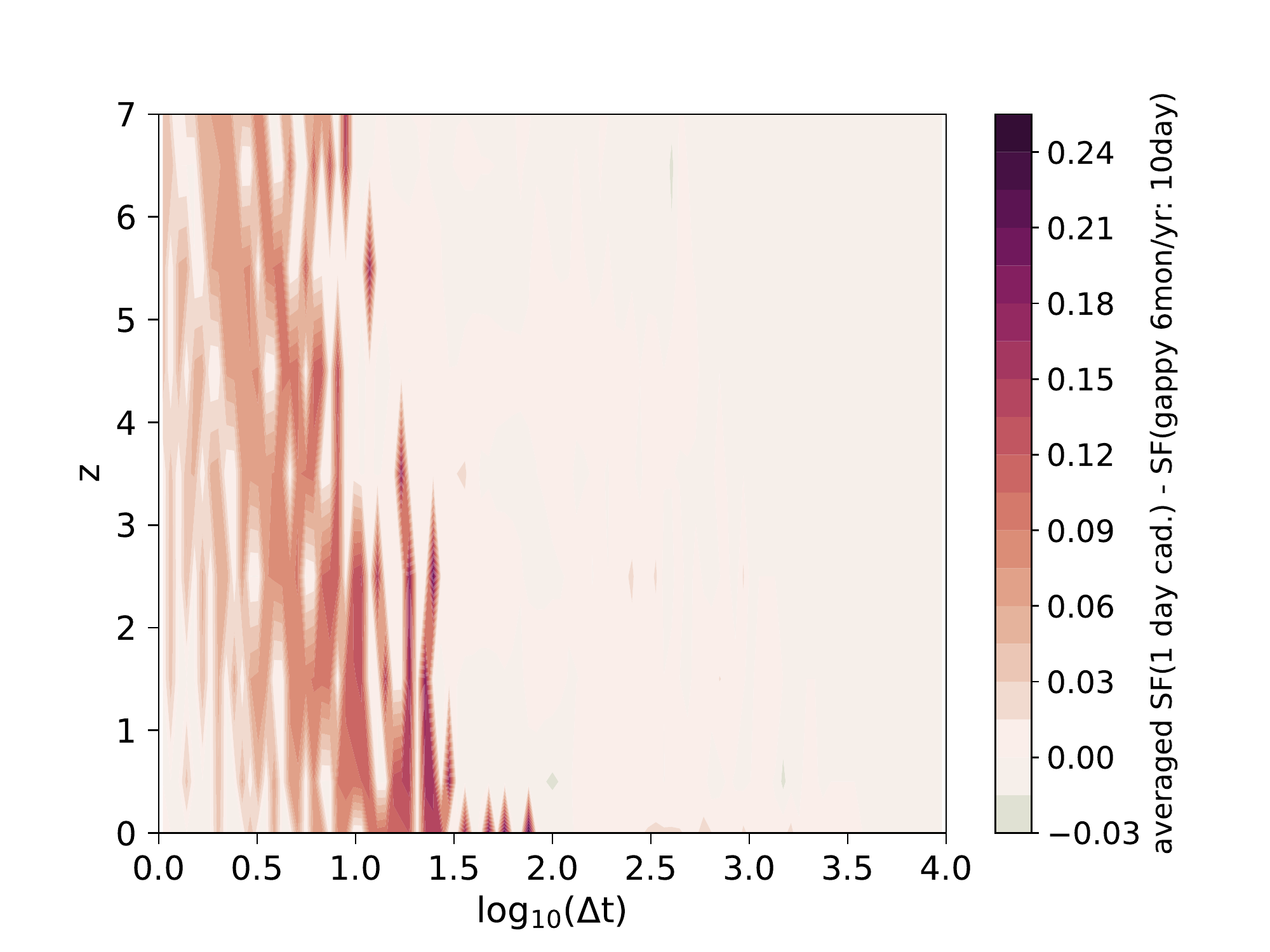}
	\caption{The same as Fig. \ref{fig:fig2} but  for  cadence 6 month/yr  in the rest frame of quasar.  }
	\label{fig:fig3}%
\end{figure*}

\begin{figure*}
	\centering
	\includegraphics[clip,trim=35 11 27 47,width=0.33\textwidth]{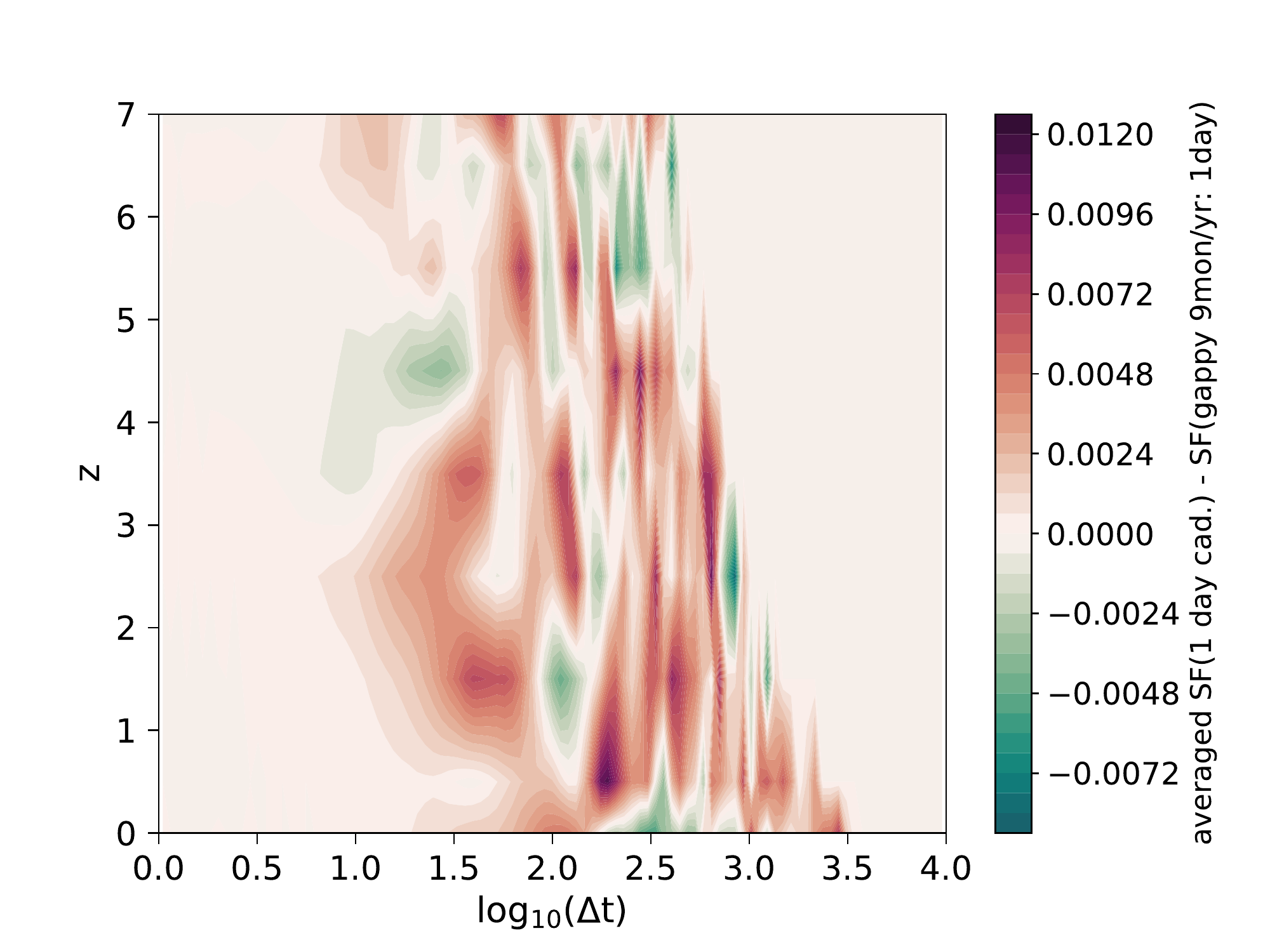}
	\includegraphics[clip,trim=35 11 27 47,width=0.33\textwidth]{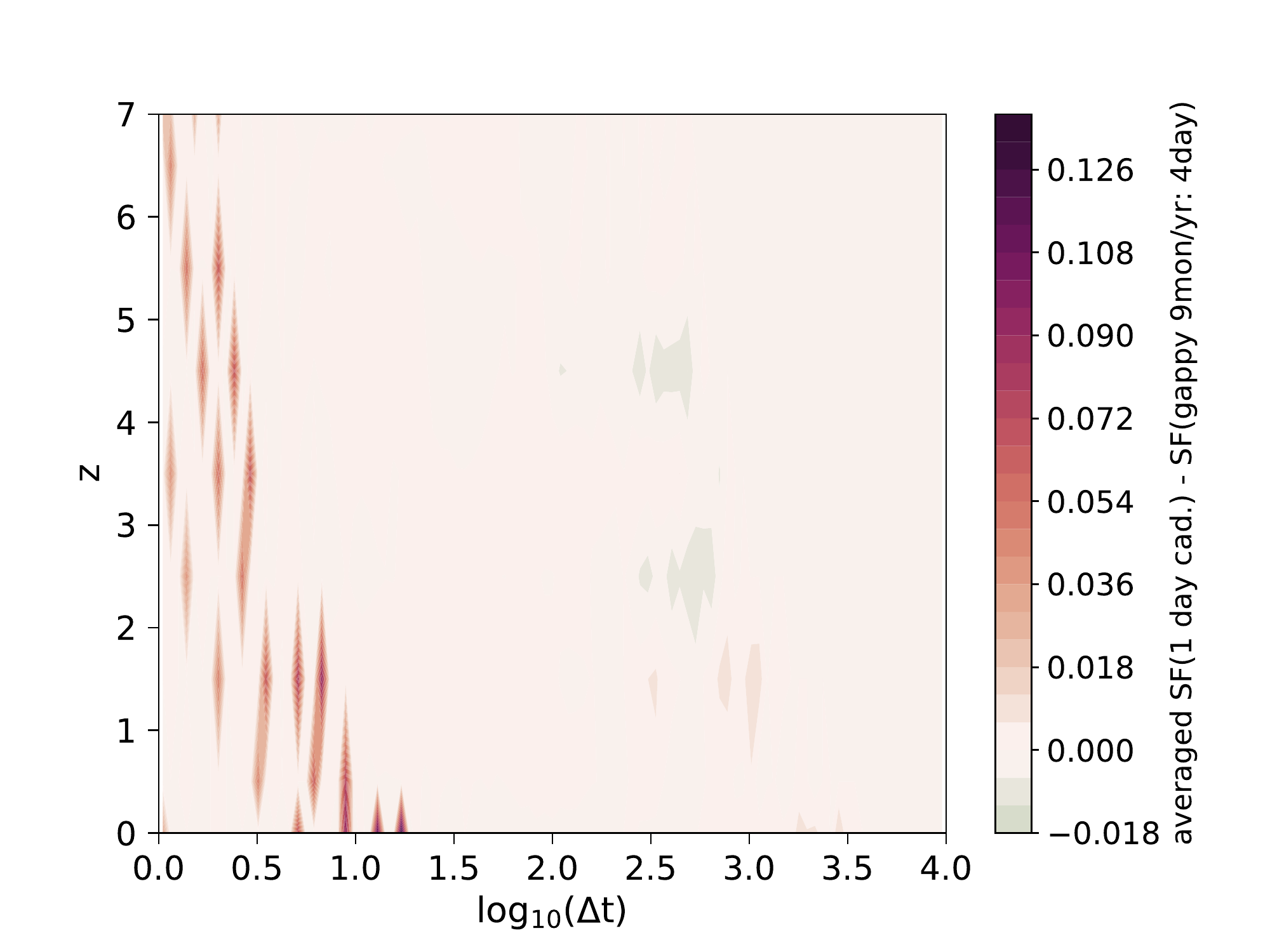}
	\includegraphics[clip,trim=35 11 27 47,width=0.33\textwidth]{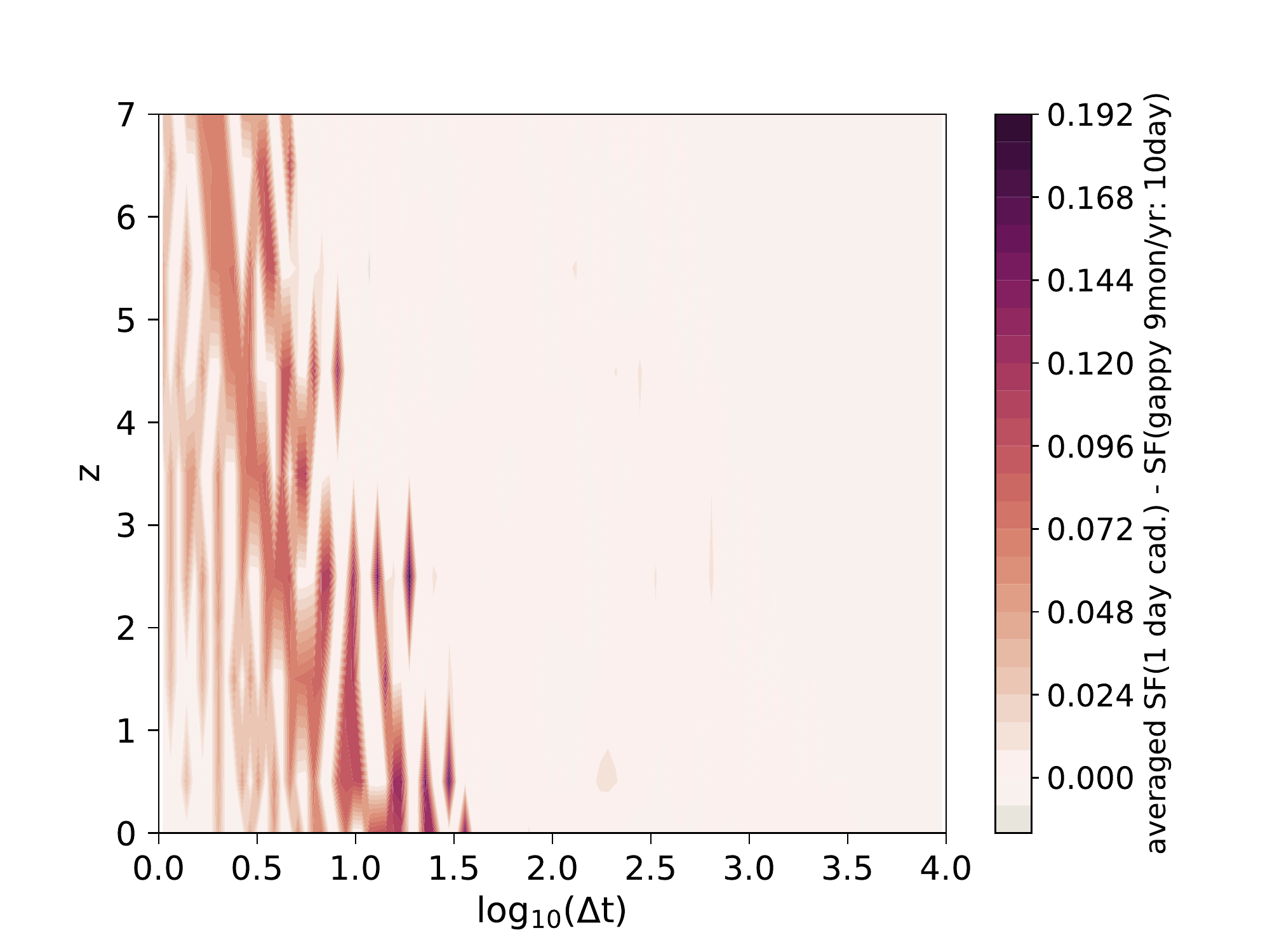}
	\caption{The same as Fig. \ref{fig:fig2} but   for cadence 9 month/yr  in the rest frame of quasar. }
	\label{fig:fig4}%
\end{figure*}
\begin{figure}
	\centering
	\includegraphics[width=\linewidth]{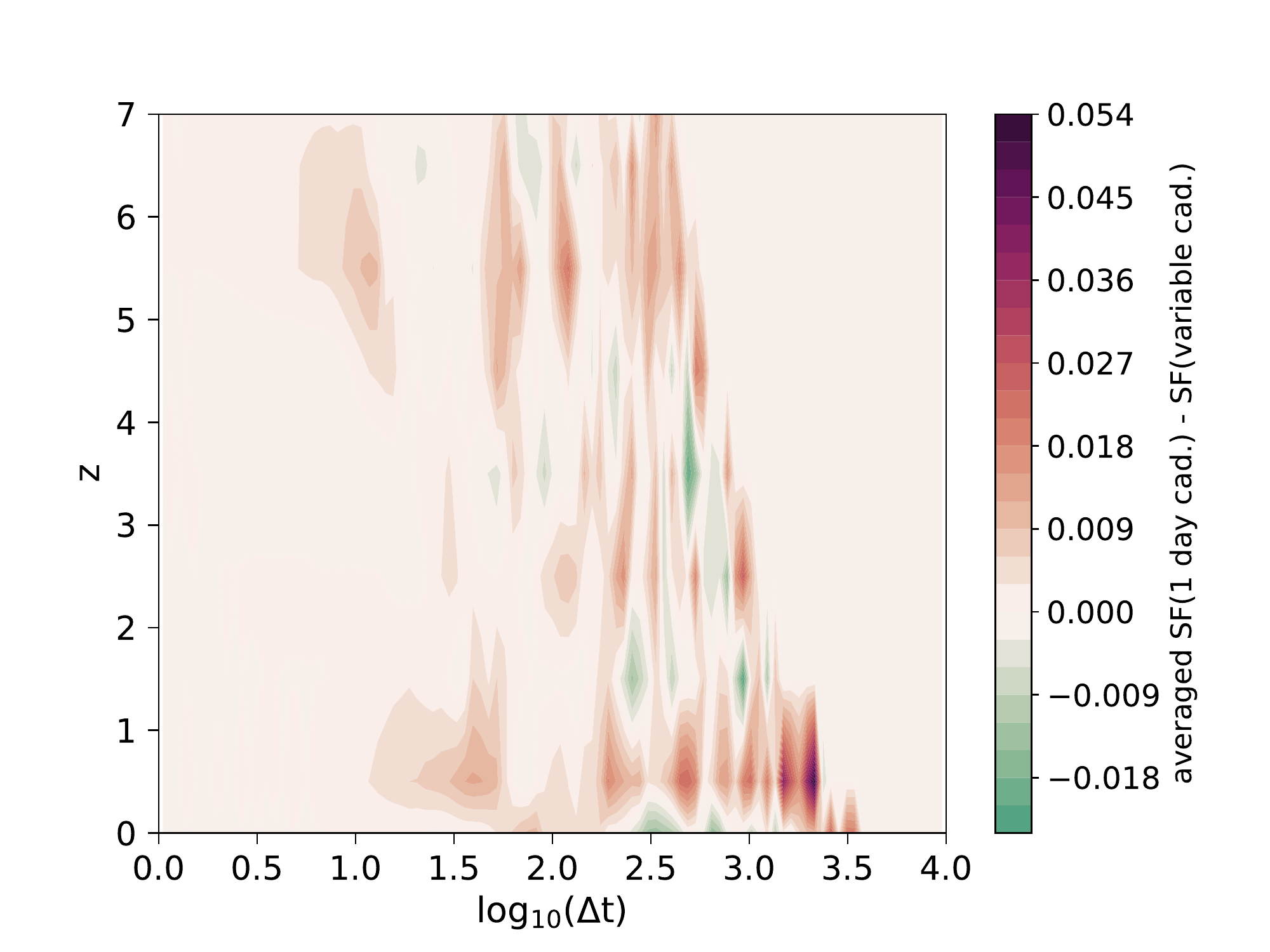}
	\caption{The same as Fig. \ref{fig:fig2} but   for  variable cadence comprising 3 months observed with 1 day sampling followed by  6 months with 30 day sampling, with exception of the first year when the cadence of 3 month is present.   }
	\label{fig:fig5}%
\end{figure}

Further, we consider deviations of SFs of an ensemble of homogeneous light curves across redshifts. For each redshift bin $(z_{bin} =\left\{i|\, i\in[0.5,6.5], \Delta i=0.5 \right\})$, we generate {   a 10-year long artificial light curve  based on DRW model, setting continuous 1-day cadence and underlying oscillatory signal according to described procedure}. 

Then, we apply to these light curves the "gappy" {   observing strategy with 3 months/yr, 6 months/yr and 9 months/yr of observations of different cadence:  1, 4  and 10 days. This would} generated 50 light curves for each redhsift bin and {   three "gappy" observing strategies}. 

For each "gappy" light curve, we estimate the $SF_{gappy}$.  For each redshift bin, we derive the averaged deviation curve $M^{i}$.
Then we plot $M^{i}$ vs. redshifts $z$ with projected SF timescales $\Delta t$ in the form of heatmaps. {   Below we discuss the results of these three observing strategies.}

 Case i) The heatmaps of SF deviations for {   observing strategy of} of 3 months/yr are given in Fig. \ref{fig:fig2} for sampling rates of 1 day (left panel), 4 days (middle panel) and 10 days (right panel). The deviations form 'evolutionary tracks' for time scales between 1 and 3.0 {   are seen}. At smaller scale (below 1.5), SF deviations  became apparent  {while it becomes} 'noisy' when {the sampling} rates are larger (4 and 10 days - middle and right panel, respectively). {Changes at larger time-scales ($\Delta t>2$) are prominent also for larger cadences (4 and 10 days) at larger redshifts (middle and right panel in Fig. \ref{fig:fig2})}. We note that deviations can be either positive or negative following mild bent tracks across redshifts. Positive deviations mean that SF values of continuous curve are larger than SF of "gappy" curve and vice versa. Deviations became more 'noisy' at intermediate scales (1,2.5) for sparse sampling of 10 days. Dominant positive deviation (values in the range (0,2)) seen as blue track is persistent across redshifts and different sampling rates.

Case ii) For comparison, heatmaps of SF  deviations for  "gappy" {   strategy of continuous observations during} 6 month/yr are given in Fig. \ref{fig:fig3}. The maps are smoother then corresponding maps  for case i) cadence (Fig. \ref{fig:fig2}). Also, the evolutionary tracks of deviations are translated to larger  time scales beyond 2.5.  The deviations at smaller scales (below 1.5) became apparent and {become noisy} with larger cadence of 4 and 10 days. Deviations are   smaller then in the case i).
The blue track dominant for case i) "gappy" {   observing strategy}, disappears from heatmaps of 6 month/yr "gappy" {   observing strategy}.

Case iii) For 9 month {   continuous observations}, heatmaps in Fig. \ref{fig:fig4} are similar to the corresponding heatmaps for case ii), but deviations are smaller.  The evolutionary tracks of differences are attenuated at larger time scales. As expected, the noise appears at smaller time scales for larger samplings.

For variable-cadence light curve comprising 3 observed months with 1-day sampling and 6 months with 30-day sampling, with exception of the first year when the {   ideal 1-day cadence during 3 months} is present, the heatmap is given in Fig. \ref{fig:fig5}.  The evolutionary track of  deviations at time scale around 2.5 is prominent. The noise is present at intermediate time scales between 1 and 2. We note that it is similar to the heatmap of 6-month "gappy" {   observing strategy} with 1-day sampling but with reversed coloring (the left plot in Fig. \ref{fig:fig4}.)
It means that denser sampling within 3 months helps to get smaller deviations of SF of gapped light curves.

\subsubsection{OpSim Rolling cadences}

\begin{figure*}
	\centering
	\includegraphics[clip,trim=11 5 40 30,width=0.49\textwidth]{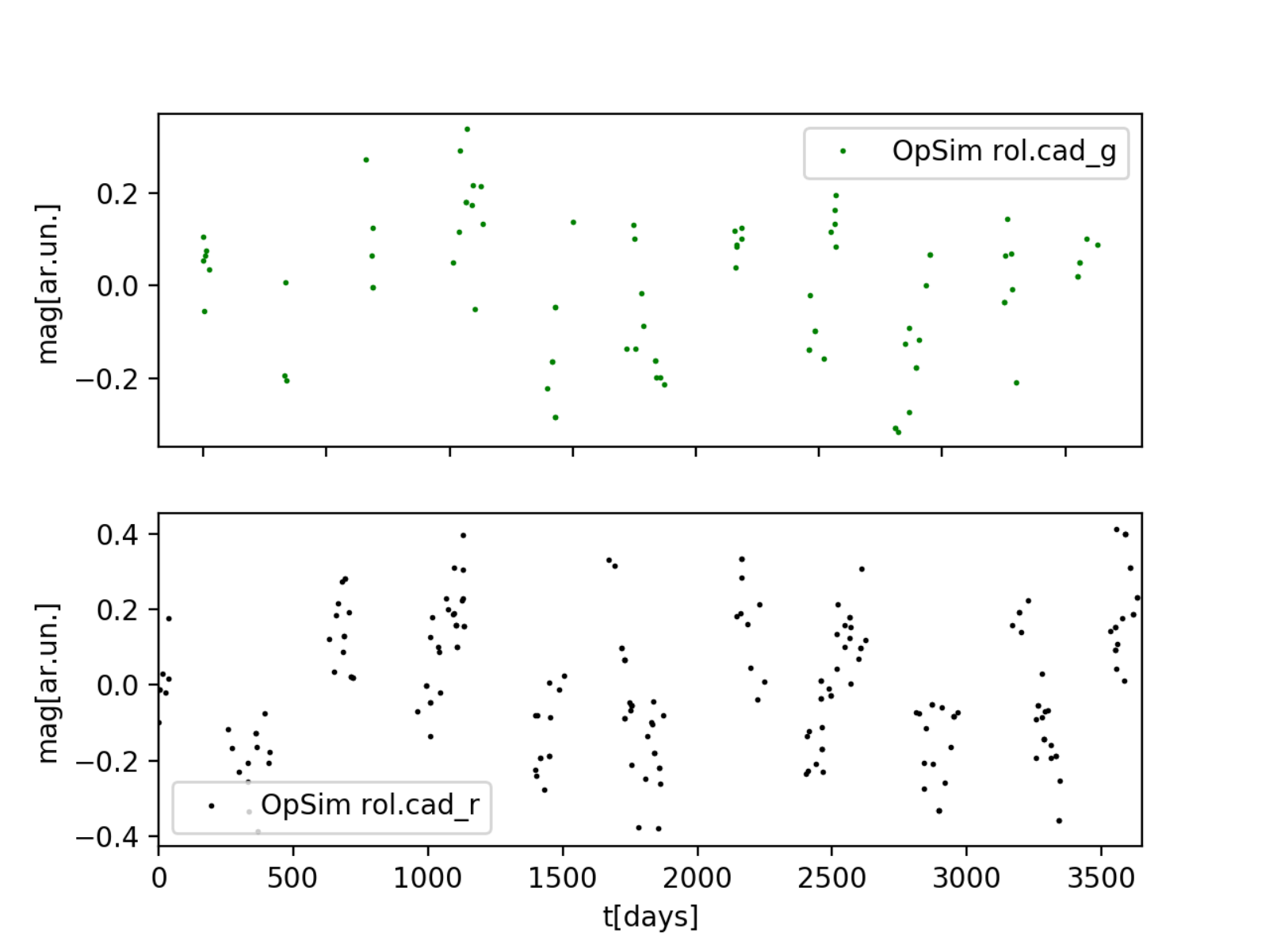}
	\includegraphics[clip,trim=5 5 40 30,width=0.49\textwidth]{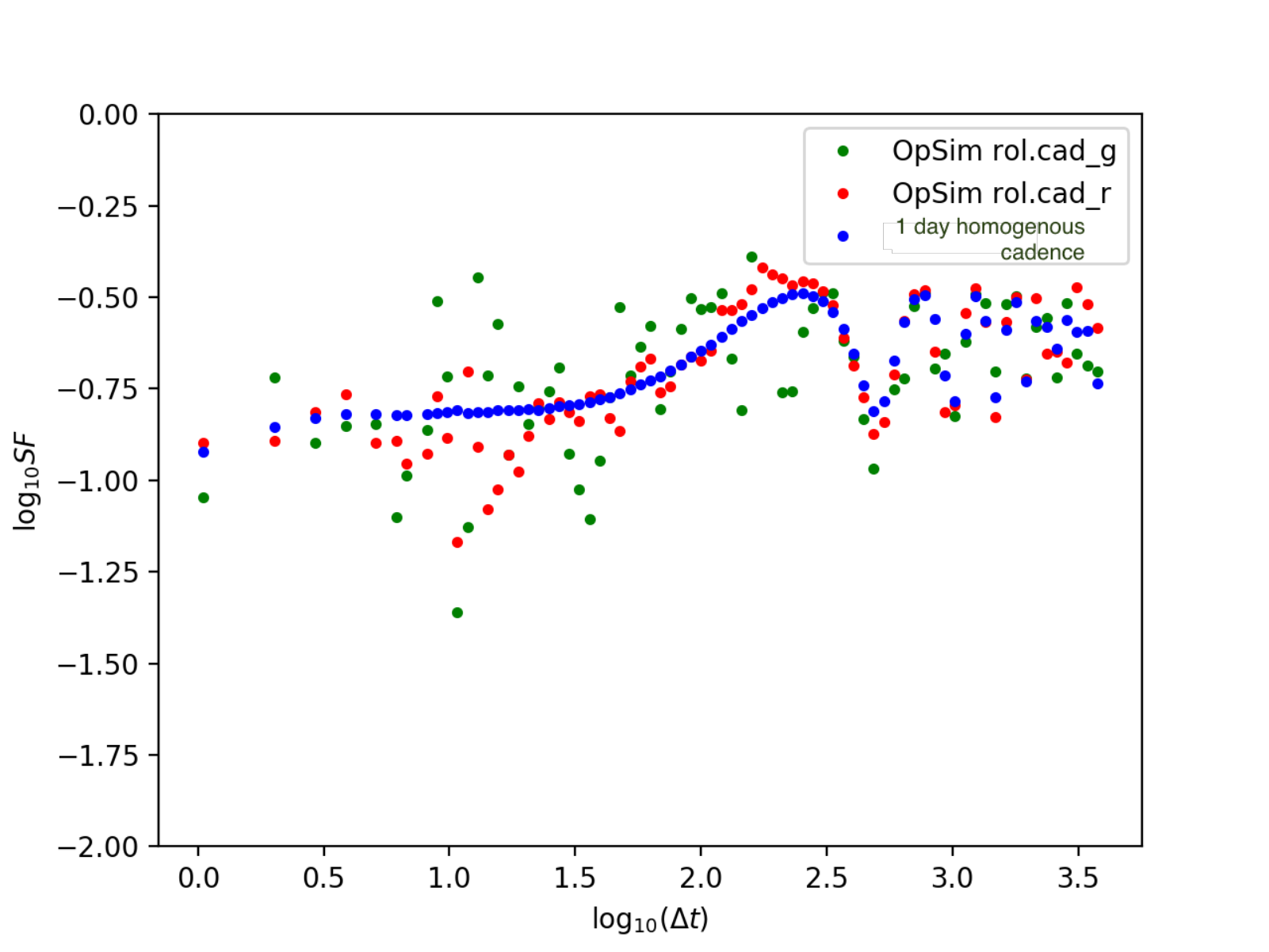}
	\caption{ \textit{Left}: Simulated AGN light curve using DRW and OpSim rolling cadences.
	From top to the bottom: light curve with	OpSim\_ roll.cad\_0.8\_g\_RA\_0.0\_D\_-10.0 cadence; light curve with OpSim\_ roll.cad\_0.8\_r\_RA\_0.0\_D\_-10.0. \textit{Right}: Structure function calculated for the light curves given in the left panel. Blue curve stands for SFs calculated for homogeneous 1 day cadence.}
	\label{fig:artifopsim}%
\end{figure*}

We generated  artificial  light curves with underlying oscillations that correspond to different cadences runs from OpSim outputs. OpSim\_ roll.cad\_0.8\_g\_RA\_0.0\_D\_-10.0   comprises 87 and OpSim\_ roll.cad\_0.8\_r\_RA\_0.0\_D\_-10.0  204 observations over 10 yr. A realization of these light curves is given in left panel of  Fig. \ref{fig:artifopsim}, for g and r filter OpSim rolling cadence. Corresponding SFs (right panel in Fig. \ref{fig:artifopsim} ) show larger deviations of SFs in g and {r} filter at time scales below 1.5, oscillations are present  after time scale 2.0.
In g filter, {a dip of the SF} is around time scale 2.25.  SF in r filter follows closely {SF$_0$}.

\begin{figure*}
	\centering
	\includegraphics[clip,trim=35 11 45 47,width=0.33\textwidth]{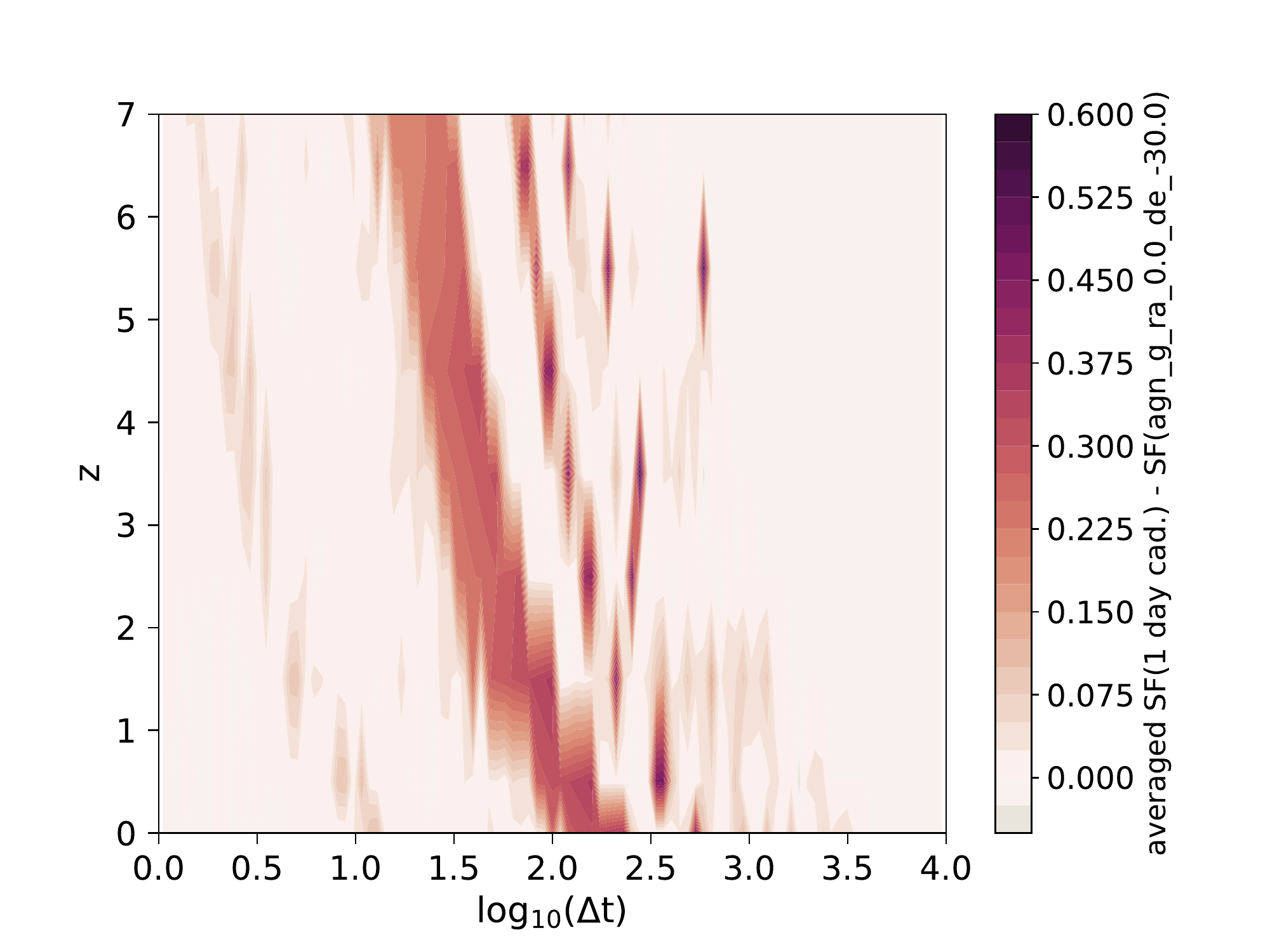}
		\includegraphics[clip,trim=35 11 45 47,width=0.33\textwidth]{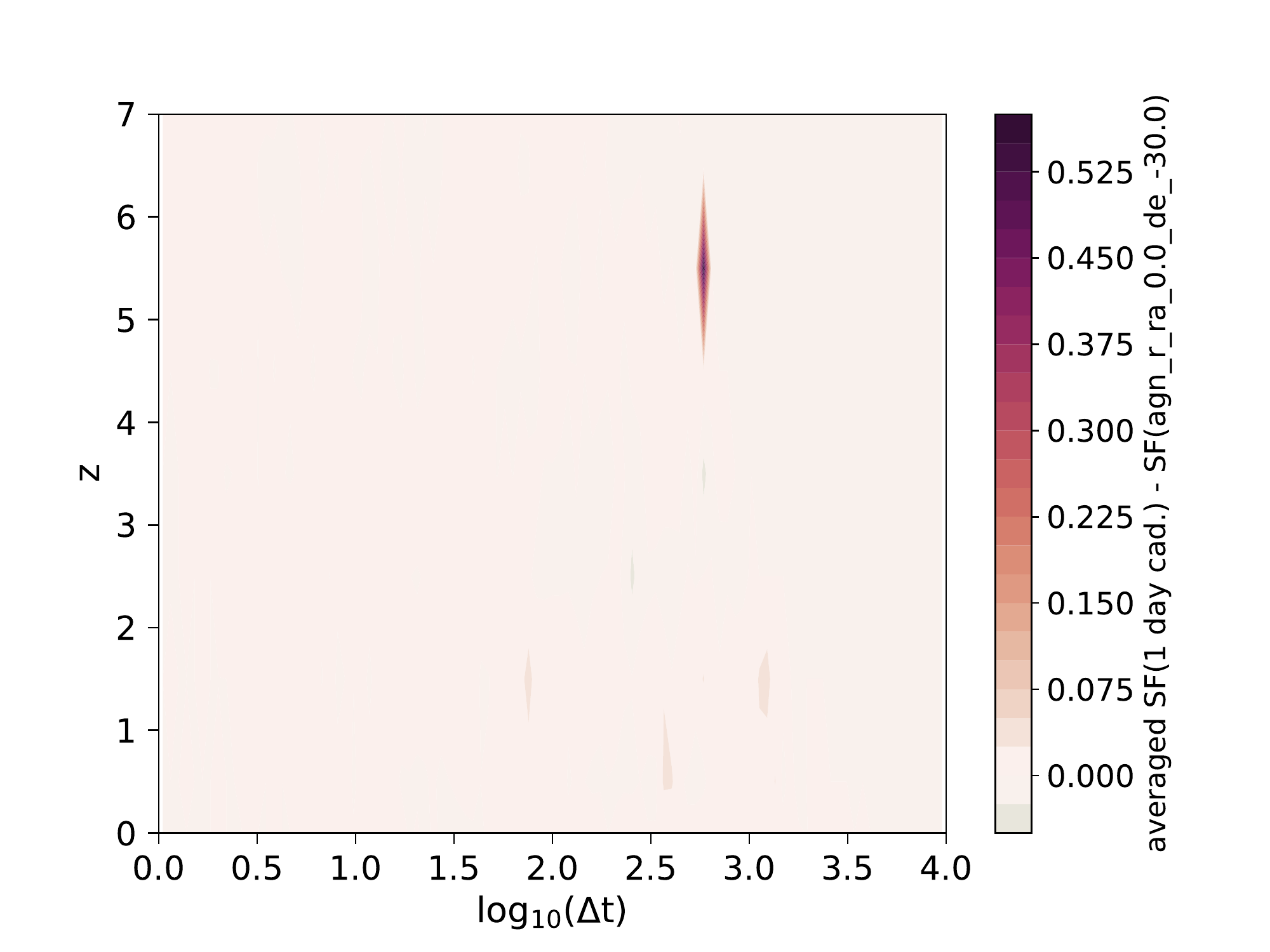}
		\includegraphics[clip,trim=35 11 36 47,width=0.33\textwidth]{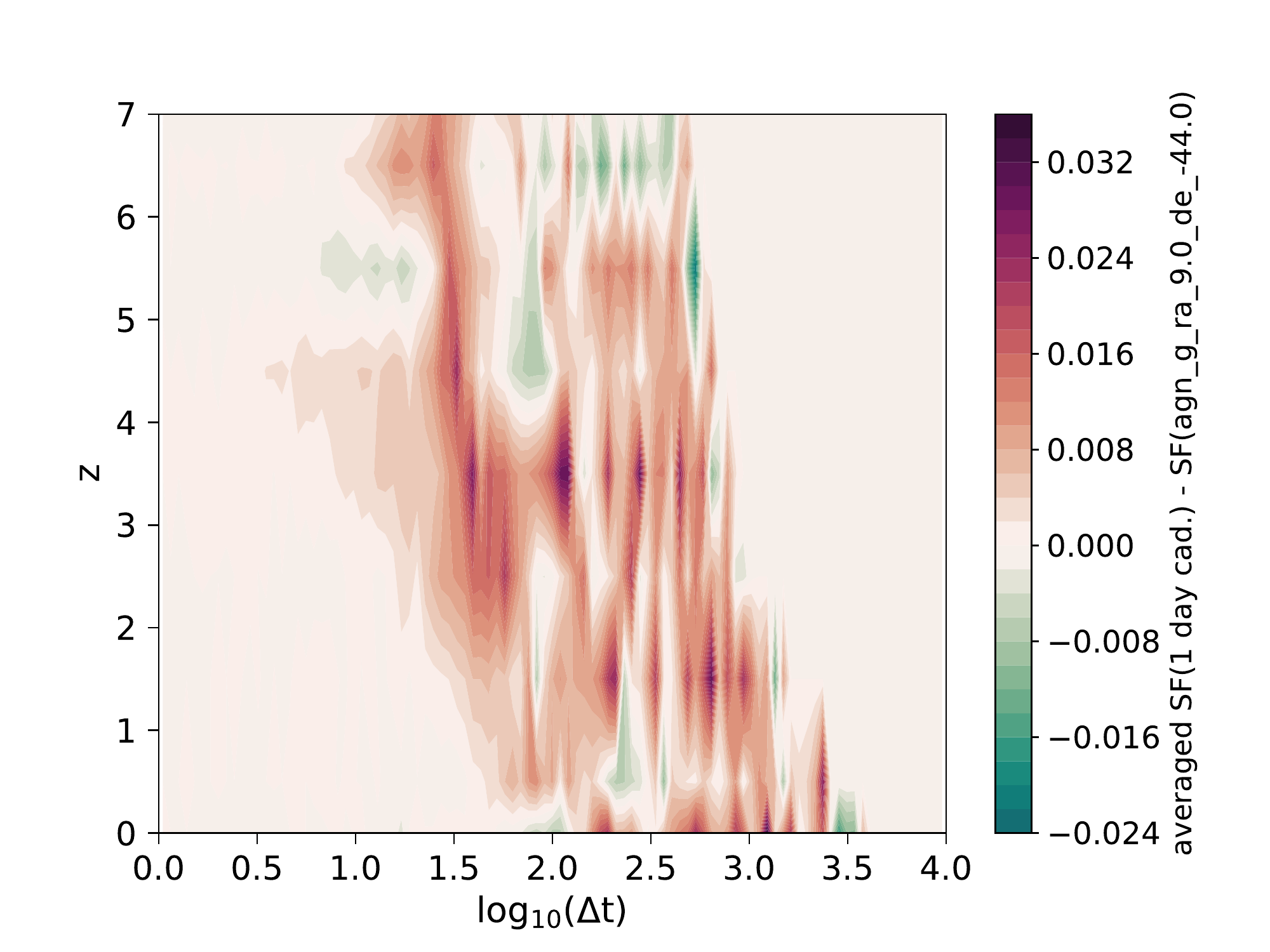}	

	\caption{Heatmap of SF-deviations for OpSim1.5 deep drilling field\_ AGN.g\_ra\_0.0\_de\_-30.0 (left panel), OpSim1.5 deep drilling field\_m\_ AGN.r\_ra\_0.0\_de\_-30.0 (middle panel) and OpSim agn\_g\_ra\_9.0\_de\_-44.0. Colorbar represents deviations. Positive deviations stand for$SF_{0}>SF_{gapped}$  in average per redshift bin and vice versa. }
	\label{fig:figopsim1}%
\end{figure*}

Averaged SF deviations for  OpSim1.5 deep drilling field (DDF)\_ AGN.g\_ra\_0.0\_de\_-30.0  and AGN.r\_ra\_0.0\_de\_-30.0   are given in
 Fig.	\ref{fig:figopsim1}.   AGN.g\_ra\_0.0\_de\_-30.0 realization is fragmented and noisy,  indicating erratic behavior of SFs. The evolutionary track close to the time scale 2.5 is present.  AGN.r\_ra\_0.0\_de\_-30.0   produces heatmap which is similar to the simulated rolling cadence of 3 months/yr with 1-day sampling 	(Fig. \ref{fig:fig2}). The deviations evolutionary tracks are less fragmented and more bent, with two times larger values then those found in heatmaps of 3 months/yr gappy cadence. {   However, OpSim agn\_g\_ra\_9.0\_de\_-44.0 cadence contains 2616 observations and  produces map (right panel) similar to map 6 months/yr cadence Fig. \ref{fig:fig3}.}

\section{Discussion}

To quantify LSST-like observing strategies' products, we focused on AGN variability-related observables (time lag, periodic oscillations, and SF) and their connection to predicting the most suitable LSST-like cadences. The importance of the first two lies in their effects on constraining reliable AGN models and the third is essential for direct measurement of the covariance function of the AGN light curves and can display oscillatory signals.
In this light, we  propose a multiple regression model 
to statistically identify the cadence-formal error pattern knowing AGN-variability observables from surveys (real-world and simulated) operations.

In order to evaluate the performance of the proposed regression model, case studies of a real-world  (pre-LSST era) and  artificial observing strategies (LSST-like)  are provided.
Assumptions made in multiple regression model abstract from the details of the real and artificial light curves, yet capture the general relationship  with AGN-variability observables. It connects the relative error of the AGN observable with light curve variability, flux errors, curve cadence and measured observable. 
It is reasonable to expect that the error of time lag and period will decrease with increasing $\frac{F_{var}}{\sigma}$ and $\frac{\mathcal{T}_{obs}}{\Delta t}$.

Fig.  \ref{fig:fig1} shows the comparative multiple regression model prediction performances between empirical and artificial LSST-like data sets. 
The appearance of outliers is due to unfavorable combinations of formal errors, cadences, and gaps in the light curves. The time lags, periodicities, and their uncertainties are determined using Gaussian Process learned light curves from real and simulated ones. A simple inspection of time lags and periodicities obtained from real (Table \ref{tab:data}) and artificial data (Table \ref{tab:Opsimgen}),  show that the formal errors in real data are more fluctuating than those found in LSST like set.
 Possibly this is a consequence of real survey cadences having more random gaps than planned LSST-like strategies. The formal errors of detected periodicities are calculated as half-width of the relevant correlation cluster in 2DHybrid method, which depends on the amplitude of correlation peaks of continuous wavelet transforms of light curves \citep{2018MNRAS.475.2051K}.
Also, some lower-luminosity objects such as NGC 4151 were, in general, targeted by the early ground-based monitoring campaigns due to their low redshift and apparent high brightness. The more considerable uncertainties in their time lag measurements partially come from the loosely constrained observational factors (such as observing cadence, spectral resolution, detector efficiency, etc.). 
Two versions of multiple regression model  predict comparable time lags and oscillation cadences,   while at $F_{\rm var}\sim 10\%$ the results  differ for oscillation detection.    We propose that this can be explained by the sensitivity of oscillation detection to the AGN variability and light curves characteristics.
The artificial set of light curves supports our general  expectation from multiple regression model  that the error of time lag and period will decrease with increasing two variables (coefficients $C_1$ and $C_2$ are negative). The empirical set of light curve  produces model realization with alternate sign of coefficients $C_1$ and $C_2$, which implies that there can be  an additional systematic variable (or error)  in the real data set. 

{
The errors of the model coefficients ($C_1, C_2$) are inversely proportional to the square root of the sample size, and the noise in the data affects the errors in the coefficient estimates. For example, four times as much data will reduce the errors of all coefficients by a factor of $\sim 2$. All estimated coefficients exceed twice their error, except for coefficient $C_1$, which accounts for $F_{var}/\sigma$ in proxy for a time lag of real-world sample (see Fig. \ref{fig:fig1}). This indicates that those coefficients  are significantly different from zero using
t-test and $\alpha=0.05$ \citep{2016Glantz}. The error of the coefficient $C_1$ for real-world time-lag metric is slightly inflated (at the level of $59\%$, Fig. \ref{fig:fig1}) by the considerable variation of fractional variability $F_{var}$ in the real-world sample. Because the model describes the general gradient in the data, this error is not likely to affect the results drastically.   Based on the previous, even the available sample is not sufficiently detailed to assess the influence of coefficient errors in detail; such errors are not likely to significantly divert the cadence requirement than the estimates given here.
}

While these findings are highly promising, our next steps will be testing the proposed regression model  in much larger sets of objects such as from the SDSS { RM campaign, which has monitoring 849 spectroscopically-confirmed quasars during 3-years long-period \citep{2019ApJS..241...34S}. Moreover, the  concept shown here could be potentially important for designing the cadence strategy for the MSE quasars reverberation mapping survey of $\sim$5000 quasars}.

Some phenomena could affect positively or negatively the detection of the underlying oscillatory signal. For example, we analyzed periodicities for PG 1302 -102  \citep{10.3847/1538-4357/aaf731} and Mrk 231 \citep{10.1093/mnras/staa737} which were observed photometrically by Catalina Real-time Transient Survey (CRTS)  and All-Sky Automated Survey for Supernovae (ASAS-SN). Analyzing Mrk 231 photometric curve, we found that adequate data sampling of ASAS-SN survey is more suitable for periodicity detection \citep{10.1093/mnras/staa737}.
Sudden changes in the target light curves can occur, as was the case of PG 1302-102 when unexpected {flare} appeared recently. This object has been considered as one of the best targets for the next generation of gravitational wave surveys. This demonstrates the importance of the information contained in {individual segments of light curve}.
{  We emphasize that input {for the artificial light curve} based on OpSim cadences differs from   the real RM monitoring  by not {including} factors  such as flares, real physical processes, observation uncertainties, jets, etc.}
{Since the magnitude of the regression coefficients is related to the light curves' parameters, mean cadence, and the formal errors of derived quantities from the data sets, then the coefficients of model runs on real and LSST-like surveys are not directly comparable and represent different model runs.}

To further expand on the topic of possible periodicity detection, one can ask the question how many close binary SMBH systems could be detected by LSST.

Assuming that detection of periodicity in the light curves is possible if  {binary mutual separation is above of anticipated value $10 \mu \rm{as}$  (e.g., corresponding to the binaries at mutual separation $\sim 0.01$ pc and distance of $\sim 200$Mpc), and the orbital period is shorter than twice of the survey lifetime.}
The minimum binary separation $a$ and the binary mass $M$ give the minimum binary SMBH orbital period for which LSST could detect orbital motion:

\begin{equation}
P_{\rm min}=\frac{2\pi a_{\rm min}^{3/2}}{\sqrt{G M}},
\end{equation}

\noindent where is assumed that 
\begin{equation}
a_{\rm min}=\theta/d\geq 10 \mu \rm{as},
\end{equation}
\noindent and $P_{\rm min}$ and $M$ depends on luminosity and redshift of targets,
assuming that  at angular- diameter distance $d$, the orbital angular radius  of binary SMBH is $\theta \sim a/d$ where $a$ is semimajor axis of binary.

{We calculate the number of binary SMBH which can be detected by analyzing  LSST light curves} up to $z\sim 5$, using {the estimated number} of binary SMBHs  per $\log z$ \citep{10.1103/PhysRevD.100.103016}:

\begin{equation}
\frac{dN}{d\log z}=4\pi \frac{d^{2} V}{dz d\Omega}
\frac{\phi_{0}}{(\frac{L}{L_{0}})^{\gamma_{1}}+((\frac{L}{L_{0}})^{\gamma_{2}  } }
min\{\frac{t_{res}}{t_{l}},1\} (1+e^{-2W}),
\end{equation}

\noindent where
\begin{equation}
\frac{d^{2} V}{dz d\Omega},
\end{equation}
\noindent  is  the co-moving volume per redshift and solid angle ($\Omega$).
 Also, 
 \begin{equation}
\frac{\phi_{0}}{(\frac{L}{L_{0}})^{\gamma_{1}}+(\frac{L}{L_{0}})^{\gamma_{2}  } }
 \end{equation}
 is the quasar luminosity function \citep[see][parameters are given in the last row of their Table 3]{10.1086/509629}, where
 \begin{equation}
   t_{res}=\frac{20}{256}\Big(\frac{P}{2\pi}\Big)^{8/3}\Big(\frac{GM}{c^3}\Big)^{-5/3} q^{-1}_{s}   
 \end{equation}
is the residence time of binary due to gravitational wave emission, $t_l\sim 10^7$ yr is the approximate AGN lifetime, $W=10yr-P_{\rm min}$ where 10 yr is a LSST mission lifetime. For simplicity, we assume that at larger redshifts we expect brighter and more massive sources.

\begin{figure*}
	\centering
	\includegraphics[clip,trim=120 30 38 40,width=0.31\textwidth]{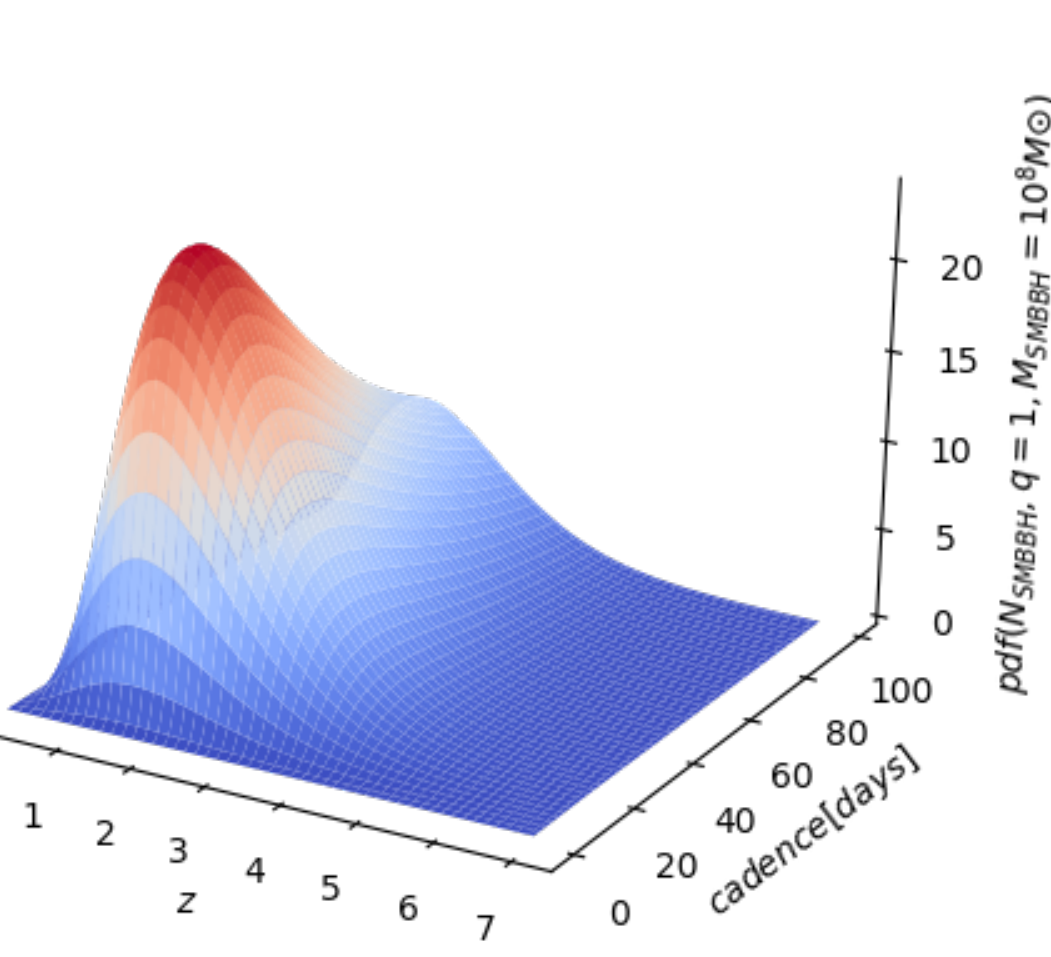}
	\includegraphics[clip,trim=120 30 38 40,width=0.31\textwidth]{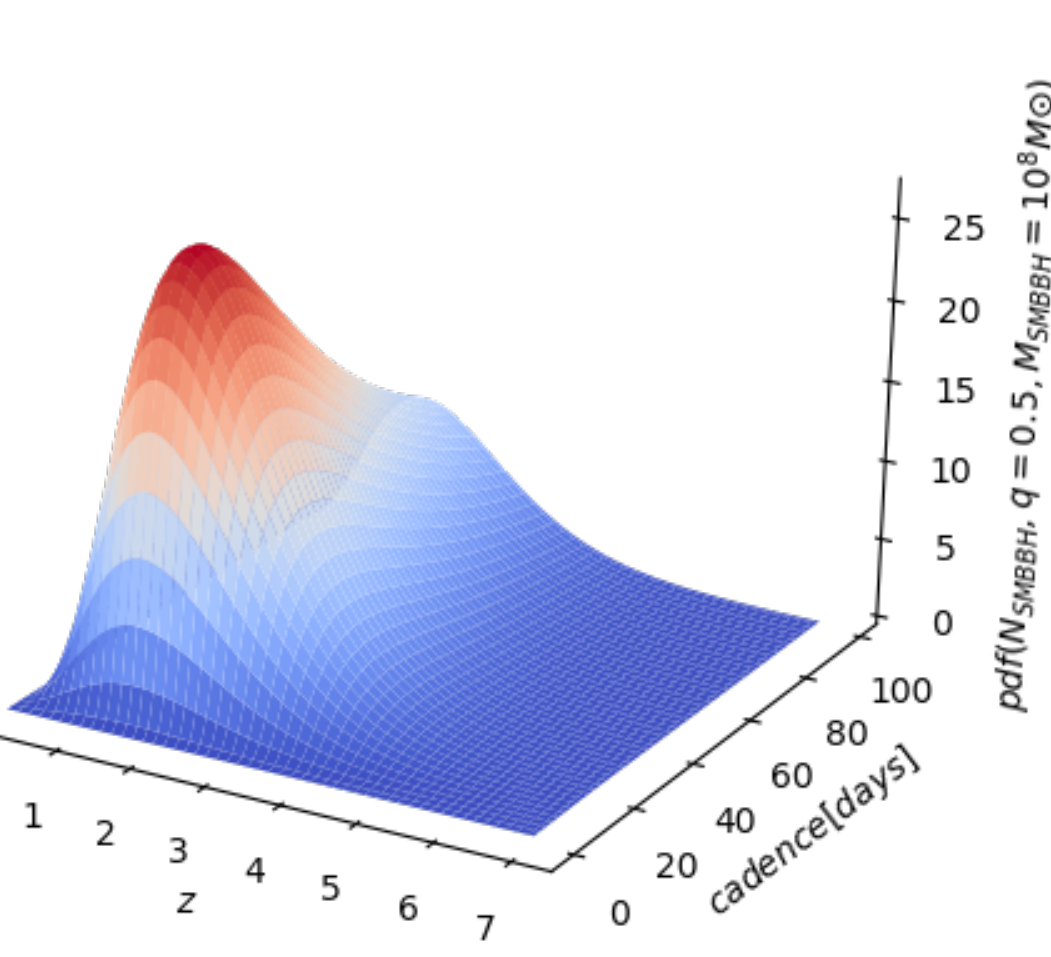}
	\includegraphics[clip,trim=120 30 38 40,width=0.31\textwidth]{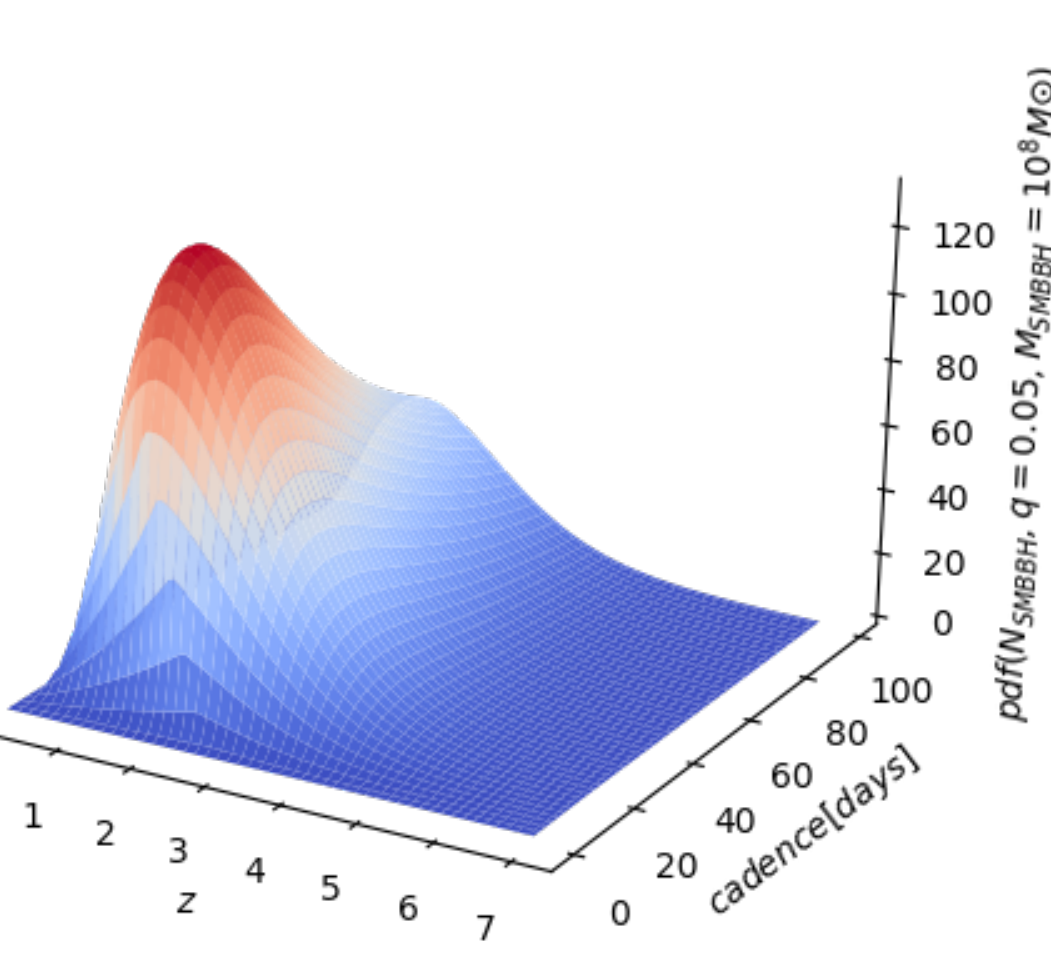}
	\caption{Probability density functions of expected number of CB-SMBH with total mass $M_{tot}=10^{8} M{\odot}$ and mass ratios $q=1,0.5,0.05$ (from left to right respectively), as the probability for finding an CB-SMHB at the orbital period resolvable by LSST. }
	\label{fig:prediction}%
\end{figure*}

Fig. \ref{fig:prediction} displays the  distribution of  detectable CB-SMBH of total mass $10^{8} M{\odot}$ and for three mass ratios $q=1,0.5,0.05$, at different redshifts and cadences. The peak for all distributions is about redshift 2 as expected for AGN, however the number of possible detections varies across CB-SMBH mass ratios. It seems that as mass ratio decreases the number of possible detections increases. For example for mass ratio $q=1$ and $q=2$ we expect probability density function (PDF) peak at 20-25 objects (see Fig. \ref{fig:prediction}), however for $q=0.05$ we can expect even 6 times more possible detections.

{{Although} the number of expected CB-SMBH seems to increase with decreasing mass ratio, this does not necessarily mean that the number of effective CB-SMBH's detection increases.
Namely, the decreasing mass ratio implies that the light curve will contain a periodicity signal with a smaller amplitude, but the small-amplitude oscillation detection is harder to perform. In reality, perhaps many factors can affect detectability. Some MHD studies \citep[e.g.,][]{10.1093/mnras/stt1787, 10.1088/0004-637X/783/2/134, 10.1088/0004-637X/807/2/131} have simulated unequal-mass $\leq 0.1$  binaries. Their accretion rates  are less bursty; and the cases for $q = 0.075$ and $q = 0.1$ binaries in \citet{10.1093/mnras/stt1787} are very similar to PG 1302-102’s light-curve, which has smooth sinusoidal appearance.   }{ Also, the cadences between 20 and 80 days (Fig. \ref{fig:prediction}) are sufficient for the most probable detections, which is in agreement with results given in Table \ref{tab:perlag}.}

\begin{figure*}
	\centering
	\includegraphics[clip,trim=11 12 50 50,width=0.42\textwidth]{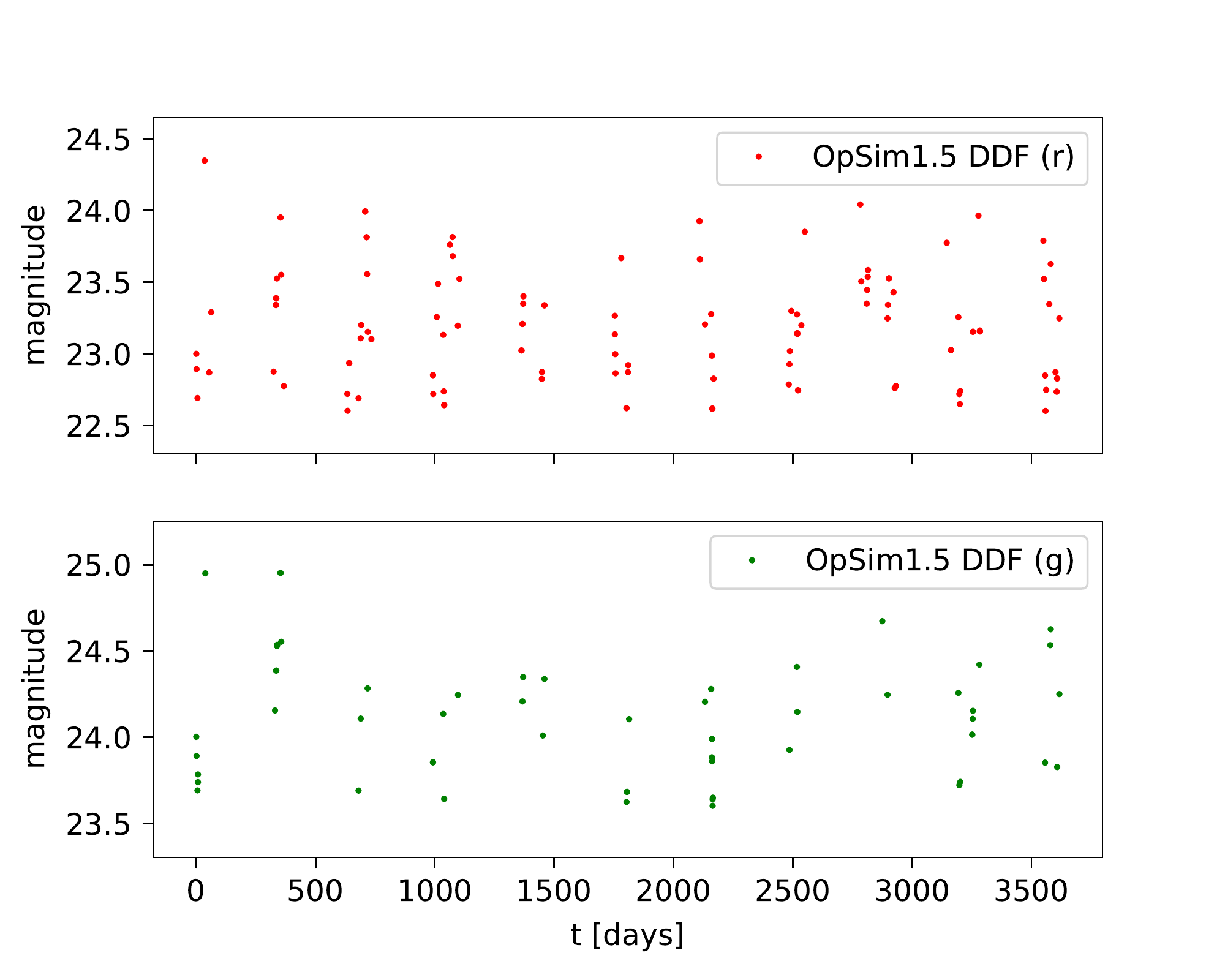}
	\includegraphics[clip,trim=1 10 50 20,width=0.43\textwidth]{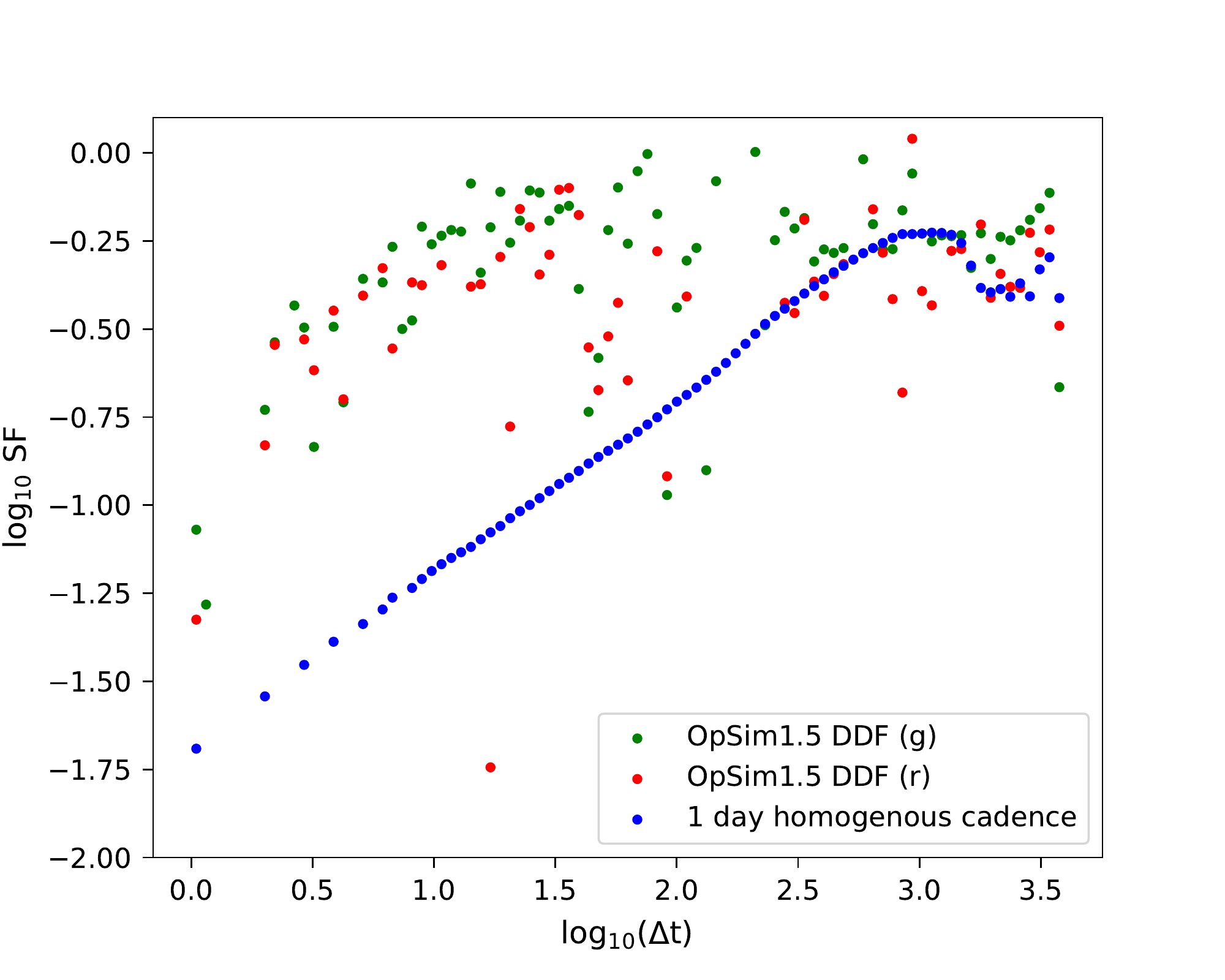}

	\caption{\textit{Left}: AGN Light curve in r and g band obtained from OpSim. \textit{Right}: Corresponding SFs compared to the SF of homogeneous  1-day cadence light curve. }
	\label{fig:figopsim2}%
\end{figure*}

The periodicity signal is present in SFs based on ideal and LSST-like cadences. The signal is prominent in the SFs for homogeneous 1-day cadence and in {   ideal surveys' "gappy" light curves with 6 and 9-months observing sets}. The separation between subsequent wiggling of SF peaks is about imparted oscillation in the light curves.
The deviation of SFs of {   light curves} with larger {gap} from the SF of the ideal series with homogeneous cadence inspired us to introduce a simple metric for SF. In logarithmic scale (right panel, Fig. \ref{fig:fig10}) oscillations are still present, but with smaller amplitudes.  SFs (right panel in Fig. \ref{fig:artifopsim} ) based on the combination of DRW and OpSim {rolling} cadences show the oscillatory pattern in contrast to plain {(non-oscillatory)} AGN light curves in g and r band obtained from OpSim1.5 DDF  in Fig. \ref{fig:figopsim2}. These cases emphasize that the detection of binary candidates could be done via SFs.

Averaged SF deviations for  OpSim1.5 (DDF, see Fig.	\ref{fig:figopsim1}) are fragmented and noisy,  indicating erratic behavior of SFs concerning homogenous SF. As much as the cadence is denser, the SFs resemble more those obtained from a homogenous light curve.

Some other essential factors could influence the relation between cadence, AGN variability observables, and their formal errors that we did not cover.
 Perhaps, the dependent and two independent variables used in our regression model are not the only critical light-curve characteristics that should be taken into the regression model.
  The more subtle influence will have light curves' nonstationarity,  trends, and peak and valleys sharpness. {For example the presence of a periodic signal in the light curve is imprinted in the periodic behavior of its cross-correlation function (CCF) see e.g., \citet[][and their Figure 4b and 4c]{10.1086/317967} and
  \citet[][and their Figure 18]{2018MNRAS.475.2051K} or 
\citet{apps.dtic.mil/dtic/tr/fulltext/u2/a094384.pdf}}. Suppose the underlying signal in the light curve is complicated, such as  
  $$signal=t\sin{(\theta t)}$$ 
  where $t$ is time. In that case, the value of any local maximum of this function is greater than the values of all previous local maxima, and localization of the maxima of the CCF cannot be done using the highest value of the CCF correlation coefficient, which will {bias the measurement of time lag}.

There are some other conceptions that we did not cover in this study. Namely, it has been reported that the characteristics of SF for AGN light curves scales with physical parameters of AGN, for example, long time scale RMS variability SF$^{\infty}$ is anti-correlated with AGN luminosity \citep{10.1088/0004-637X/721/2/1014}. However, its characteristic time scale is correlated with AGN luminosity \citep{10.3847/1538-4357/aae208} and short term RMS variability is sensitive to AGN bolometric luminosity \citep{10.3847/1538-4357/aae208}. Thus it will be possible to characterize how much bias in the SF measurement is introduced by the LSST observing cadence. We plan to focus on such analysis in our future work when the Rubin Observatory releases Data Preview  (DP) with AGN data.
 Based on three characteristic shape parameters of SF (SF$^{\infty}$, its associated characteristic time scale $\tau$ and power-law slope $\beta$ of SF  defined as $SF\sim (\frac{\Delta t}{\tau})^{\beta}$ the observed SF can be fitted with  \citep[see][]{10.3847/1538-4357/aae208}:

\begin{equation}
SF(\Delta t|\tau,\tilde{\sigma})=\tilde{\sigma}^{2}\tau \Big(1-e^{(\frac{-\Delta t}{\tau})^{\beta}}\Big)+\sigma_{p}    
\end{equation}

\noindent {where $\sigma_p$ is the uncertainty of the magnitude difference between two observations at time distance $\Delta t$. From the ‘ideal’ light curve, one can measure the ‘reference’ parameters: $Q=(SF^\infty, \tau, \beta)$ by fitting the observed SF using the above model.
Then the shape parameters for the ‘gappy’ SFs obtained by the same fitting procedure for the reference SF can be compared with the reference shape parameters }
\begin{equation}
    \delta Q(i)=\frac{Q(i)^{ref}-Q(i)^{gappy}}{Q(i)^{ref}}.
\end{equation}

\section{Conclusion}

{To assess the observing strategies of the present and future spectroscopic and photometric surveys, we examined AGN variability-related observables (time lag, periodic oscillations, and SF) and their relation  to predicting the most suitable LSST-like cadences. The  first two  observables affect constraining reliable AGN models and the third is essential for determination of the covariance function of the AGN light curves and can display oscillatory signals.}

{From this perspective, we constructed a multiple regression model 
to statistically identify the cadence-formal error pattern knowing AGN-variability observables from different surveys.
We tested the performance of the proposed regression model on case studies of real  (pre-LSST era) and an artificial observing strategies (LSST-like).
Multiple regression model abstracts from the details of the real and artificial light curves, but  establishes the general relationship  with AGN-variability observables. 
We employed two different observing strategies}:  the optically uniform dataset including decade-long reverberation mapping campaigns of eight type 1 AGN, with distinct variability and optical spectra characteristics which is difficult to simulate; and the artificial data set,   simulated according to the DRW method with an added periodic oscillation. 
The artificial data sets are constructed  on  several idealized  and LSST-like observing strategies. { For examination of cadence effects on SF we used only the idealized observing strategy.}
For time lag extraction we used
the Gaussian process light curve modeling, and for the periodicity detection our 2DHybrid method.
The results of our analysis are as follows:
\begin{enumerate}
    \item   {
    The two model versions (based on real and LSST-like observing strategy) predict comparable cadences for time lags and  oscillation  detection,  whereas  at $F_{var}\sim10\%$  cadences for oscillation detection differ. The difference might be explained by sensitivity of oscillation detection to the light curve variability, because artificial set of objects contains only one object with $F_{var}\sim 10\%$.
     In general for time-lag and periodicity, for objects with higher ($\sim20\%$) variability the predicted cadences  are larger then those estimated for $F_{var}\sim 10\%$. As expected the predicted cadences for time-lag and periodicity are decreasing with assumed redshift of the object. The proposed multiple regression has shown promising potential for predicting AGN time-lag and periodicity cadences, but before such estimation can eventually meet observable practice, the regressions should be tested further in larger sets of object samples such as SDSS RM campaign. }

    \item {{We find that, for SFs constructed on both idealized and LSST OpSim cadences,  if the light curves contains periodic signal, the same oscillatory signal is seen in the large SFs time scales. 
       We defined the simple metric to measure the properties of the SF, accounting for the deviation of the observed SF with respect to the ideal light curves. We showed that light curves with reasonable gaps would preserve the SF shape, and that even with larger gaps, some strategies of denser sampling could help to get smaller deviations of SF of "gappy" light curves from the SF constructed on homogenous 1-day cadence.}}
    \item The {smallest deviations of} gapped SFs  {from idealized SF} are {observed} {when cadences are highly idealized  or } very dense LSST DDF cadences, having about 1500 observations in r filter. However, sparse DDF cadences in g filter indicate that gappy SFs would {significantly deviate from homogenous SF}.
\item We predict that the PDF of number of CB-SMBH, that LSST will detect on average  during its lifetime, would have a peak at two dozens of objects for $q=1$ and {$q=0.5$}. However, the PDF would peak at about 100 objects for $q=0.05$. {Based on constructed PDFs the cadences between $\sim 20$ and $\sim 80$ days are required for majority detection of binary candidates which is alike to multiple regression model prediction}.
\end{enumerate}

 The multiple regression model presented may be used in assessing observing strategies of the present and  future photometric and spectroscopic surveys, such as the  LSST, MSE, SDSS-V, and many other to come.
{For the purpose of granting scientists to easily review and analyse the method described in this paper, we have developed various Jupyter notebooks. }
Our code is publicly available as open-source code on GitHub (\url{https://github.com/LSST-sersag/agn_cadences}).

\section*{Acknowledgements}
    We  thank the anonymous Referee for their thoughtful comments  towards improving our work. The authors also thank the  members of the Vera C. Rubin LSST Active galactic nuclei (AGN) and the Transients and variable stars (TVS) Scientific collaborations  for their feedback that improved the presentation of our results.
    The authors acknowledge funding provided by  Faculty of Mathematics University of Belgrade (the contract 451-03-9$/$2021-14$/$200104),  Astronomical Observatory (the contract 451-03-68$/$2020-14/200002) and Faculty of Science University of Kragujevac (the contract 451-03-9/2021-14/200122), through the grants by the Ministry of Education, Science, and Technological Development of the Republic of Serbia. A. K. and L. {\v C}. P. acknowledge the support by  Chinese Academy of Sciences President's International Fellowship Initiative (PIFI) for visiting scientist.  D. I. acknowledges the support of the Alexander von Humboldt Foundation.

\section*{Data Availability}

 The data underlying this article were provided by Vera C. Rubin Observatory Legacy Survey of Space and Time  by permission. Data will be shared on request to the corresponding author with permission of Vera C. Rubin Observatory Legacy Survey of Space and Time.




\begin{thebibliography}{99}

\bibitem[\protect\citeauthoryear{Allevato et al.}{2013}]
{10.1088/0004-637X/771/1/9}
 Allevato, V., Paolillo, M., Papadakis, I., Pinto, C. 2013, ApJ, 771, 9pp
 


\bibitem[\protect\citeauthoryear{Barth et al.}{2011}]{2011ApJ...743L...4B}
Barth A. J., Pancoast A., Thorman S. J. et al., 2011, ApJ, 743, 4

\bibitem[\protect\citeauthoryear{Barth et al.} {2013}]{2013ApJ...769..128B}
Barth A. J., Pancoast A., Bennert V. N. et al., 2013, ApJ, 769, 128

\bibitem[\protect\citeauthoryear{[Barth et al.} {2015}]{2015ApJS..217...26B}
Barth A. J., Bennert V. N., Canalizo G. et al., 2015, ApJS, 217, 26

\bibitem[\protect\citeauthoryear{Bentz et al.} {2013}]{10.1088/0004-637X/767/2/149}
Bentz, M. C.,  Denney, K. D.,  Grier, C. J., Barth, A. J.,  Peterson, B. M.,  Vestergaard, M. et al., 2013, ApJ, 767,  149

\bibitem[\protect\citeauthoryear{Bentz et al.} {2008}]{2008ApJ...689L..21B}
Bentz M. C., Walsh J. L., Barth A. J. et al., 2008, ApJL, 689, L21

\bibitem[\protect\citeauthoryear{Bentz et al.} {2009a}]{2009ApJ...705..199B}
Bentz M. C., Walsh J. L., Barth A. J. et al., 2009a, ApJ, 705, 199


\bibitem[\protect\citeauthoryear{Biswas et al.} {2020}]{10.3847/1538-4365/ab72f2}
Biswas, R.,  Daniel, S.F.,  Hlo{\v z}ek, R.,  Kim, A.G. et al. 2020, ApJS, 247, 60


\bibitem[\protect\citeauthoryear{Bon et al.}{2012}]{2012ApJ...759..118B}
Bon, E.,  Jovanovi{\'c}, P., Marziani,  P.,  Shapovalova, A. I.,  Bon,  N., Borka Jovanovi{\'c}, V. et al.,  2012, \apj, 759, 2, id.118 


\bibitem[\protect\citeauthoryear{Bon et al.}{2016}]{bon16} Bon, E., Zucker, S., Netzer, H. et al.,  2016,  \apjs, 225, 29


\bibitem[\protect\citeauthoryear{Brandt et al.}{2018}]{brandt18}
Brandt, W. N.,  Ni, Q., Yang, G.,  Anderson, S. F.,  Assef, R. J., Barth, A. J. et al. 2018, white paper on LSST cadence optimization,  arXiv:1811.06542

\bibitem[\protect\citeauthoryear{Burke-Spolaor et al.}{2019}]{10.1007/s00159-019-0115-7}
Burke-Spolaor, S., Taylor, S.R., Charisi, M. et al., 2019 Astron Astrophys Rev 27, 5 

\bibitem[\protect\citeauthoryear{Caplar et al.}{2017}]{10.3847/1538-4357/834/2/111}
Caplar, N., Lilly, S., J., Trakhtenbrot, B., 2017, ApJ, 834, 111

\bibitem[\protect\citeauthoryear{Chelouche \& Daniel} {2012}]{10.1088/0004-637X/747/1/62}
Chelouche, D., Daniel, E., 2012, ApJ,  747, 62

{
\bibitem[\protect\citeauthoryear{Chelouche et al.} {2012}]{10.1088/2041-8205/750/2/L43}
Chelouche D., Daniel E., Kaspi S., 2012, ApJ, 750, 43
}




\bibitem[\protect\citeauthoryear{Chen} {2017}]{10.1080/24709360.2017.1396742}
 Chen, Y. C., 2017, Biostatistics \& Epidemiology, 1, 161 



\bibitem[\protect\citeauthoryear{Charisi et al.}{2018}]{2018MNRAS.476.4617C}
Charisi, M.,  Haiman, Z.,  Schiminovich, D., D' Orazio, D. J., 2018, \mnras,  476, 4617


\bibitem[\protect\citeauthoryear{Denney et al.} {2009}]{2009ApJ...704L..80D}
Denney, K. D., Peterson, B. M., Pogge, R. W. et al., 2009, ApJL, 704, L80

\bibitem[\protect\citeauthoryear{Denney et al.} {2014}]{10.1088/0004-637X/796/2/134}
Denney, K. D., De Rosa, G.,  Croxall, K., Gupta, A.,  Bentz, M. C. et al., 2014, ApJ,  796,  id. 134 

\bibitem[\protect\citeauthoryear{D' Orazio et al.}{2013}]{10.1093/mnras/stt1787}
D’Orazio, D. J., Haiman, Z. MacFadyen, A., 2013,  MNRAS, 436, 2997


\bibitem[\protect\citeauthoryear{D' Orazio et al.}{2015}]{2015Natur.525..351D}	D' Orazio, D. J., Haiman, Z., Schiminovich, D.,  2015, \nat, 525, 351

\bibitem[\protect\citeauthoryear{D' Orazio \& Haiman}{2017}]{2017MNRAS.470.1198D}
D' Orazio, D. J.,  Haiman, Z.,  2017, \mnras, 470, 1198


\bibitem[\protect\citeauthoryear{D' Orazio \& Loeb }{2019}]{10.1103/PhysRevD.100.103016}
D' Orazio, D., Loeb, A., 2019, Phys. Rev. D., 100, 103016

\bibitem[\protect\citeauthoryear{Du et al.} {2014}]{2014ApJ...782...45D}
Du, P., Hu, C., Lu K.-X., Wang, F., Qui, J., Li, Y.-R., Bai, J.-M., Kaspi, S., Netzer, H., Wang, J.-M., 2014, ApJ, 782, 45

\bibitem[\protect\citeauthoryear{Du et al.} {2015}]{10.1088/0004-637X/806/1/22}
Du, P., Hu C., Lu, K.-X., Huang, Y.-K., Cheng, C., Qiu, J., Li, Y.-R. et al., 2015, ApJ, 806, 22



\bibitem[\protect\citeauthoryear{Du et al.} {2016}]{2016ApJ...825..126D}
Du, P., Lu, K.-X., Zhang, Z.-X., Huang, Y.-K., Wang, K., Hu, C., Qiu, J., Li, Y.-R. et al., 2016, ApJ, 825, 126

\bibitem[\protect\citeauthoryear{Du et al.} {2018}]{2018ApJ...856....6D}
Du, P., Zhang, Z.-X., Wang, K.,  Huang, Y.-K., Zhang, Y., Lu, K.-X., Hu, C., Li, Y.-R. et al., 2018, ApJ, 856, 6

\bibitem[\protect\citeauthoryear{Edelson et al.} {1990}]{10.1086/169036}
Edelson, R., Pike, G.F., Krolik, J. H., 1990, {\apj}, 359, 86



\bibitem[\protect\citeauthoryear{Edelson et al.} {2019}]{2019ApJ...870..123E}
Edelson, R., Gelbord, J., Cackett, E., Peterson, B. M., Horne, K., Barth, A. J. et al., 2019,  ApJ,  870,  2, id. 123

{
\bibitem[\protect\citeauthoryear{Edri et al.} {2012}]{10.1088/0004-637X/756/1/73}
Edri H., Rafter S. E., Chelouche D., Kaspi, S., Behar, E., 2012, ApJ, 756, 73}





\bibitem[\protect\citeauthoryear{Elvis} {2001}]
{10.1007/978-94-010-0320-9}
Elvis, M., 2001,  in The Century of Space Science  (eds J.A. Bleeker, J. Geiss, and  M. Huber),  Kluwer Academiic Publishers, 529

\bibitem[\protect\citeauthoryear{Farris et al.}{2014}]{10.1088/0004-637X/783/2/134}
Farris, B. D., Duffell, P., MacFadyen, A. I., Haiman, Z., 2014, ApJ, 783, 134 

\bibitem[\protect\citeauthoryear{Faisst et al.}{2019}]{10.3847/2041-8213/ab3581}
Faisst, A. L., Prakash, A.,  Capak, P. et al., 2019, ApJL, 881, L9

\bibitem[\protect\citeauthoryear{Graham et al.}{2015}]{2015Natur.518...74G}
Graham, M. J.,  Djorgovski, S. G., Stern, D., Glikman, E., Drake, A. J. et al., 2015, \nat, 518, 74

\bibitem[\protect\citeauthoryear{Glantz et al.}{2016}]{2016Glantz}
Glantz, S., Slinker, B., Neilands, T. B., 2016,  Primer of Applied Regression and Variance analysis (Third Edition),  McGraw-Hill



\bibitem[\protect\citeauthoryear{Graham et al.}{2018}]{10.3847/1538-3881/aa99d4}
Graham, M. L., Connolly, A.J.,  Ivezi{\'c}, {\v Z} ,  Schmidt, S. J., Lynne Jones, R.,  Juri{\'c}, M. et al., 2018, AJ, 155, 1
\bibitem[\protect\citeauthoryear{Grier et al.} {2012}]{2012ApJ...755...60G}
Grier, C. J., Peterson, B. M., Pogge, R. W. et al., 2012, ApJ, 755, 60

\bibitem[\protect\citeauthoryear{Grier et al.} {2017}]{2017ApJ...851...21G}
Grier, C. J., Trump, J. R., Shen, Y. et al., 2017, ApJ, 851, 21

{
\bibitem[\protect\citeauthoryear{Hieftje \& Horlick} {1981}]{apps.dtic.mil/dtic/tr/fulltext/u2/a094384.pdf}
Hieftje, G. M., Horlick, G., 1981, correlation methods in the chemistry laboratory}




\bibitem[\protect\citeauthoryear{Hopkins, Richards \& Hernquist} {2007}]{10.1086/509629}
 Hopkins, P. F.,  Richards, G. T.,  Hernquist, L., 2007,  ApJ, 654, 731 

\bibitem[\protect\citeauthoryear{Horne et al.} {2004}]{10.1086/420755}
Horne, K.,  Peterson, B. M.,  Collier, S. J.,  Netzer, H., 2004, PASP, 116, 819, 465


\bibitem[\protect\citeauthoryear{Ili{\'c} et al.} {2020}]{Ilic20}
Ili{\'c}, D.,  Oknyansky, V.,  Popovi{\'c}, L. {\v C}.,  Tsygankov, S. S.,  Belinski, A. A., Tatarnikov, A. M.,  Dodin, A. V.,  Shatsky, N. I.,  Ikonnikova, N. P., Raki{\'c}, N.,  Kova{\v c}evi{\'c}, A.,  Mar{\v c}eta-Mandi{\'c}, S. et al., 2020, A\&A, 638, A13 

\bibitem[\protect\citeauthoryear{Ivezi{\'c} et al.} {2019}]{10.3847/1538-4357/ab042c}
Ivezi{\'c}, {\v Z}., Kahn, S. M.,  Tyson, A.,  Abel, B., Acosta, E. et al., 2019, ApJ, 873, 44


\bibitem[\protect\citeauthoryear{Jones et al.} {2020}]{2020DPS....5211002J} Jones, L., Yoachim, P., Ivezi{\' c}, {v Z}., Juri{\' c}, M., Eggl, S., Chesley, S., Fraser, W., et al., 2020, DPS

\bibitem[\protect\citeauthoryear{Jones et al.} {2021}]{pstn-051.lsst.io/PSTN-051.pdf}
Jones, R. L., Yoachim, P., Ivezi{\'c}, {\v Z}., Neilsen, E. H.,  Ribeiro, T.,  2021,  Survey Strategy and Cadence Choices for the Vera C. Rubin Observatory Legacy Survey of Space and Time

\bibitem[\protect\citeauthoryear{Jun et al.}{2015}]{2015ApJ...814L..12J}
Jun, H. D., Stern, D., Graham, M. J., Djorgovski, S. G., Mainzer, A., Cutri, R., M., Drake, A. J., Mahabal, A. A.,  2015, \apjl, 814, L12
\bibitem[\protect\citeauthoryear{Kelly et al.} {2009}]{10.1088/0004-637X/698/1/895}
Kelly, B. C., Bechtold, J.,  Siemiginowska, A., 2009, ApJ, 698, 895
\bibitem[\protect\citeauthoryear{Kelly et al.} {2013}]{10.1088/0004-637X/779/2/187}
Kelly, B. C., Treu, T., Malkan, M., Pancoast, A.,  Woo, J.-H., 2013, ApJ, 779, 187


\bibitem[\protect\citeauthoryear{Kim et al.} {2019}]{10.3847/1538-4357/ab40cd}
Kim, J.,  Im, M.,  Choi, C., Hwang, S., 2019, ApJ, 884, 103 


\bibitem[\protect\citeauthoryear{Kollmeier et al.}{2006}]{10.1086/505646/pdf}
Kollmeier, J. A.,   Onken, C. A.,  Kochanek,  C. S.,  Gould, A. et al., 2006, ApJ,  648, 128



\bibitem[\protect\citeauthoryear{Kollatschny \& Fricke} {1985}]{1985A&A...146L..11K}
Kollatschny, W.,  Fricke, K. J., 1985, A\&A,  146, L11


\bibitem[\protect\citeauthoryear{Kova{\v c}evi{\' c} et al.}{2018}]{2018MNRAS.475.2051K}
Kova{\v c}evi{\' c}, A.~B., P{\'e}rez-Hern{\'a}ndez, E., Popovi{\' c}, L. {\v C}., Shapovalova, A. I.,  Kollatschny, W.,  Ili{\' c}, D.,  2018, \mnras, 475, 2051

\bibitem[\protect\citeauthoryear{Kova{\v c}evi{\' c} et al.}{2017}]{2017Ap&SS.362...31K}
Kova{\v c}evi{\' c}, A., Popovi{\' c}, L. {\v C}., Shapovalova, A. I., Ili{\' c}, D.,  2017,  Ap\& SS, 362, id. 31

\bibitem[\protect\citeauthoryear{Kova{\v c}evi{\' c} et al.}{2020a}]{10.1515/astro-2020-0007}
Kova{\v c}evi{\' c}, A., Popovi{\' c}, L. {\v C}., Ili{\' c}, D.,  2020, Open astronomy, 29,  51


\bibitem[\protect\citeauthoryear{Kova{\v c}evi{\' c} et al.}{2019}]{10.3847/1538-4357/aaf731}
Kova{\v c}evi{\' c}, A.~B.,  Popovi{\' c}, L. {\v C}., Simi{\'c}, S.,  Ili{\' c}, D., 2019, ApJ, 871,  article id. 32

\bibitem[\protect\citeauthoryear{Kova{\v c}evi{\' c} et al.}{2020b}]{10.1093/mnras/staa737}
Kova{\v c}evi{\' c}, A.~B.,  Yi, T.,  Dai, X.,  Yang, X.,  {\v C}vorovi{\'c}-Hajdinjak, I., Popovi{\' c}, L. {\v C}., 2020, MNRAS, 494, 4069

\bibitem[\protect\citeauthoryear{Koz{\l}owski} {2016}]{10.3847/0004-637x/826/2/118}
Koz{\l}owski, S., 2016, ApJ, 826, 118
\bibitem[\protect\citeauthoryear{Koz{\l}owski} {2017}]{10.1051/0004-6361/201629890}
Koz{\l}owski, S. 2017, A\&A, 597, A128




\bibitem[\protect\citeauthoryear{Li et al.}{2016}]{li16}
Li, Y.-R., Wang, J.-M., Ho, L. C., Lu, K.-X., Qiu, J., Du, P., Hu, C., Huang, Y.-K., Zhang, Z.-X., Wang, K., Bai, J.-M.,  2016, \apj, 822, 4

\bibitem[\protect\citeauthoryear{Liu et al.}{2018}]{2018ApJ...859L..12L}Liu, T.,  Gezari, S., Coleman Miller, M.,  2018, \apjl, 859,  id. L12





\bibitem[\protect\citeauthoryear{Lyutyj et al.}{1984}]{1984PAZh...10..803L}
Lyutyj, V. M., Oknyanskij, V. L.,  Chuvaev, K. K., 1984,  Soviet Astronomy Letters, 10, 335 


\bibitem[\protect\citeauthoryear{MacLeod et al.}{2010}]{10.1088/0004-637X/721/2/1014} 
MacLeod, C. L.,  Ivezi{\'c}, {\v Z}.,  Kochanek, C. S.,  Koz{\l}owski, S., Kelly, B. et al., 2010, ApJ, 721, 1014  


\bibitem[\protect\citeauthoryear{LSST Science Collaborations et al.}{2017}]{1708.04058}
Marshall, P., Anguita, T., Bianco, F., Bellm, E. C., Brandt, W. N. et al., 2017, preprint,arXiv:1708.04058


\bibitem[\protect\citeauthoryear{Moreno et al.} {2019}]{2019PASP..131f3001M} Moreno, J., Vogeley, M.~S., Richards, G.~T., Yu, W., 2019, PASP, 131, 063001. 

\bibitem[\protect\citeauthoryear{The MSE Science Team et al.}{2019}]{2019arXiv190404907T} The MSE Science Team, Babusiaux, C., Bergemann, M., et al., 2019, arXiv:1904.04907

\bibitem[\protect\citeauthoryear{Neira et al.} {2020}]{10.1093/mnras/staa1208}
 Neira, F., Anguita, T., Vernardos, G., 2020, MNRAS, 495, 544

\bibitem[\protect\citeauthoryear{Netzer \& Trakhtenbrot} {2007}]{10.1086/509650}
Netzer, H., Trakhtenbrot B., 2007, ApJ, 654, 754 


\bibitem[\protect\citeauthoryear{Nu\~nez Pozo et al.}{2014}]{10.1051/0004-6361/201322736}
  Nu\~nez Pozo, F., Haas, M., Ramolla, M., Bruckmann, C. et al., 2014, A\&A, 568, A36


\bibitem[\protect\citeauthoryear{Peterson et al.} {1998}]{1998ApJ...501...82P}
Peterson, B. M., Wanders, I., Bertram, R. et al., 1998, ApJ, 501, 82

\bibitem[\protect\citeauthoryear{Peterson et al.} {2002})]{2002ApJ...581..197P}
Peterson, B. M., Berlind, P., Bertram, R. et al., 2002, ApJ, 581, 197



\bibitem[\protect\citeauthoryear{Peterson et al.} {2004}]{10.1086/423269}
Peterson, B. M., Ferrarese, L., Gilbert, K. M., Kaspi, S. et al., 2004, ApJ, 613,682

\bibitem[\protect\citeauthoryear{Popovi{\'c} et al.} {2011}]{2011A&A...528A.130P}
Popovi{\'c}, L. {\v C}., Shapovalova, A. I., Ili{\'c}, D., Kova{\v c}evi{\'c}, A., Kollatschny, W. et al., 2011, A\&A, 528, A130 

\bibitem[\protect\citeauthoryear{Popovi{\'c} et al.} {2014}]{2014A&A...572A..66P}
Popovi{\'c}, L. {\v C}., Shapovalova, A. I.,  Ili{\'c}, D., Burenkov, A. N., Chavushyan, V. H.; Kollatschny, W, Kova{\v c}evi{\'c}, A. et al., 2014, A\&A, 572, A66




\bibitem[\protect\citeauthoryear{Rodriguez-Pascual et al.} {1997}]{10.1086/312996}
Rodriguez-Pascual, R.M., Alloin, D., Clavel, J., Crenshaw, D. M., Horne, K. et al., 1997, ApJS, 110, 9




\bibitem[\protect\citeauthoryear{Shankar et al.} {2009}]{10.1088/0004-637X/690/1/20}
 Shankar, F., Weinberg, D. H., Miralda-Escud{e}, J., 2009, ApJ, 690, 20

\bibitem[\protect\citeauthoryear{Shapovalova et al.} {2001}]{2001A&A...376..775S}
Shapovalova, A. I., Burenkov, A. N., Carrasco, L., Chavushyan, V. H. et al., 2001, A\&A, 376, p. 775

\bibitem[\protect\citeauthoryear{Shapovalova et al.} {2004}]{2004A&A...422..925S}
Shapovalova, A. I., Doroshenko, V. T., Bochkarev, N. G., Burenkov, A. N. et al., 2004, A\&A, 422, 92

\bibitem[\protect\citeauthoryear{Shapovalova et al.} {2008}]{2008A&A...486...99S}
Shapovalova, A. I., Popovi{\'c} L. {\v C}., Collin, S., Burenkov, A. N. et al., 2008, A\&A,486,  99


\bibitem[\protect\citeauthoryear{Shapovalova et al.} {2010a}]{2010A&A...509A.106S}
Shapovalova, A. I., Popovi{\'c}, L. {\v C}., Burenkov, A. N., Chavushyan, V. H., Ili{\'c}, D., Kova{\v c}evi{\'c}, A. et al., 2010a, A\&A,  509, A106

\bibitem[\protect\citeauthoryear{Shapovalova et al.} {2010b}]{2010A&A...517A..42S}
Shapovalova, A. I., Popovi{\'c}, L. {\v C}., Burenkov, A. N., Chavushyan, V. H., Ili{\'c}, D., Kollatschny, W., Kova{\v c}evi{\'c}, A. et al., 2010b, A\&A, 517, A42

\bibitem[\protect\citeauthoryear{Shapovalova et al.} {2012}]{2012ApJS..202...10S}
Shapovalova, A. I., Popovi{\'c}, L. {\v C}., Burenkov, A. N., Chavushyan, V. H., Ili{\'c}, D., Kova{\v c}evi{\'c}, A, Kollatschny, W., Kova{\v c}evi{\'c},  J. et al., 2012, ApJS, 202, 10

\bibitem[\protect\citeauthoryear{Shapovalova et al.} {2013}]{2013A&A...559A..10S}
Shapovalova, A. I., Popovi{\'c}, L. {\v C}., Burenkov, A. N., Chavushyan, V. H.,  Ilii{\'c},, D., Kollatschny, W.,  Kova{\v c}evi{\'c}, A. et al., 2013, A\&A, 559, A10

\bibitem[\protect\citeauthoryear{Shapovalova et al.} {2016}]{2016ApJS..222...25S}
Shapovalova, A. I., Popovi{\'c}, L. {\v C}, Chavushyan, V. H. Burenkov, A. N., Ili{\'c}, D., Kollatschny, W., Kova{\v c}evi{\'c}, A. et al., 2016, ApJS, 222, id. 25

\bibitem[\protect\citeauthoryear{Shapovalova et al.} {2017}]{2017MNRAS.466.4759S} 
Shapovalova, Alla I, Popovi{\'c}, L. {\v C}., Chavushyan, V. H., Afanasiev, V. L., Ili{\'c},  D., Kova{\v c}evi{\'c}, A. et al., 2017, MNRAS,  466, 4759

\bibitem[\protect\citeauthoryear{Shapovalova et al.} {2019}]{2019MNRAS.485.4790S}
Shapovalova, A. I., Popovi{\'c}, L. {\v C}, Afanasiev, V. L., Ili{\'c}, D.,Kova{\v c}evi{\'c}, A., Burenkov, A. N., Chavushyan, V. H., Mar{\v c}eta-Mandi{\'c}, S. et al., 2019, MNRAS,  485, 4790 

\bibitem[\protect\citeauthoryear{Shen et al.}{2019}]{2019BAAS...51c.274S} Shen, Y., Anderson, S., Berger, E. et al., 2019, BASS, 51, 274


\bibitem[\protect\citeauthoryear{Shen et al.} {2016}]{2016ApJ...818...30S}
Shen, Y., Horne, K., Grier, C. J., Peterson, B. M., Denney, K. D. et al., 2016, ApJ, 818, 30


\bibitem[\protect\citeauthoryear{Shen et al.}{2019}]{2019ApJS..241...34S} Shen, Y., Hall, P. B., Horne, K., et al., 2019, \apjs, 241, 34

\bibitem[\protect\citeauthoryear{Shi \& Krolik} {2015}]{10.1088/0004-637X/807/2/131}
Shi, J.-M., Krolik, J. H., 2015, ApJ, 807, 131 

\bibitem[Simonetti et al. (1985)]{10.1086/163418}
Simonetti, J. H., Cordes, J. M., Heeschen, D. S. 1985, ApJ, 296, 46


\bibitem[Suberlak et al. (2021)]{10.3847/1538-4357/abc698/pdf}
Suberlak, K. L., Ivezi{\'c}, {\v Z}, MacLeod, C. 2021, ApJ, 907, 96



\bibitem[\protect\citeauthoryear{Sun et al.} {2018}]{10.3847/1538-4357/aae208}
Sun, M.,  Xue, Y., Wang, J.,  Cai, Z.,  Guo, H., 2018, ApJ, 866, 74


\bibitem[\protect\citeauthoryear{Wang et al.} {2014}]{2014ApJ...793..108W}
Wang, J.-M., Du, P., Hu, C., Hu, C., Netzer, H. et al., 2014, ApJ, 793, 108


\bibitem[\protect\citeauthoryear{Wang et al.} {2020}]{10.1038/s41550-019-0979-5}
Wang, J., Songsheng, Y., Li, Y. R., Du P., Zhang, Z. X., 2020, Nat. Astron., 4, 517


\bibitem[\protect\citeauthoryear{Wang et al.} {2018}]{10.3847/1538-4357/aab88b}
Wang, J. M., Xu, D., W., Wei, J. Y., 2018, ApJ, 858, 49

\bibitem[\protect\citeauthoryear{Wang et al.} {2017}]{10.1007/s10509-017-3079-y}
Wang, H., Yin, C., Xiang, F., 2017, ApSS, 362, 99  

\bibitem[\protect\citeauthoryear{Woo \& Urry} {2002}]{10.1086/342878}
Woo, J. H, Urry, C. M., 2002, ApJ, 579, 530
 

\bibitem[\protect\citeauthoryear{Vaughan} {2010}]{2010MNRAS.402..307V} Vaughan, S., 2010, \mnras, 402, 307 

\bibitem[\protect\citeauthoryear{Vaughan et al.} {2016}]{10.1093/mnras/stw1412}
Vaughan, S.,  Uttley, P.,  Markowitz,  A. G., Huppenkothen, D. et al., 2016, MNRAS, 461, 3145

{
\bibitem[\protect\citeauthoryear{Vio \& Wamsteker} {2001}]{10.1086/317967}
Vio, R. Wamsteker, W. 2001, PASP, 113, 86}



\end{thebibliography}




\appendix

\section{Example of the LSST rolling cadence}

\begin{figure*}
	\centering
	\includegraphics[clip,trim=60 200 60 40,width=0.8\textwidth]{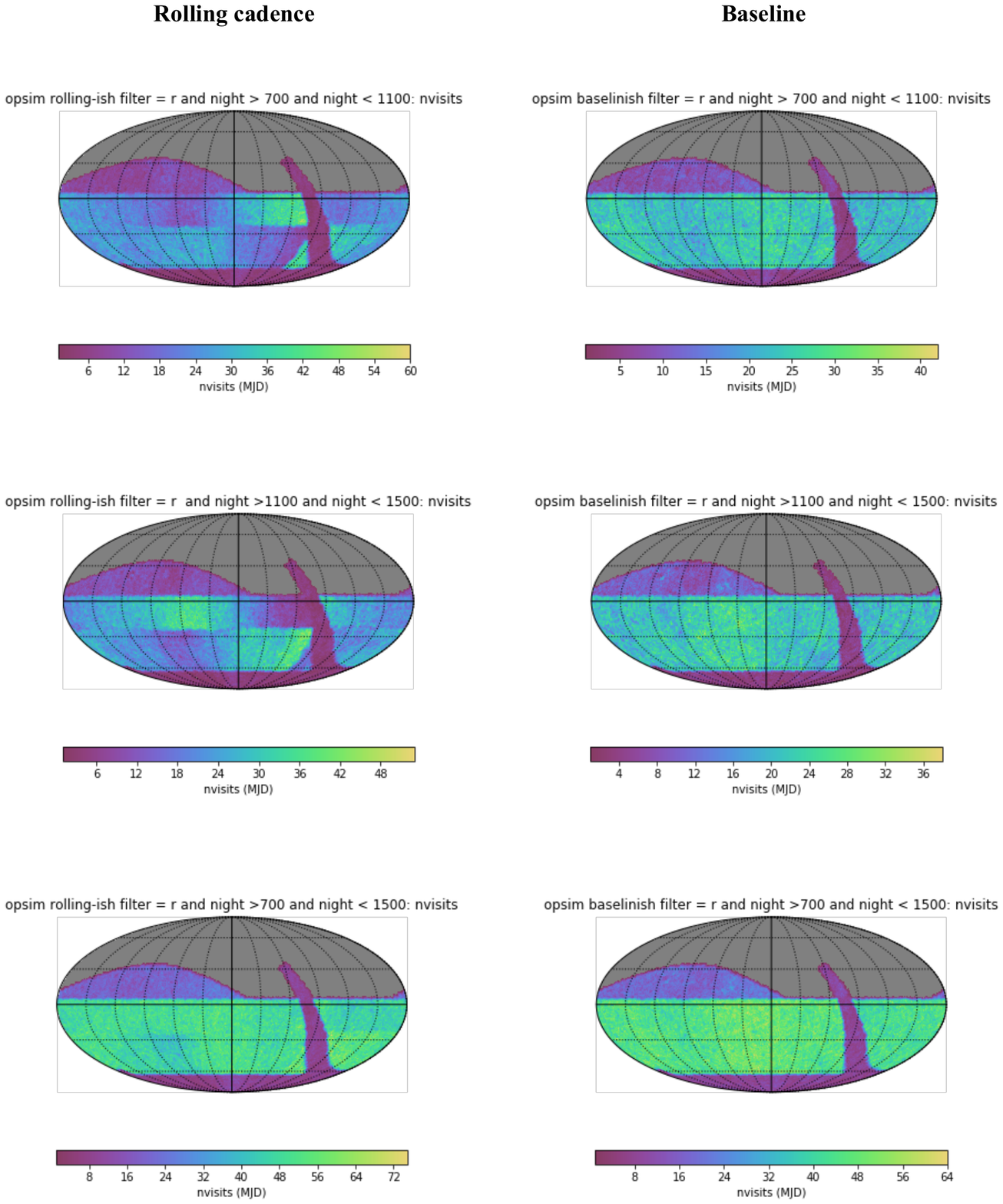}
	\caption{ LSST FBS 1.6 simulations of rolling (left panels) and baseline cadence (right plots). The rolling cadence divides the  WFD ($-62^{\circ}$
< $\delta$ < $2^{\circ}$) into two distinct declination bands which are alternated in different years (top and middle row).}
	\label{fig:rollbase}%
\end{figure*}

Figure \ref{fig:rollbase} compares  the rolling cadence (left plots) and the
baseline cadence (right plots). The simulations were created with FBS 1.6 code \citep{pstn-051.lsst.io/PSTN-051.pdf} which produces smoother rolling cadences.
The shown rolling cadence 
rolling\_fpo\_2nslice1.0  breakes the WFD ($-62^{\circ}$
< $\delta$ <  $2^{\circ}$) into two  declination sectors and then “roll” those areas around the sky.
The rolling weight for this cadence is 99\%.  At the end of the
10 years operations (bottom plots), the sum is very similar to the
baseline survey strategy.

\bsp	
\label{lastpage}
\end{document}